\documentclass{sn-jnl}% Math and Physical 

\usepackage{graphicx}%
\usepackage{multirow}%
\usepackage{amsmath,amssymb,amsfonts}%
\usepackage{amsthm}%
\usepackage{mathrsfs}%
\usepackage[title]{appendix}%
\RequirePackage[dvipsnames]{xcolor}%
\usepackage{textcomp}%
\usepackage{manyfoot}%
\usepackage{booktabs}%
\usepackage{algorithm}%
\usepackage{algorithmicx}%
\usepackage{algpseudocode}%
\usepackage{listings}%

\usepackage{enumerate}
\usepackage{natbib} 
\usepackage{url} 
\usepackage{pdflscape}
\usepackage{setspace}
\usepackage{hyperref}
\usepackage{bbm} 
\usepackage{bm}
\usepackage{mathptmx}
\usepackage{algcompatible}
\usepackage{algpseudocode}
\usepackage{mathtools}
\mathtoolsset{showonlyrefs}
\usepackage{caption}
\usepackage{subcaption}
\usepackage{placeins}

\usepackage{etoolbox}
\usepackage{environ} 
\usepackage{soul}

\NewEnviron{myequation}{%
    \begin{equation}
    \setstretch{1.5}
    \BODY
    \end{equation}
}
\NewEnviron{myequation*}{%
    \begin{equation*}
    \setstretch{1.5}
    \BODY
    \end{equation*}
}

\numberwithin{equation}{section}

\newcommand{\NN}{\mathbb{N}}

\newcommand{\RR}{\mathbb{R}}

\newcommand{\Tau}{\mathrm{T}}
\newcommand\independent{\protect\mathpalette{\protect\independenT}{\perp}}
\def\independenT#1#2{\mathrel{\rlap{$#1#2$}\mkern2mu{#1#2}}}

\theoremstyle{remark}

\theoremstyle{definition}

\begin{document}

\title{ \centering \textbf{Extreme value methods for estimating rare events in Utopia} \\
\; \vspace{-0.5em} \\
\large EVA (2023) Conference Data Challenge: Team Lancopula Utopiversity}

\author[1]{\fnm{Lídia Maria} \sur{André}}

\author[2]{\fnm{Ryan} \sur{Campbell}}

\author[3]{\fnm{Eleanor} \sur{D'Arcy}}

\author[2]{\fnm{Aiden} \sur{Farrell}}

\author[4]{\fnm{Dáire} \sur{Healy}}

\author*[2]{\fnm{Lydia} \sur{Kakampakou}}\email{l.kakampakou1@lancaster.ac.uk}

\author[1]{\fnm{Conor}\sur{Murphy}}

\author[5,6]{\fnm{Callum John Rowlandson} \sur{Murphy-Barltrop}}

\author[1]{\fnm{Matthew} \sur{Speers}}

\affil[1]{\orgdiv{STOR-i Centre for Doctoral Training}, \orgname{Lancaster University}, \orgaddress{\city{Lancaster}, \postcode{LA1 4YR}, \country{United Kingdom}}}

\affil[2]{\orgdiv{Department of Mathematics and Statistics}, \orgname{Lancaster University}, \orgaddress{\city{Lancaster}, \postcode{LA1 4YF}, \country{United Kingdom}}}

\affil[3]{\orgname{Environment Agency}, \orgaddress{\street{Lutra House, Dodd Way Off Seedlee Road, Walton Summit Centre}, \city{Preston}, \postcode{PR5 8BX}, \country{United Kingdom}}}

\affil[4]{\orgdiv{Dipartimento di Scienze Ambientali, Informatica e Statistica}, \orgname{Università Ca’ Foscari Venezia}, \orgaddress{\street{Campus Scientifico, via Torino 155, Mestre} \city{Venezia}, \postcode{30172}, \country{Italia}}}

\affil[5]{\orgdiv{Institut Für Mathematische Stochastik}, \orgname{Technische Universität Dresden}, \orgaddress{\street{Helmholtzstraße 10} \city{Dresden}, \postcode{01069}, \country{Germany}}}

\affil[6]{\orgname{Center for Scalable Data Analytics and Artificial Intelligence (ScaDS.AI)}, \orgaddress{\city{Dresden/Leipzig}, \country{Germany}}}

\abstract{To capture the extremal behaviour of complex environmental phenomena in practice, flexi\-ble techniques for modelling tail behaviour are required. In this paper, we introduce a variety of such methods, which were used by the Lancopula Utopiversity team to tackle the EVA (2023) Conference Data Challenge. This data challenge was split into four challenges, labelled C1-C4. Challenges C1 and C2 comprise univariate problems, where the goal is to estimate extreme quantiles for a non-stationary time series exhibiting several complex features. For these, we propose a flexible modelling technique, based on generalised additive models, with diagnostics indicating generally good performance for the observed data. Challenges C3 and C4 concern multivariate problems where the focus is on estimating joint extremal probabilities. For challenge C3, we propose an extension of available mo\-dels in the multivariate literature and use this framework to estimate extreme probabilities in the presence of non-stationary dependence. Finally, for challenge C4, which concerns a 50 dimensional random vector, we employ a clustering technique to achieve dimension reduction and use a conditional modelling approach to estimate extremal probabilities across independent groups of variables. }




\maketitle
\section{Introduction} \label{sec:main_intro}
% use last data challenge (March '21), editorial
%\CB{Note to all - to avoid confusion when editing, I think we need a better file structure. Perhaps just four folders with Introduction, C1+C2, C3, C4. It's confusing having misc, Univariate, MV etc. This also makes life much much easier when we submit, and I've had a journal complain about a confusing file structure before.} \ED{Done for C1+C2 :)} \CB{Thank you!} \LA{This is done for C3 and C4}

This paper details an approach to the data challenge organised for the Extreme Value Analysis (EVA) 2023 Conference. The objective of the challenge was to estimate extremal probabilities, or their associated quantiles, for simulated environmental data sets for various locations in a fictitious country called Utopia. %, where a year consists of 12 months of 25 days each, for a total of 300 days in a year. 
The data challenge is split into 4 challenges; challenges C1 and C2 focus solely on the univariate setting, where data is obtained from a single location while challenges C3 and C4 concern multivariate data sets, where data is obtained simultaneously from multiple locations.

%The data challenge can be split into 4 challenges; challenges C1 and C2 where focus lays solely on the univariate case using environmental data from a single location, and challenges C3 and C4, where we are concerned with the multivariate case using data at multiple locations to estimate the probabilities of observing data within extreme sets/regions.

% Both univariate and multivariate EVA have been widely applied in the environmental setting; see for instance \citet{Eastoe2009,Keef2013a} and \citet{Jonathan2014b}. Mainly, the concern lies in estimating the frequency of rare events to mitigate against risks. %In many such cases, EVA techniques are used to estimate frequencies of rare, high impact events. In practice, these estimates act as a guide for informed decision-making and risk assessment. 
% In this data challenge, we are interested in estimating extreme quantities for simulated environmental data sets. \LA{I think this paragraph is not needed; seems like we are repeating ourselves. I suggest cutting it or move the first 2 lines to after ``Utopia" in the first paragraph}

Challenge C1 requires estimation of the 0.9999-quantile of the distribution of the environmental response variable $Y$ conditional on a covariate vector $\boldsymbol{X}$, for 100 realisations of covariates. To do so, we model the tail of $Y\mid\boldsymbol{X}=\boldsymbol{x}$ using a generalised Pareto distribution \cite[GPD;][]{Pickands1975} and employ the extreme value generalised additive modelling (EVGAM) framework, first introduced by \cite{Youngman2019}, to account for the non-stationary data structure.
We consider a variety of model formulations and select our final model using cross-validation. Furthermore, central 50\% confidence intervals are estimated via a non-stationary bootstrapping technique, and the final model performance is assessed using the number of times the true conditional quantile lies in the confidence intervals \citep{Rohrbeck2023}.
For Challenge C2, we are interested in estimating the value of $q$ that satisfies $\Pr(Y>q)={1}/{(300T)}$, where $T = 200$. %\AF{expand on this, seems a little rushed.}
% \AF{CB can you check this new wording please?}
% Moreover the government agency has determined that over-estimation of $q$ is preferential to under-estimation. To ensure this is the case, our estimate $\hat{q}$ is obtained via the minimisation of a specific loss function provided by the agency \CB{minimisation of a specific objective function that accounts for the aforementioned scoring criteria?}. Estimates $\hat{q}$ of $q$ are obtained from a data set of observed responses $\boldsymbol{y}_t$ and observed covariates $\boldsymbol{x}_t$ via fitting a GPD with a penalty defined by the provided loss function $\mathcal{L}(q,\hat{q})$ using an EVGAM-formulated model, after which a quantile $q$ can be estimated. 

Challenges C3 and C4 concern the estimation of probabilities for extreme multivariate regions, subsets of $\mathbb{R}^d$, where some or all of the values are so large that we seldom observe any data in them. Such estimates require techniques for modelling and extrapolating within the joint tail. For challenge C3, we want to estimate two joint tail probabilities for three unknown non-stationary environmental variables. 
% Specifically, 
% \begin{align*}
%     &p_1 = \Pr\left(Y_1>y,Y_2>y,Y_3>y\right)\\
%     &p_2 = \Pr\left(Y_1>v,Y_2>v,Y_3<m\right)
% \end{align*}
% for given values $y=6$, $v=7$, $m=-\log(\log 2)$.
To achieve this, we propose a non-stationary extension of the model introduced by \cite{Wadsworth2013}. Lastly, for challenge C4, 
we wish to estimate the probability that 50 variables (locations) jointly exceed prespecified extreme thresholds. 
% This is a very high-dimensional problem to employ classical multivariate extremes techniques; therefore, we calculate this joint probability as a product of lower-dimensional probabilities. This is done by separating groups of variables into near-independent clusters using an empirically estimated extremal dependence coefficient, and then estimating within-cluster probabilities. Each probability in the product is estimated via the conditional extremes method of \cite{Heffernan2004}. \LA{Idk if it reads better this way, I was finding this paragraph a bit confusing} \CB{Think this paragraph could be shortened a bit to something like: ... some extreme thresholds. 
Based on an initial analysis, we separate the variables into five independent groups, and obtain distinct extremal probability estimates for each group using the conditional extremes approach of \citet{Heffernan2004}.% Under the assumption of independence between the groups, we obtain an estimate for the joint probability by multiplying the probability estimates for each group.}

The remainder of the paper is structured as follows. A suitable background to EVA is provided in Section~\ref{subsec::background_EVA}, introducing concepts required throughout our work. Section~\ref{sec:c1andc2} covers our approach to the univariate challenges C1 and C2, and the multivariate challenges C3 and C4 are considered in Sections \ref{sec:c3} and \ref{sec:c4}, respectively. The paper ends with a discussion of the results of all challenges in Section \ref{sec:discussion} 

% Section~\ref{sec:c1andc2} covers all aspects of the univariate challenges C1 and C2; we outline our exploratory data analysis (EDA) and detail our methodology based on EVGAM, introducing tools for model selection and comparison. In Section~\ref{sec:c3} we cover the first multivariate challenge C3. After establishing the presence of non-stationarity in the underlying data through EDA, we detail our extension of the \citet{Wadsworth2013} model, alongside inferential techniques for this extended framework. Given this non-stationarity, an overview of quantile regression and model fitting is presented. Challenge C4 is covered in Section~\ref{sec:c4}. Given the high-dimensional nature of this problem, our data analysis provides the basis for clustering the variables into independent subgroups, and the conditional extremes approach is used to approximate probabilities for each subgroup. The paper ends with a discussion of the results of all challenges.
\section{EVA background}\label{subsec::background_EVA}

\subsection{Univariate modelling}

Univariate EVA methods are concerned with capturing the behaviour of the tail of a distribution which allows for extreme quantities to be estimated. The most common univariate approach is the peaks-over-threshold framework. Consider a continuous, independent and identically distributed (IID) random variable $Y$ with distribution function $F$ and upper endpoint $y^F := \text{sup}\{y : F(y) < 1\}$. \citet{Pickands1975} shows that, for some high threshold $v < y^F$, the excesses $(Y - v) \mid Y>v$, after suitable rescaling, converge in distribution to a GPD as $v \rightarrow y^F$. 
% In practice, this limit is taken to hold exactly for an appropriately chosen high threshold $v$ such that
% \begin{myequation}
%     \Pr(Y > y + v \mid Y > v) = 
%     \begin{cases}
%     \left(1 + \xi y/\sigma\right)_+^{-1/\xi} & \text{if }\xi \neq 0,\\
%    \exp\left(-y/\sigma\right) & \text{if }\xi = 0,
%     \end{cases}
%     \label{eqn:gpd}
% \end{myequation} 
% for $y > 0$, $w_+ = \text{max}(w,0)$, shape parameter $\xi \in \mathbb{R}$ and threshold-dependent scale parameter $\sigma > 0$. Note that the case when $\xi = 0$ is taken in the limit as $\xi \to 0$. We write $(Y - v) \mid Y>v \sim \text{GPD}(\sigma, \xi)$. For $\xi < 0$, the distribution has a finite upper end-point at $v - \sigma/\xi$ but is unbounded above for $\xi \geq 0$. 
\citet{Davison1990} provide an overview of the properties of the GPD, and also propose an extension of this framework to the non-stationary setting: given a non-stationary process $Y$ with associated covariate(s) $\boldsymbol{X}$, the authors propose the following model
\begin{myequation} \label{eqn:gpd_non_stationary}
    \Pr(Y > y + v \mid Y > v, \boldsymbol{X}=\boldsymbol{x}) = \left(1 +  \frac{y\xi(\boldsymbol{x})}{\sigma(\boldsymbol{x})}\right)_+^{-1/\xi(\boldsymbol{x})},
\end{myequation}
for $y>0$, where $\sigma(\cdot), \xi(\cdot)$ are the covariate dependent scale and shape parameters, respectively. Recent extensions of the \citet{Davison1990} framework include allowing the threshold to be covariate-dependent, i.e., $v(\boldsymbol{x})$ \citep{Kysely2010,Northrop2011}, and capturing the covariate functions in a flexible manner using generalised additive models \citep[GAMs;][]{Chavez-Demoulin2005,Youngman2019}.

\subsection{Extremal dependence measures}
In addition to analysing marginal tail behaviours, multivariate EVA methods are concerned with quantifying the dependence between extremes of multiple observations. An important classification of this dependence is obtained through the measure $\chi$~\citep{Joe1997}: given a $d$-dimensional random vector $\boldsymbol{X}$, with $d\geq 2$ and $X_i \sim F$ for all $i \in \{1,\hdots,d\}$,
\begin{myequation} \label{eq:chi}
    \chi(u) := \left(\frac{1}{1-u}\right)\Pr( F(X_1) > u, \dots, F(X_d) > u),
\end{myequation} 
with $u \in [0,1)$. Where the limit exists, we set $\chi := \lim_{u \to 1}\chi(u) \in [0,1]$. When $\chi > 0$, we say that the variables in $\bm{X}$ exhibit asymptotic dependence, i.e., can take their largest values simultaneously, with the strength of dependence increasing as $\chi$ approaches $1$. If $\chi = 0$, the variables cannot all take their largest values together. In particular, for $d=2$, we refer to the case $\chi = 0$ as asymptotic independence.

We also consider the coefficient of tail dependence proposed by \citet{Ledford1996}. Using the formulation given in \citet{Resnick2002}, let 
\begin{myequation}\label{eq:etar}
    \eta(u):=\frac{\log\left(1-u\right)}{\log \Pr\left(F(X_1) > u, \dots, F(X_d) > u\right)},
\end{myequation}
with $u \in [0,1)$. When the limit exists, we set $\eta:=\lim_{u\rightarrow 1}\eta(u) \in (0,1]$. The cases $\eta = 1$ and $\eta < 1$, corres\-pond to $\chi > 0$ and $\chi = 0$, respectively. For $\eta<1$, this coefficient quantifies the form of dependence for random vectors that do not take their largest values simultaneously. 

Since $\chi$ and $\eta$ are limiting values, they are unknown in practice and must be approximated using numerical techniques. Therefore, when quantifying extremal dependence, we appro\-ximate $\chi$ ($\eta$) using empirical estimates of $\chi(u)$ $\big(\eta(u)\big)$ for some high threshold $u$.

\section{Challenges C1 and C2} \label{sec:c1andc2}
Both challenges concern 70 years of daily data for the capital city of Amaurot. Each year has 12 months of 25 days and two seasons (season 1 for months 1-6, and season 2 for months 6-12). Suppose $Y$ is an unknown response variable, and $\boldsymbol{X}=(V_{1},\ldots, V_{8})$ is a vector of covariates, $(V_1, V_2, V_3, V_4)$ denoting unknown environmental variables and $(V_5,V_6,V_7,V_8)$ denoting season, wind direction (radians), wind speed (unknown scale), and atmosphere (recorded monthly), respectively. 

For C1, we build a model for $Y \mid \boldsymbol{X}$ and estimate the 0.9999-quantile, with associated 50\% confidence intervals, for 100 different covariate combinations denoted $\boldsymbol x_i$ for $i \in \{1,\ldots,100\}$. Note $\boldsymbol{x}_i$ are not covariates observed within the data set, but new observations provided by the challenge organisers.

For C2, we estimate the marginal quantile $q$ such that $\Pr(Y>q)=(6\times10)^{-4}$, which corresponds to a once in 200 year event in the IID setting; in particular, $q$ is obtained subject to a predefined loss function. We first estimate the marginal distribution $F_{Y}(y)$ using Monte-Carlo techniques; see for instance,~\cite{Eastoe2009}. Since we have a large sample size, $n=21,000$, it is reasonable to assume that the observed covariate sample is representative of $\boldsymbol{X}$. Thus, we can approximate the marginal distribution $F_Y(y)$ as follows,
\begin{myequation}
    \hat F_{Y}(y)=\int_{\boldsymbol{X}} F_{Y\mid{\boldsymbol{X}}}(y\mid\boldsymbol{x})f_{\boldsymbol{X}}(\boldsymbol{x})\mathrm{d}\boldsymbol{x}\approx \frac{1}{n}\sum\limits_{t=1}^{n} F_{Y_t\mid{\boldsymbol{X}_t}}(y_t\mid\boldsymbol{x}_t).\label{eq:MC_dist}
\end{myequation} 
%\AF{Abrupt introduction of the loss function. Can we introduce this first when we say that we are calculating the quantile} 
To achieve this, we first re-estimate the GPD parameters, now using a penalised log-likelihood which incorporates the loss function provided by the challenge organisers,
\begin{myequation}
\mathcal{L}(q,\hat{q}) = 
    \begin{cases}
        0.9(0.99q - \hat{q})&\text{if }\, 0.99q>\hat{q},\\
        0 &\text{if }\, \left|q-\hat{q}\right|\leq 0.01q,\\
        0.1(\hat{q}-1.01q)&\text{if }\, 1.01q<\hat{q},
    \end{cases}
    \label{eq:loss_fn}
\end{myequation} where $q$ and $\hat{q}$ are the true and estimated marginal quantiles, respectively. This loss function penalises under-estimation more heavily than an over-estimation. 

We conduct the same exploratory data analysis for both challenges given the same covariates are used; this is outlined in Section~\ref{subsec:edac1andc2}. In Section~\ref{section:model_development} we introduce our techniques for modelling $Y \mid \boldsymbol{X}$, which is then used for modelling $Y$ via~\eqref{eq:MC_dist}. Our approach for uncertainty quantification is outlined in Section~\ref{section: uncertaintyC1C2}, and we give our results for both challenges in Section~\ref{section:resultsc1andc2}.

\subsection{Exploratory data analysis}\label{subsec:edac1andc2}
%This section details our exploratory analysis for challenges C1 and C2. 
The environmental response variable $Y_{t},\; t \in \{1,\hdots,n\},$ is independent over time \citep{Rohrbeck2023}, but is affected by the covariate vector $\boldsymbol{X_t}=\{V_{1,t},\ldots,V_{8,t}\}$. However, it is not clear which covariates affect $Y$, and what form these covariate-response relationships take. In what follows, we aim to explore these relationships so we can account for them in our modelling framework. 

%Both challenges concern 70 years of daily data for the city of Amaurot. Note that years have 12 months of 25 days. \textcolor{red}{The environmental response variable $Y$ is denoted $Y_{1,t}$ for site 1 (Amaurot) and day $t=1,\ldots,n$; for the remainder of this section we use $Y_t$ for simplicity as we are only concerned with Amaurot.} We have eight potential explanatory variables \CM{Move to intro}$\bm X_{t}=\{V_{1,t},\ldots,V_{8,t}\}$ where $V_1,\ldots,V_4$ are unnamed covariates that are assumed to be IID, and $V_5,\ldots,V_8$ correspond to season ($S1$ for months $1,\ldots,6$ and $S2$ for months $7,\ldots,12$), wind speed, wind direction and atmosphere. In figures and expressions, we henceforth denote as $S$, $WS$, $WD$ and $A$, respectively. This notation follows from~\cite{Rohrbeck2023}.

\begin{figure}[h]
    \centering
    \includegraphics[width=\textwidth]{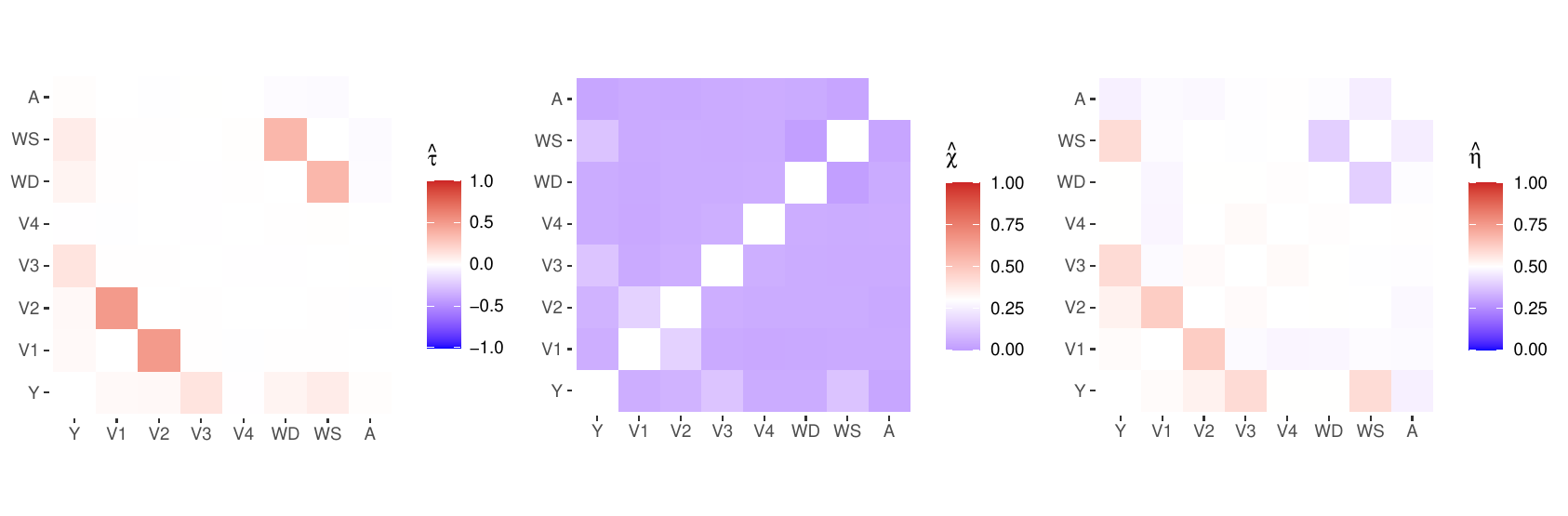}
    \caption{Heat maps for dependence measures for each pair of variables: Kendall's $\tau$ (left), $\chi$ (middle) and $\eta$ (right). Note the scale in each plot varies, depending on the support of the measure, and the diagonals are left blank, where each variable is compared against itself.}
    \label{fig:dep_heat}
\end{figure}

To begin, we explore the dependence between all variables to understand the relationships between covariates, as well the relationships between individual covariates and the response variable. We investigate dependence in the main body of the data using Kendall's $\tau$ measure, while for the joint tails, we use the pairwise extremal dependence coefficients $\chi$ and $\eta$ defined in Section~\ref{subsec::background_EVA}; values for all pairs are shown in Figure~\ref{fig:dep_heat}, with the threshold $u$ set at the empirical 0.95-quantile for the extremal measures. %Recall that for $\chi\in(0,1)$, the two variables are asymptotically dependent (AD) whilst $\chi=1$ corresponds to asymptotic independence (AI). Whilst, $\overline\chi=1$ and $\overline\chi<1$ correspond to asymptotic dependence \AF{AD now we have defined} and asymptotic independence \AF{AI}, respectively. 

The response variable $Y$ has the strongest dependence with $V_3$ in the body of the distribution, followed by $V_6$ (wind speed) then $V_7$ (wind direction). Similarly, $Y$ has strong dependence with $V_2$, $V_3$ and $V_6$ in the tail. We also find strong dependence between $V_6$ and $V_7$ in the body, but evidence of weak dependence in the tail (dark blue for $\hat\chi$ and $\hat\eta$). There is also strong dependence between $V_1$ and $V_2$ in both the body and tail (see dark red for $\hat\eta$). We find very similar dependence relationships when the data are split into seasons. In the Supplementary Material, we show scatter plots of each covariate against the response variable; these demonstrate a highly non-linear relationship for each explanatory variable with $Y$.

%Since $V_6$ and $V_7$ have strong relationships with the response variable $Y$, we explore these variables in more detail. The most notable feature of both variables is that they have a significant shift in distribution. We use the \verb|changepoint| package~\citep{Killick2014} in \verb|R| to estimate the difference in mean for both variables; just 5 observations separate the estimated changepoints, so we assume the changes occur simultaneously. Before the changepoint, winds typically occur in the southwest direction and are greater in magnitude, while after the changepoint, they occur mainly in the northeast direction and are lower in magnitude; see the Supplementary Material.~\citet{Rohrbeck2023} state this non-stationarity was created unintentionally when designing the data challenge.

Next, we explore temporal relationships for the response variable $Y$. We first find temporal non-stationarity as the distribution of $Y$ varies significantly with $V_5$ (season); see the Supplementary Material for more detail. The magnitude of $Y$ is higher for season 1 than season 2, in both the main body and tail of the distribution. However, within each season, across months, there is little temporal variation in the distribution of $Y$. We also find that $Y$ exhibits temporal independence at all lags, with auto-correlation function (acf) values close to zero; see the Supplementary Material. 

%We also explore temporal dependence in the covariates.~\cite{Rohrbeck2023} states that $V_1,\ldots,V_4$ are temporally independent. We find that for $V_6$ and $V_7$, the acf remains significantly different from zero at all time lags (see the Supplementary Material) due to the changepoint discussed earlier. Finally, $V_8$ has high acf values at the earliest lags that decrease rapidly until lag 25 (i.e., the length of a month) and then continue to decrease at a slower rate until $\sim50$ time lags. This is because atmosphere is recorded monthly.

As noted in \citet{Rohrbeck2023}, 11.7\% of the observations have at least one possible predictor variable missing completely at random (MCAR). A detailed breakdown of the pattern of missing predictor observations is provided in the Supplementary Material. Since we can assume the data are MCAR, ignoring the observations that have a missing predictor covariate will not bias our inference, however, a complete case analysis is undesirable due to the amount of data loss. To mitigate against this, we attempted to impute the observations where predictors are missing but ultimately could not find an imputation method that satisfactorily retained the dependence structure between the response and potential covariates, particularly in the tails of the distribution. Therefore, we use a case analysis approach, whereby an observation is only removed if a predictor covariate of interest is missing. This results in only 4\% of observations being removed for our final model.

\subsection{Methods}\label{section:model_development}

Due to the complex nature of the data, we consider various non-stationary GPD models, as in equation~\eqref{eqn:gpd_non_stationary}, that are formulated as GAMs to fit $Y\mid\boldsymbol{X}$. For threshold selection, we extend the method proposed by \citet{murphy2024automated} to select a threshold for non-stationary, covariate-dependent GPD models; the details are provided in Section~\ref{subsubsec:general_methods}. Our inference and model selection procedures are then provided in Sections~\ref{subsubsec:inf} and~\ref{subsubsec:model_sel}, respectively. We note that the same model formulation is used for both C1 and C2 with a small adjustment to the parameter estimation procedure for C2 in order to incorporate the provided loss function given in~\eqref{eq:loss_fn}. We utilise equation~\eqref{eq:MC_dist} to obtain the marginal distribution of $Y$.

\subsubsection{General model formulation}
\label{subsubsec:general_methods}
Let $\Tilde{\boldsymbol{X}_t}$ denote the set of predictor covariates, for all $t \in \{1,\hdots,n\}$. Then $y_{t}$ and $\Tilde{\boldsymbol{x}}_t$ denote the observations of the response variable and predictive covariates, respectively. We consider models with the following form, 
\begin{myequation}
    F_{Y_t|\Tilde{\boldsymbol{X}}_t}(y_{t} | \Tilde{\boldsymbol{X}}_t = \Tilde{\boldsymbol{x}}_t) = 1-\lambda(\Tilde{\boldsymbol{x}}_t)\left[1+\xi(\Tilde{\boldsymbol{x}}_t)\left(\frac{y_{t}-v(\Tilde{\boldsymbol{x}}_t)}{\sigma(\Tilde{\boldsymbol{x}_t})}\right)\right]_+^{-1/\xi(\Tilde{\boldsymbol{x}}_t)}, \label{eq:C1_model}
\end{myequation} where $v(\Tilde{\boldsymbol{x}}_t)$ and $\lambda(\Tilde{\boldsymbol{x}}_t)$ are a covariate dependent threshold and rate parameter, respectively. 

Our analysis in Section \ref{subsec:edac1andc2} indicates that $V_{3}$, $V_{5}$ (season), and $V_{6}$ (wind speed) exhibit non-trivial dependence relationships with the response variable. Therefore we assume these variables can be used as predictor variables for modelling $Y$, and set $\Tilde{\boldsymbol{X}} := (\boldsymbol{V}_{j})_{j\in \{3,5,6\}}$. Although $V_{7}$ (wind direction) also exhibits predictor power, we do not consider it here since it is highly correlated with wind speed so would involve adding complex interaction terms to the model formulation, and $V_6$ has a stronger relationship with $Y$ compared to $V_7$ (see Figure~\ref{fig:dep_heat}). %Even after accounting for seasonal variability in the threshold, we still explore including seasonal variability, as well as other covariate dependence, in the scale function formulation for the remainder of the section. 

Owing to the complex covariate structure observed in the data, as described in Section \ref{subsec:edac1andc2}, we employ the flexible EVGAM framework proposed in \citet{Youngman2019} for modelling GPD tail behaviour. Under this framework, GAM formulations are used to capture non-stationarity in the threshold, scale and shape functions introduced given in equation \eqref{eq:C1_model}. Without loss of generality, consider the scale function $\sigma(\boldsymbol{x})$. We assume that 
\begin{myequation}
    h(\sigma(\boldsymbol{x})) = \psi_\sigma(\boldsymbol{x}), \quad
\text{with} \quad    \psi_{\sigma}(\boldsymbol{x}) = \beta_0 + \sum\limits_{\kappa=1}^{K}\sum\limits_{p=1}^{P_{\kappa}} \beta_{\kappa p}b_{\kappa p} (\boldsymbol{x}),
    \label{eqn:general_gam_form}
\end{myequation}
where $h(x) := \log(x)$ denotes the link function which ensures the correct support, with coefficients $\beta_0,\beta_{\kappa p} \in \RR$ and basis functions $b_{\kappa p}$ for $p \in \{1, \hdots, P_{\kappa}\}, \kappa \in \{1, \hdots, K\}$. The basis functions can be in terms of individual covariates, i.e., $b_{\kappa p}:\RR \mapsto \RR$, or multiple covariates, i.e., $b_{\kappa p}:\RR^m \mapsto \RR$, $1 < m \leq 8$. Analogous forms can be taken for $v(\boldsymbol{x})$ and $\xi(\boldsymbol{x})$, adjusting the link function $h$ as appropriate, although these are not considered here for reasons detailed below.

To select an appropriate threshold, we employ the threshold selection method of \citet{murphy2024automated} and extend this approach to select a threshold for non-stationary, covariate-dependent GPD models. The method selects a threshold based on minimising the expected quantile discrepancy (EQD) between the sample quantiles and fitted GPD model quantiles. When fitting a non-stationary model, the excesses will not be identically distributed across covariates. Thus, to utilise the EQD method in this case, we use the fitted non-stationary GPD parameter estimates to transform the excesses to common standard exponential margins and compare sample quantiles against theoretical quantiles from the standard exponential distribution. 

We use a stepped-threshold according to season as there is clear variation in  the distribution, and thereby the extremes, of $Y$ between seasons; see the Supplementary Material for more details. Specifically, we set $v(\Tilde{\boldsymbol{x}}_t):= \mathbbm{1}(\Tilde{x}_{2,t} = 1)v_1 + \mathbbm{1}(\Tilde{x}_{2,t} = 2)v_2$, $v_1,v_2 \in \RR$, with corresponding rate parameter $\lambda(\Tilde{\boldsymbol{x}}_t):= \mathbbm{1}(\Tilde{x}_{2,t} = 1)\lambda_1 + \mathbbm{1}(\Tilde{x}_{2,t} = 2)\lambda_2$, where $\lambda_1, \lambda_2 \in [0,1]$ denote the non-exceedance probabilities for seasons $1$ and $2$, respectively, and $\Tilde{x}_{r,t}$ are realisations of the $r^{\text{th}}$ component of $\Tilde{\boldsymbol{X}}$ for $r \in \{1,2,3\}$. This seasonal threshold significantly improves model fits; see the Supplementary Material for further details. GAM forms for the threshold were also explored, but did not offer significant improvement. Furthermore, the smooth GAM formulation of the GPD scale parameter adequately captures any residual variation in the response arising due to covariate dependence. 

\subsubsection{Inference}\label{subsubsec:inf}
For all GAM formulations, we only consider basis functions of singular covariates, since specifying basis functions of multiple variables requires a detailed understanding of covariate interactions and can significantly increase the computational complexity of the modelling procedure \citep{Wood2017}. We keep the shape function $\xi(\boldsymbol{x}):=\xi \in \RR$ constant across covariates; this is common in non-stationary analyses, since this parameter is difficult to estimate \citep{Chavez-Demoulin2005}. Within the GAM formulation, we consider several parametric forms to account for the predictive covariates in the scale parameter using linear models, indicator functions and splines. 

When using splines, we are required to select a basis dimension $B \in \NN$; this determines the number of coefficients to be estimated. Basis dimension is the most important choice within spline modelling procedures and directly corresponds with the flexibility of the framework \citep{Wood2017}. We only consider splines for $V_3$ and $V_6$. To determine $B$ for each $\Tilde{X}_{r}$, $r \in \{1,3\}$, we first build a model for $Y_t \mid \Tilde{X}_{r,t}$, allowing us to consider the effect of this predictor on the response directly. We vary the basis dimension, and compare the resulting models using cross validation (CV), detailed in the following section. We set $B=4$ and $B=3$ for $V_3$ and $V_6$, respectively.

For C2, we incorporate the loss function of equation~\eqref{eq:loss_fn} into the estimation procedure. Let $\mathcal{I}_v := \{ t \in \{1,\hdots,n\} \mid y_t > v(\Tilde{\boldsymbol{x}}_t) \}$ denote the set of temporal indices corresponding to threshold exceedances and $n_v := | \mathcal{I}_v |$. We consider the objective function
\begin{myequation}
    S(\boldsymbol{\theta}) := - l_R(\boldsymbol{\theta}) + \sum\limits_{i\in\mathcal{I}_v} \mathcal{L}(q_i^*, \hat{q}_i)/n_v,
    \label{eq:penalised_likelihood}
\end{myequation}
where $l_R(\boldsymbol{\theta})$ denotes the penalised log-likelihood function of the restricted maximum likelihood estimation (REML)  approach \citep{Wood2017}, $\boldsymbol{\theta}$ denotes the parameter vector associated with the GPD formulation of equation~\eqref{eqn:general_gam_form}, and $\sum_{i\in\mathcal{I}_v} \mathcal{L}(q_i^*, \hat{q}_i)/n_v$ denotes the average loss between the sample quantiles of the transformed excesses and the theoretical standard exponential quantiles. Specifically, we transform the excesses, $(y_{t} - v(\Tilde{\boldsymbol{x}}_t))_{t \in \mathcal{I}_v}$, to standard exponential margins using the fitted non-stationary GPD parameter estimates and compare the ordered excesses, $\bm{q}^*$, to the theoretical quantiles, $\hat{\bm{q}}$, from a standard exponential distribution evaluated at probabilities $\{p_i = i/(n_v + 1), i=1, \ldots, n_v\}$. Minimising the objective function $S(\boldsymbol{\theta})$ ensures that the parameter estimates also account for and minimise the loss function, $\mathcal{L}$. We use this formulation to adjust the GPD parameters for challenge C2 once a threshold is selected.

%if we denote the mapping between $\mathcal{I}_v$ and the order statistics of $(y_{t} - v(\Tilde{\boldsymbol{x}}_t))_{t \in \mathcal{I}_v}$ by $\pi,$ then $q_i^*$ is the $\pi(i)$\textsuperscript{th} order statistic of $(y_{t} - v(\Tilde{\boldsymbol{x}}_t))_{t \in \mathcal{I}_v}$ and $\hat{q}_i =  \sigma(\bm{x}_{\pi(i)})[\{ 1 - \pi(i)/(n_v + 1)  \}^{-\xi}-1]/\xi$.
%Minimising the objective function $S(\boldsymbol{\theta})$ ensures that the parameter estimates also account for and minimise the loss function, $\mathcal{L}$. We use this formulation to adjust the GPD parameters for challenge C2 once a threshold is selected.

\subsubsection{Model selection}
\label{subsubsec:model_sel}
To determine the best-fitting model, we use a forward selection process and aim to minimise the model's CV score. For each model, we apply $k$-fold CV \citep[Ch 7.]{Hastie2008} utilising the continuous ranked probability score \citep[CRPS,][]{Gneiting2014} as our goodness-of-fit metric. CRPS describes the discrepancy between the predicted distribution function and observed values without the specification of empirical quantiles. We explore model ranking by taking both $k = 10$ and $50$, and find that both give an equivalent ranking; we present results for the latter. We also provide the Akaike Information Criterion (AIC) and Bayesian Information Criterion (BIC) values to aid in model selection. A subset of models used in the forward selection process are detailed in Table \ref{tab:C1_Models} where, for each model, we provide the change in the CRPS, AIC and BIC relative to model 1. The parameterisation of model 7 achieves the largest reduction for all three metrics relative to the baseline model.

\begin{table}[h!]
\caption{Table of selected models considered for challenge C1. $\mathbbm{1}(\cdot)$ denotes an indicator function, $s_{i}(\cdot)$ for $i \in \{1,2\}$ denote thin-plate regression splines, $\beta_0,\beta_1$ are coefficients to be estimated, and $\tilde{{x}}_{r,t}$ is defined as in the text. All values have been given to one decimal place.}
\centering
\vspace{0.5em}
\begin{tabular}{l|l|l|l|l}
\toprule
Model & $\sigma(\tilde{\boldsymbol{x}}_{t})$ & $\Delta$CRPS & $\Delta$AIC & $\Delta$BIC \\
\hline
1 & $\beta_{0}$ & 0 & 0 & 0 \\
\hline
2 & $\beta_{0} + \beta_{1} \mathbbm{1}(\tilde{x}_{2,t} = 1)$ & -0.5 & -33.4 & -26.1 \\
\hline
3 & $\beta_{0} + s_{1}(\tilde{x}_{1,t})$ & -0.9 & -408.5 & -379.2 \\
\hline
4 & $\beta_{0} + s_{2}(\tilde{x}_{3,t})$ & -0.5 & -284.3 & -276.8 \\
\hline
5 & $\beta_{0} + \beta_{1} \mathbbm{1}(\tilde{x}_{2,t} = 1) + s_{1}(\tilde{x}_{1,t})$ & -0.9 & -425.8 & -388.1 \\
\hline
6 & $\beta_{0} + s_{1}(\tilde{x}_{1,t}) + s_{2}(\tilde{x}_{3,t})$ & -1.0 & -752.7 & -717.2 \\
\hline
7 & $\beta_{0} + \beta_{1} \mathbbm{1}(\tilde{x}_{2,t} = 1) + s_{1}(\tilde{x}_{1,t}) + s_{2}(\tilde{x}_{3,t})$ & \textbf{-1.1} & \textbf{-780.0} & \textbf{-735.3} \\
\bottomrule 
\end{tabular}

\label{tab:C1_Models}
\end{table}

\subsection{Uncertainty}\label{section: uncertaintyC1C2}

For each of the 100 different covariate combinations, $\boldsymbol x_i$ for $i\in\{1,\ldots,100\}$, we need to construct central 50\% confidence intervals. We use a bootstrapping procedure to avoid making potentially inaccurate assumptions such as the asymptotic normality approximation of maximum likelihood estimates for example. Traditional bootstrap approaches are non-parametric and randomly resample the data with replacement. However, in Section~\ref{subsec:edac1andc2} we find that the response variable is dependent on covariates, and these covariates exhibit temporal dependence. A standard bootstrap procedure would therefore not retain this dependence. Instead, we preserve the temporal dependence structure of covariates and their relationship with the response variable by approximating our confidence intervals using the stationary, semi-parametric bootstrapping procedure adopted by~\cite{DArcy2023}.

First, the response variable $Y_t$ is transformed to Uniform(0,1) margins; denote this sequence $U_t^Y=F_{Y_t|\tilde{\boldsymbol X}_t}(Y_t|\tilde{\boldsymbol{X}}_t=\boldsymbol \tilde x_t)$ where $F_{Y_t|\tilde{\boldsymbol X}_t}$ is the estimated model given in equation~\eqref{eq:C1_model}. We then adopt the stationary bootstrap procedure of~\cite{Politis1994} to retain the temporal dependence in the response and explanatory variables. The block length $L$ is simulated from a Geometric$(1/l)$ distribution, where the mean block length $l\in\mathbb{N}$ is carefully selected based on the autocorrelation function. This was selected at 50 days, the maximum lag for which the autocorrelation was significant across all variables; see the Supplementary Material. Denote this bootstrapped sequence on Uniform margins by $U_t^B$. We transform  $U_t^B$ back to the original scale using our fitted model, preserving the original structure of $Y_t$; we denote this series $Y_t^B$. Then we fit our model to $Y_t^B$ to re-estimate all of the parameters and thus the quantile of interest. We repeat this procedure to obtain 200 bootstrap samples. %\LA{do we need to say why we chose 200 samples?}

\subsection{Results}\label{section:resultsc1andc2}

\begin{figure}
     \centering
     \begin{subfigure}[h]{0.45\textwidth}
         \centering
         \includegraphics[width=\textwidth]{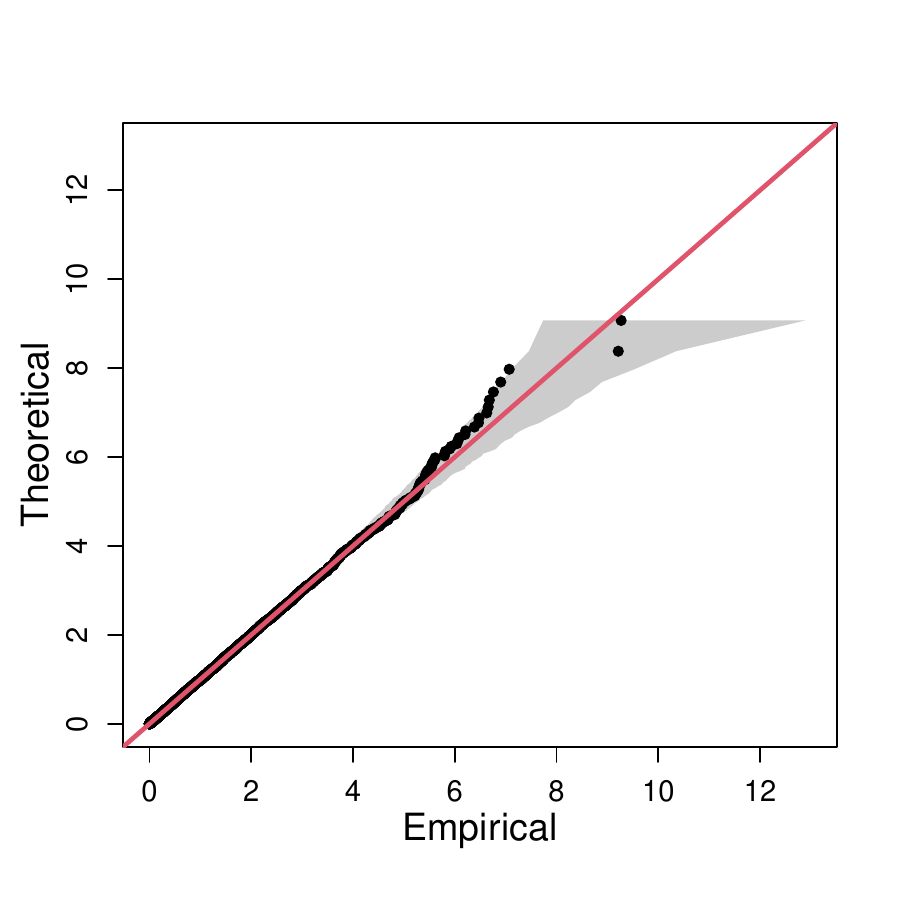}
         \label{fig:231120_QQ_plot_Model_13}
     \end{subfigure}
     \begin{subfigure}[h]{0.45\textwidth}
         \centering
         \includegraphics[width=\textwidth]{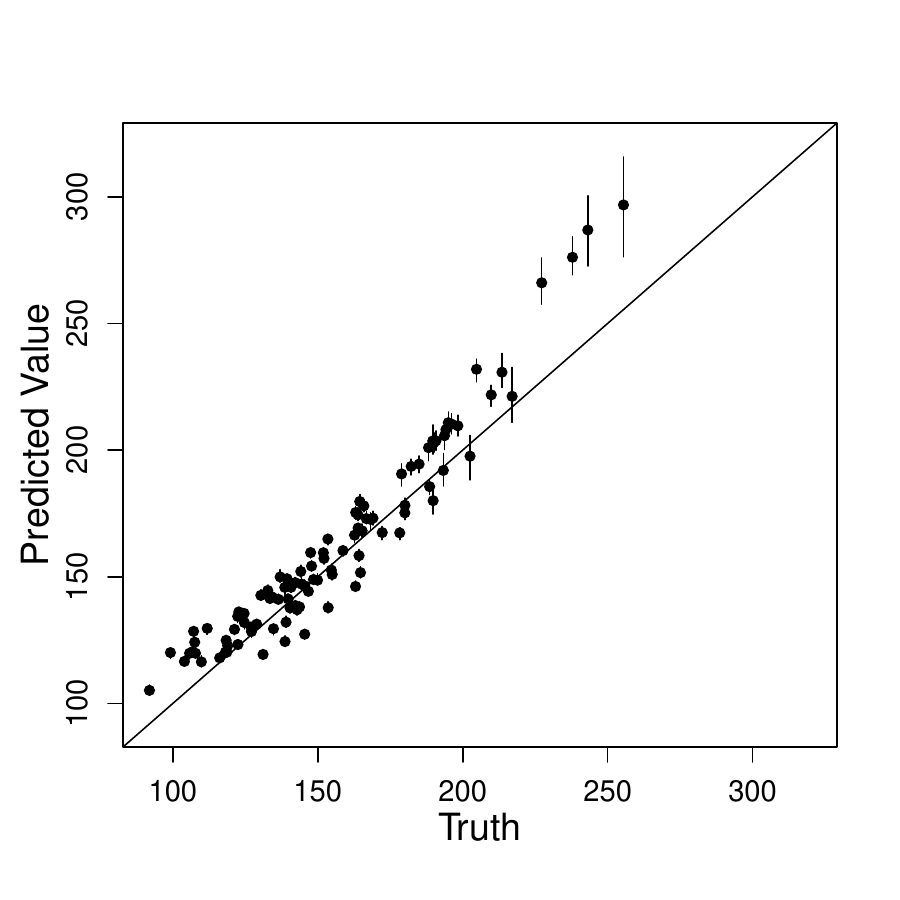}
         \label{fig:231123_C1_Results}
     \end{subfigure}
     \caption{QQ plot for our final model (model 7 in Table \ref{tab:C1_Models}) on standard exponential margins. The $y=x$ line is given in red and the grey region represents the 95\% tolerance bounds (left). Predicted $0.9999-$quantiles against true quantiles for the 100 covariate combinations. The points are the median predicted quantile over 200 bootstrapped samples and the vertical error bars are the corresponding 50\% confidence intervals. The $y = x$ line is also shown (right).} 
    \label{fig:C1_Results}
\end{figure}

For C1, we use our final model of Section~\ref{subsubsec:model_sel} to estimate the 0.9999-quantile of $Y\mid\tilde{\boldsymbol{X}} = \tilde{\boldsymbol{x}_{i}}$, $i \in \{1,\hdots,100\}$, for the set of 100 covariate combinations. The left panel of Figure \ref{fig:C1_Results} shows the quantile-quantile (QQ) plot for our model. There is general alignment between the model and empirical quantiles; however, there is some over-estimation in the upper tail, and our 95\% tolerance bounds do not contain some of the most extreme response values. The right panel of Figure \ref{fig:C1_Results} shows our predicted quantiles, and their associated confidence intervals, compared to their true quantiles. As expected, our predictions tend to over-estimate the true quantiles. We note this figure is different from the one presented by \citet{Rohrbeck2023} due to an error in our code being fixed after submission. In this scenario, our estimated confidence intervals lead to a $14\%$ coverage of the true quantiles, which does not alter our ranking for this challenge. Our performance and model improvements are discussed in Section \ref{sec:discussion}.

For challenge C2, we estimate the quantile of interest as %$\hat q= 212.6 \; (208.4, 246.1)$ 
$\hat{q} = 213.1 \; (209.3, 242.1)$. A 95\% confidence interval for the estimate is given in parentheses based on the bootstrapping procedure outlined in Section~\ref{subsubsec:general_methods}. Due to a coding error, this value differs from the original estimate submitted for the EVA (2023) Conference Data Challenge. The updated value over-estimates compared to the truth ($q=196.6$).  %Our estimate overestimates the true quantile and ranks 4$^{\text{th}}$ for the C2 subchallenge~\citep{Rohrbeck2023}.
% \input{Sections/C1C2/intro}

% \subsection{Exploratory data analysis}\label{subsec:edac1andc2}
% \input{Sections/C1C2/exploratory_analysis}

% \subsection{Methods}\label{section:model_development}

% \AF{I think this section might need a restructure with some subsections. Let's discuss in the meeting.}

% \input{Sections/C1C2/model_development}

% %\subsection{Model selection}\label{section:model_select}
% \input{Sections/C1C2/model_selection}

% \input{Sections/C1C2/penalised_likelihood}

% \subsection{Uncertainty quantification}\label{section:uncertainty}

% %\AF{The reviewer has asked for a little more justification of this method. ED can you advise? reviewers specific questions: What are the advantages and disadvantages of this procedure?; Are there alternative methods for constructing these confidence intervals?; and interesting to see whether the ranking can be improved if a different approach for confidence interval construction is used.}

% \input{Sections/C1C2/bootstrapping}

% \subsection{Results} \label{section:resultsc1andc2}
% \input{Sections/C1C2/results}

\section{Challenge C3} \label{sec:c3}
\subsection{Exploratory data analysis} \label{subsec:edac3}

For challenge C3, we are provided with 70 years of daily data of an environmental variable for three towns on the island of Coputopia. These data are denoted by $Y_{i,t}$, $i \in \{1,2,3\}$, $t \in \{1,\hdots,n\}$, where $i$ is the index of each town and $t$ is the point in time. Each year consists of 12 months, each lasting $25$ days, resulting in $n=21,000$ observations for each location. 

% We let $\mathcal{M}_t\in \{1,\hdots,12\}$ and $\mathcal{Y}_t \in \{1,\hdots,70\}$ denote the month and year, respectively, for all $t\in \{1,2,\hdots,n\}$. \CB{If the statement below is worth cutting, we don't need this bit}

We are also provided with daily covariate observations $\boldsymbol{X}_t = (S_t,A_t)$, where $S_t$ and $A_t$ denote seasonal and atmospheric conditions, respectively. Season is a binary variable, taking values in the set $\{1,2\}$, with each year of observations exhibiting both seasons for exactly $150$ consecutive days. Atmospheric conditions are piecewise constant over months, with large variation in the observed values between months. A descriptive figure of both covariates is given in the Supplementary Material. 

% , i.e., $F_{i,t}(x) = \exp(-\exp(-x))$, $x \in \RR$ for all $i \in \{1,2,3\}$, $t \in \{1,\hdots,n\}$ vary between different months, while staying constant within a fixed month, i.e., given any $a \in \{1,\hdots,12\}$ and $b \in \{1,\hdots,70\}, \; A_t$ is constant for $t \in \{ t \mid \mathcal{M}_t = a, \mathcal{Y}_t = b\}$.As such, $A_t$ is discrete in nature, but also highly variable between months.

In \citet{Rohrbeck2023}, we are informed that $Y_{i,t}$ are distributed identically across all sites and over time, with standard Gumbel margins. However, it is not known whether the covariates $\boldsymbol{X}_t$ influence the dependence structure of $\boldsymbol{Y}_t := (Y_{1,t},Y_{2,t},Y_{3,t})$. We are also informed that, conditioned on covariates, the process is independent over time, i.e., $(\boldsymbol{Y}_t \mid \boldsymbol{X}_t)\independent (\boldsymbol{Y}_{t'} \mid \boldsymbol{X}_{t'})$ for any $t \neq t'$. In this section, we examine what influence, if any, the covariate process $\boldsymbol{X}_t$ may have on the dependence structure of $\boldsymbol{Y}_t$. 
% If such relationships exist, they need to be accounted for when estimating joint tail probabilities. 

%We begin by transforming the time series to standard exponential margins via $Z_{i,t} := -\log (1-F(Y_{i,t}))$, $i \in \{1,2,3\}$, $t \in \{1,\hdots,n\}$, where $F(y) = \exp(-\exp(-y))$, $y \in \RR$. 
We begin by transforming the time series $Y_{i,t}$ to standard exponential margins, denoted by $\boldsymbol{Z}_{i,t}$, via the probability integral transform. This transformation is common in the study of multivariate extremes and can simplify the description of extremal dependence \citep{Keef2013a}. To explore the extremal dependence in the Coputopia time series, we consider all 2- and 3-dimensional subvectors of the process, i.e., $\{ Z_{i,t}, i \in I, t \in \{1, \hdots, n\}\}, \; I \in \mathcal{I}:= \{ \{1,2\}, \{ 1,3\} , \{ 2,3\} , \{ 1,2,3\} \}.$ This separation is important to ensure the overall dependence structure is fully understood, since intermediate scenarios can exist where a random vector exhibits $\chi=0$, but $\chi>0$ for some 2-dimensional subvector(s) \citep{Simpson2020}. 
% Moreover, we know through Sklar's theorem \citep{Sklar1959} that such transformation do not alter the underlying dependence structure. The choice of exponential margins is therefore one of convenience. 

%An important classification of extremal dependence is obtained through the measure $\chi$ \citep{Joe1997}. For any d-dimensional random vector $\boldsymbol{X}$ on exponential margins, with $d\geq 2$, consider the probability 
% \begin{equation} \label{eq:chi}
%     \chi(u) := \frac{\Pr( X_i > u, i = 1, \hdots, d)}{\exp(-u)}.
% \end{equation}
% Where this limit exists, we have $\chi := \lim_{u \to \infty}\chi(u)$. When $\chi > 0$, we say that the variables exhibit asymptotic dependence, i.e., can take their largest values simultaneously. If $\chi = 0$, the variables cannot all take their largest values together. Specifically for $d=2$, we refer to the case $\chi = 0$ as asymptotic independence.

Furthermore, to explore the impact of covariates on the dependence structure, we partition the time series into subsets using the covariates. For the seasonal covariate, let $G^S_{I,j} := \{ Z_{i,t}, i \in I, S_t = j \}$ for $j = 1,2,$ and for the atmospheric covariate, let $\pi : \{1,\hdots,n\} \to \{1,\hdots,n\}$ denote the permutation associated with the order statistics of $A_t$, defined so that ties in the data are accounted for. We then split the data into $10$ equally sized subsets corresponding to the atmospheric order statistics, i.e., $G^A_{I,k} := \left\{ Z_{i,t}, i \in I, t \in \Sigma^k \right\}$ for $k = 1,2,\hdots,10,$ where $\Sigma^k := \{ t \mid (k-1)n/10 + 1 \leq \pi(t) \leq kn/10 \}$. Thus, the atmospheric values associated with each subset $G^A_{I,k}$ will increase over $k$. %, each of size $n/10$,%, i.e., $\pi(t)$ gives the ranking of $A_t$ in the set $\{A_t \mid t = 1, \hdots, n\}$ and satisfies $\pi(t) < \pi(t')$ when $A_t=A_{t'}$, $t<t'$. \LA{I'm with Aiden, I will cut from i.e} 

\begin{figure}[h!]
    \centering
    \includegraphics[width=.8\textwidth]{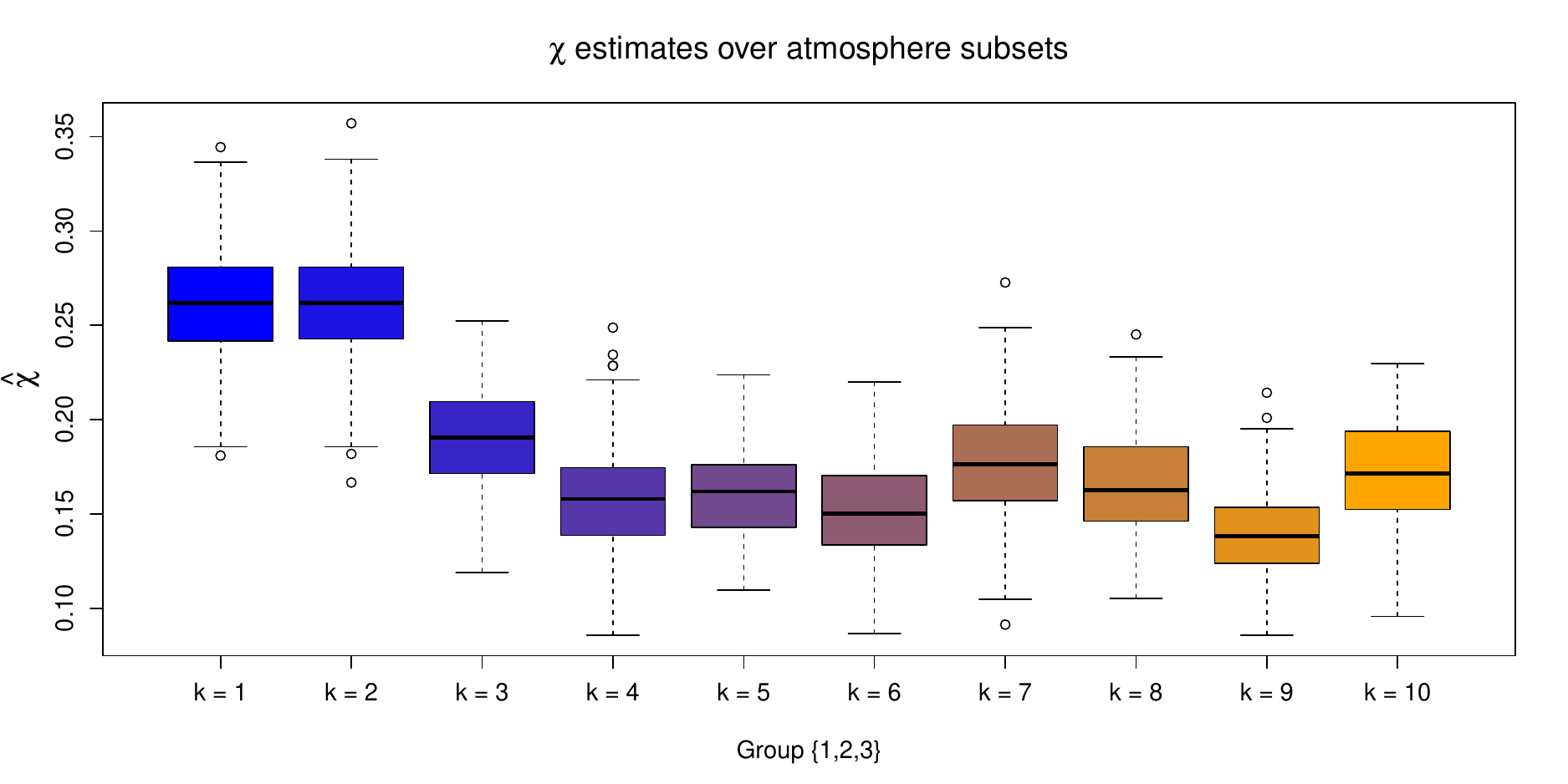}
    \caption{Boxplots of empirical $\chi$ estimates obtained for the subsets $G^A_{I,k}$, with $k = 1, \hdots, 10$ and $I=\{1,2,3\}$. The colour transition (from blue to orange) over $k$ illustrates the trend in $\chi$ estimates as the atmospheric values are increased.}
    \label{fig:c3_chi_estimates_all}
\end{figure}

The idea behind these subsets is to examine whether altering the values of either covariate impacts the extremal dependence structure. Consequently,  we set $u=0.9$ and estimate $\chi(u)$ using the techniques outlined in Section \ref{subsec::background_EVA}, with uncertainty quantified through bootstrapping with $200$ samples. The bootstrapped $\chi$ estimates for $G^A_{I,k}$ with $I = \{1,2,3\}$ are given in Figure \ref{fig:c3_chi_estimates_all}. The plots for the remaining index sets in $\mathcal{I}$, along with the subsets associated with the seasonal covariate, are given in the Supplementary Material. The estimates of $\chi$ appear to vary, in the majority of cases, across both subset types (seasonal and atmospheric), suggesting both covariates have an impact on the dependence structure. For the atmospheric process in particular, the values of $\chi$ tend to decrease for higher atmospheric values, suggesting a negative association between the strength of positive extremal dependence and atmosphere. We also observe that across all subsets, $\chi$ appears consistently low in magnitude, suggesting the extremes of some, if not all, of the sub-vectors are unlikely to occur simultaneously. As such, for modelling the Coputopia time series, we require a framework that can capture such forms of dependence. 

% see \citet{Murphy-Barltrop2023a} for a detailed example. %\LA{do we need to say why we chose 200 samples?} we consider the measure $\chi$ for each subset in turn. However, since $\chi$ is a limiting value, it is unknown in practice and must be approximated using numerical techniques. We therefore approximate $\chi$ using empirical estimates of $\chi(u)$

% In unreported results, we also obtained consistent findings when this analysis was repeated using an alternative dependence measure $\eta\in (0,1]$ \citep{Ledford1996,Ledford1997}. \LA{Could alternatively put the $\eta$ results in the SM, but maybe not necessary} Consequently, for C3, it was decided to only consider multivariate modelling frameworks that are practicably applicable for $\chi=0$; see \citet{Ledford1996} and \citet{Heffernan2004} for further discussion. 

We also consider pointwise estimates of the function $\lambda$, as defined later in equation \eqref{eq:wadsworthtawn}, over $G^S_{I,j}$ and $G^A_{I,k}$ for fixed simplex points; these results are given in the Supplementary Material. Similar to $\chi$, estimates of $\lambda$ vary significantly across subsets, providing additional evidence of non-stationarity within extremal dependence structure.

\subsection{Modelling of joint tail probabilities under asymptotic independence} \label{subsec:depmeasures}
%\CB{I think this name sounds a little bit weird - maybe we should change to something like `Modelling joint tail probabilities under asymptotic independence'?} \LA{done!}

%\CB{I think this section will need a re-structure - my gut feeling is that it would be better to first recall the probabilities we are required to estimate, then showing how both can be written as a joint survivor probability - it feels weird having this stuff at the end of the section since it provides motivation for the approach. After, it's then easier to introduce/motivate the models that we use, as well as the model extension.} \LA{I guess what is confusing me with this suggestion, is the results section, I feel like we either just give the point estimates there and the section looks a bit naked, or we just don't have a results section for C3. Idk} \CB{OK so how about if we have the MC computation of the probability, followed by our estimates? It will still be short, but I don't think there's anything wrong with that. We shouldn't try to make it arbitrarily longer given how tight we are for space:)}

%\LA{If anyone reads this before Thursday pls ignore my english. I'm writing as I'm thinking and will rewrite it better before the meeting. Thanks}

%The region of interest for $p_1$ comprises all the variables being large together, whereas for $p_2$ we are only concerned when two of the variables are extreme simultaneously. 
For challenge C3, we are required to estimate probabilities $p_1:=\Pr\left(Y_1>y, Y_2>y, Y_3>y\right)$ and $p_2:=\Pr\left(Y_1>v, Y_2>v, Y_3<m\right)$, with $y=6$, $v = 7$ and $m = -\log(\log(2))$. Note that $p_1$ and $p_2$ are independent of the covariate process and correspond to different extremal regions in $\mathbb{R}^3$; we refer to $p_1$ and $p_2$ as parts 1 and 2 of the challenge, respectively. For the remainder of this section we will consider the transformed exponential variables $(Z_1,Z_2,Z_3)$, omitting the subscript $t$ for ease of notation. Observe that $F_{(-Z_3)}(z)=e^z,$ for $z<0;$ setting $\tilde{Z}_3\coloneqq -\log\left(1-\exp\{-Z_3\}\right),$ we have
\begin{myequation*}
    p_2 = \Pr\left(Z_1>\tilde v, Z_2>\tilde v, Z_3<\tilde m\right)=\Pr\left(Z_1>\tilde v, Z_2>\tilde v, \tilde{Z}_3>\tilde m\right),
\end{myequation*} 
% \LA{which one do you guys prefer? I took one of the probabilities in the first one}
% \begin{myalign*}
%     p_2 &=& \Pr\left(Z_1>\tilde v, Z_2>\tilde v, Z_3<\tilde m\right)=\Pr\left(Z_1>\tilde v, Z_2>\tilde v, -Z_3>-\tilde m\right) \\
%     &=&\Pr\left(Z_1>\tilde v, Z_2>\tilde v, \tilde{Z}_3>\tilde m\right),
% \end{myalign*} 
where $\tilde v$ and $\tilde m$ denote the values $v$ and $m$ transformed to the standard exponential scale, e.g., $\tilde v:=-\log\left(1-\exp\{-\exp\{-v\}\}\right)$. Similarly, we have $p_1 = \Pr\left(Z_1>\tilde y, Z_2>\tilde y, Z_3>\tilde y\right)$. Consequently, both $p_1$ and $p_2$ can be considered as joint survivor probabilities. 

Since not all extremes of $Z_1$, $Z_2$ and $Z_3$ are observed simultaneously, we
employ the framework by \citet{Wadsworth2013}, which is a generalisation of the approach proposed in \citet{Ledford1996}.
The model of \citet{Wadsworth2013} assumes that for any ray $\boldsymbol{\omega}\in \boldsymbol{S}^2:= \left\{(w_1,w_2,w_3)\in [0,1]^3: w_1+w_2 +w_3=1\right\},$ where $\boldsymbol{S}^2$ denotes the standard 2-dimensional simplex, 
% \begin{myalign}
%     \text{Pr}\left(Z_1\slash w_1>r,Z_2\slash w_2>r, Z_3\slash w_3>r\right)&=&\text{Pr}\left(\min\{Z_1\slash w_1,Z_2\slash w_2, Z_3\slash w_3\}>r\right) \nonumber \\
%     &=&\mathcal{L}(e^r;\boldsymbol{\omega})e^{-r\lambda(\boldsymbol{\omega})}, \label{eq:wadsworthtawn}
% \end{myalign}
\begin{myequation}
\begin{array}{rcl}
    \text{Pr}\left(Z_1 >w_1r,\,Z_2 >w_2r,\, Z_3 >w_3r\right)&=&\text{Pr}\left(\min\{Z_1\slash w_1,\,Z_2\slash w_2,\, Z_3\slash w_3\}>r\right) \\
    &=&\mathcal{L}(e^r;\boldsymbol{\omega})e^{-r\lambda(\boldsymbol{\omega})}, \label{eq:wadsworthtawn}
\end{array}    
\end{myequation}
% \begin{equation*}
%     \text{Pr}\left(Z_1\slash w_1>r,Z_2\slash w_2>r, Z_3\slash w_3>r\right)=\text{Pr}\left(\min\{Z_1\slash w_1,Z_2\slash w_2, Z_3\slash w_3\}>r\right) =\mathcal{L}(e^r;\boldsymbol{\omega})e^{-r\lambda(\boldsymbol{\omega})}, 
% \end{equation*}
% \CB{One one line, we have no room for the equation number}
as $r \to \infty$, where $\lambda(\boldsymbol{\omega}) \geq \max(\boldsymbol{\omega})$ is known as the angular dependence function (ADF). Asymptotic dependence occurs at the lower bound, i.e., $\lambda(\boldsymbol{\omega}) = \max(\boldsymbol{\omega})$ for all $\boldsymbol{\omega} \in \boldsymbol{S}^2$, and the coefficient of tail dependence is related to the ADF via $\eta = 1/\{3\lambda(1/3,1/3,1/3)\}$. In practice, equation \eqref{eq:wadsworthtawn} can be used to evaluate extreme joint survivor probabilities; in particular, probabilities $p_1$ and $p_2$ can be identified with the rays $\boldsymbol{\omega}^{(1)}:=(\tilde{u}, \tilde{u},\tilde{u})/r^{(1)}$ and $\boldsymbol{\omega}^{(2)}:=(\tilde{v}, \tilde{v},\tilde{m})/r^{(2)}$ in $\boldsymbol{S}^2$, respectively, where $r^{(1)} := \tilde{u}+\tilde{u}+\tilde{u}$ and $r^{(2)} := \tilde{v}+\tilde{v}+\tilde{m}$. See Section \ref{subsec:resultsc3} for further details.

\subsection{Accounting for non-stationary dependence} \label{subsec:nonstat}

% \subsection{EVGAM and tuning parameter selection} \label{subsec:evgammult}

% \subsubsection{Non-stationary modelling of joint tail probabilities}
In the stationary setting, pointwise estimates of $\lambda$ can be obtained via the Hill estimator \citep{Hill1975}, from which tail probabilities can be approximated. However, alternative procedures are required for data exhibiting trends in dependence, such as the Coputopia data set. Existing approaches for capturing non-stationary dependence structures are sparse in the extremes literature, and most approaches are limited to asymptotically dependent data structures. For the case when data are not asymptotically dependent, \citet{Mhalla2019} and \citet{Murphy-Barltrop2022} propose non-stationary extensions of the \citet{Wadsworth2013} framework, while \citet{Jonathan2014b} and \citet{Guerrero2021} propose non-stationary extensions of the \citet{Heffernan2004} model (see \citet{Murphy-Barltrop2022} for a detailed review).

To account for non-stationary dependence in C3, we propose an extension of the \citet{Wadsworth2013} framework. With $\boldsymbol{Z}_t = (Z_{1,t},Z_{2,t},Z_{3,t})$ and $\boldsymbol{X}_t$, defined as in Section \ref{subsec:edac3}, we define the structure variable $T_{\boldsymbol{\omega},t} := \min\{Z_{1,t}/w_1,Z_{2,t}/w_2,Z_{3,t}/w_3\}$,
for any $\boldsymbol{\omega} \in \boldsymbol{S}^2$; we refer to $T_{\boldsymbol{\omega},t}$ as the min-projection variable at time $t$. From Section \ref{subsec:edac3}, we know that the joint distribution of $\boldsymbol{Z}_t$ is not identically distributed over $t$; which implies non-stationarity in the distribution of $T_{\boldsymbol{\omega},t}$. To account for this, \citet{Mhalla2019} and \citet{Murphy-Barltrop2022} assume the following model
\begin{myequation} \label{eqn:ns_wt_model}
\operatorname{Pr}\left( T_{\boldsymbol{\omega},t}>u \mid \boldsymbol{X}_t=\boldsymbol{x}_t\right)=\mathcal{L}\left(e^u \mid \boldsymbol{\omega}, \boldsymbol{x}_t\right) e^{-\lambda\left(\boldsymbol{\omega} \mid \boldsymbol{x}_t\right) u} \text { as } u \rightarrow \infty,
\end{myequation}
for all $t$. %, where $\mathcal{L}(\cdot)$ denotes a slowly varying function. 
Note that this assumption is very similar in form to equation \eqref{eq:wadsworthtawn}, with the primary difference being the function $\lambda$ is non-stationary over $t$. From equation \eqref{eqn:ns_wt_model}, we have 
\begin{myequation} \label{eqn:wt_model_exp} 
\operatorname{Pr}\left( T_{\boldsymbol{\omega},t}-u>z \mid T_{\boldsymbol{\omega},t}>u, \boldsymbol{X}_t=\boldsymbol{x}_t\right)= e^{-\lambda\left(\boldsymbol{\omega} \mid \boldsymbol{x}_t\right) z} \text { as } u \rightarrow \infty, 
\end{myequation}
for $z>0$. Consequently, equation \eqref{eqn:ns_wt_model} is equivalent to assuming $(T_{\boldsymbol{\omega},t} - u) \mid \{ T_{\boldsymbol{\omega},t} > u,\boldsymbol{X}_t=\boldsymbol{x}_t\} \sim \text{Exp}(\lambda\left(\boldsymbol{\omega} \mid \boldsymbol{x}_t\right))$ as $u\to \infty$. 

% In practice, equation \eqref{eqn:ns_wt_model} has been successfully applied for estimating non-stationary joint survivor probabilities in the bivariate setting \citep{Mhalla2019a,Murphy-Barltrop2022}. For such analyses, one first requires a means to estimate $\lambda\left(\boldsymbol{\omega} \mid \boldsymbol{z}_t\right)$ over all $t \in \{1,\hdots,n\}$. Subsequently, one can evaluate joint survivor probabilities for any ray $\boldsymbol{\omega} \in \boldsymbol{S}^2$. To capture non-stationarity for C3, we propose an extension of the framework in equation \eqref{eqn:ns_wt_model}. This extension was motivated by the fact equation \eqref{eqn:ns_wt_model} was not flexible enough to capture the tail of $T_{\boldsymbol{\omega},t}$ for the Coputopia data set. The extended model is as follows: given any $z>0$ and a fixed $\boldsymbol{\omega} \in \boldsymbol{S}^2$, we assume 

We found that equation \eqref{eqn:ns_wt_model} was not flexible enough to capture the tail of $T_{\boldsymbol{\omega},t}$ for the Coputopia data; see Section \ref{subsubsec:diagnostics_C3} for further discussion. Thus, we propose the following model: given any $z>0$ and a fixed $\boldsymbol{\omega} \in \boldsymbol{S}^2$, we assume 
\begin{myequation} \label{eqn:wt_model_gpd} 
\operatorname{Pr}\left( T_{\boldsymbol{\omega},t}-u>z \mid T_{\boldsymbol{\omega},t}>u, \boldsymbol{X}_t=\boldsymbol{x}_t\right)= \left( 1 + \frac{\xi\left(\boldsymbol{\omega} \mid \boldsymbol{x}_t\right) z}{\sigma\left(\boldsymbol{\omega} \mid \boldsymbol{x}_t\right)} \right)^{-1/\xi\left(\boldsymbol{\omega} \mid \boldsymbol{x}_t\right)} \text { as } u \rightarrow \infty.
\end{myequation}
This is equivalent to assuming $(T_{\boldsymbol{\omega},t} - u) \mid \{ T_{\boldsymbol{\omega},t} > u,\boldsymbol{X}_t=\boldsymbol{x}_t\} \sim \text{GPD}(\sigma\left(\boldsymbol{\omega} \mid \boldsymbol{x}_t\right),\,\xi\left(\boldsymbol{\omega} \mid \boldsymbol{x}_t\right))$ as $u\to \infty$, and equation \eqref{eqn:wt_model_exp} is recovered by taking the limit as $\xi\left(\boldsymbol{\omega} \mid \boldsymbol{x}_t\right) \to 0$ for all $t$. 
%\pagebreak

Our proposed formulation in equation \eqref{eqn:wt_model_gpd} allows for additional flexibility within the modelling framework by including a GPD shape parameter $\xi\left(\boldsymbol{\omega} \mid \boldsymbol{x}_t\right)$, which quantifies the tail behaviour of $T_{\boldsymbol{\omega},t}$. Given the wide range of distributions in the domain of attraction of a GPD \citep{Pickands1975}, it is reasonable to assume that the tail of $T_{\boldsymbol{\omega},t}$ can be approximated by equation \eqref{eqn:wt_model_gpd}. For the Coputopia time series, this assumption appears valid, as demonstrated by the diagnostics in Section \ref{subsubsec:diagnostics_C3}.  

% However, when applied to the trivariate Coputopia dataset, the resulting diagnostics indicated poor quality model fits.  In particular, the assumption of an exponential tail for the conditioned variable $(T_{\boldsymbol{\omega},t} - u) \mid \{ T_{\boldsymbol{\omega},t} > u,\boldsymbol{Z}_t=\boldsymbol{z}_t\}$ did not appear to be valid for the two rays of interest.

\subsubsection{Model fitting}\label{subsubsec:modelfit_c3}

To apply equation \eqref{eqn:wt_model_gpd}, we first fix $\boldsymbol{\omega} \in \boldsymbol{S}^2$ and assume that the formulation holds approximately for some sufficiently high threshold level from the distribution of $T_{\boldsymbol{\omega},t}$; we denote the corresponding quantile level by $\tau \in (0,1)$. For simplicity, the same quantile level is considered across all $t$. Further, let $v_{\tau}(\boldsymbol{\omega},\boldsymbol{x}_t)$ denote the corresponding threshold function, i.e., $\Pr(T_{\boldsymbol{\omega},t} \leq v_{\tau}(\boldsymbol{\omega},\boldsymbol{x}_t) \mid \boldsymbol{X}_t=\boldsymbol{x}_t) = \tau$ for all $t$. Under our assumption, we have $(T_{\boldsymbol{\omega},t} - v_{\tau}(\boldsymbol{\omega},\boldsymbol{x}_t)) \mid \{ T_{\boldsymbol{\omega},t} > v_{\tau}(\boldsymbol{\omega},\boldsymbol{x}_t),\boldsymbol{X}_t=\boldsymbol{x}_t\} \sim \text{GPD}(\sigma\left(\boldsymbol{\omega} \mid \boldsymbol{x}_t\right),\xi\left(\boldsymbol{\omega} \mid \boldsymbol{x}_t\right))$. We emphasise that $v_{\tau}(\boldsymbol{\omega},\boldsymbol{x}_t)$ is not constant in $t$, and we would generally expect $v_{\tau}(\boldsymbol{\omega},\boldsymbol{x}_t) \neq v_{\tau}(\boldsymbol{\omega},\boldsymbol{x}_{t'})$ for $t \neq t'$. 

% By our assumption, we have that $(T_{\boldsymbol{\omega},t} - v_{\tau}(\boldsymbol{\omega},\boldsymbol{x}_t)) \mid \{ T_{\boldsymbol{\omega},t} > v_{\tau}(\boldsymbol{\omega},\boldsymbol{x}_t),\boldsymbol{X}_t=\boldsymbol{x}_t\} \sim \text{GPD}(\sigma\left(\boldsymbol{\omega} \mid \boldsymbol{x}_t\right),\xi\left(\boldsymbol{\omega} \mid \boldsymbol{x}_t\right))$, thus allowing us to associate a likelihood function for observed threshold exceedances.
% , i.e., $\{ T_{\boldsymbol{\omega},t} - v_{\tau}(\boldsymbol{\omega},\boldsymbol{x}_t) \mid t \in \{1,\hdots,n\}, T_{\boldsymbol{\omega},t} > v_{\tau}(\boldsymbol{\omega},\boldsymbol{x}_t)\}$. 

As detailed in Section \ref{subsec:depmeasures}, both $p_1$ and $p_2$ can be associated with points on the simplex $\boldsymbol{S}^2$, denoted by $\boldsymbol{\omega}^{(1)}$ and $\boldsymbol{\omega}^{(2)}$, respectively. Letting $\boldsymbol{\omega} \in \{\boldsymbol{\omega}^{(1)},\boldsymbol{\omega}^{(2)}\}$, our estimation procedure consists of two stages: estimation of the threshold function $v_{\tau}(\boldsymbol{\omega},\boldsymbol{z}_t)$ for a fixed $\tau \in (0,1)$, followed by estimation of GPD parameter functions $\sigma\left(\boldsymbol{\omega} \mid \boldsymbol{x}_t\right),\xi\left(\boldsymbol{\omega} \mid \boldsymbol{x}_t\right)$. For both steps, we take a similar approach to Section \ref{section:model_development} and use GAMs to capture these covariate relationships. To simplify our approach, we falsely assume that the atmospheric covariate $A_t$ is continuous over $t$; this step allows us to utilise GAM formulations containing smooth basis functions. Given the significant variability in $A_t$ between months, discrete formulations for this covariate would significantly increase the number of model parameters and result in higher variability. 
% In practice, we found this step to significantly increase the flexibility of the modelling procedure. Furthermore, , treating this covariate as continuous is not unreasonable. 
% In general, GAMs provide flexible functional forms that allow us to capture multiple covariate interactions. Moreover, both discrete and continuous covariates can be incorporated in the GAM framework. \CB{This sentence should probably be incorporated well before this when we introduce GAMs}

% Without loss of generality, we fix $\boldsymbol{\omega} = \boldsymbol{\omega}^{(1)}$ for the remainder of this section, noting that the methodology for $\boldsymbol{\omega}^{(2)}$ is identical. 

% \CB{Big change don't set to one or other....}
% To estimate the required probability, denoted $p_1$, we must first estimate the GPD parameter functions $\sigma\left(\boldsymbol{\omega} \mid \boldsymbol{z}_t\right),\xi\left(\boldsymbol{\omega} \mid \boldsymbol{z}_t\right)$.  

Let $\log( v_{\tau}(\boldsymbol{\omega},\boldsymbol{x}_t)) = \psi_v(\boldsymbol{x}_t)$, $\log( \sigma\left(\boldsymbol{\omega} \mid \boldsymbol{x}_t\right)) = \psi_{\sigma}(\boldsymbol{x}_t)$ and $\xi\left(\boldsymbol{\omega} \mid \boldsymbol{x}_t\right) = \psi_{\xi}(\boldsymbol{x}_t)$ denote the GAM formulations of each function, where $\psi_{-}$ denotes the basis representation of equation \eqref{eqn:general_gam_form}. Exact forms of basis functions are specified in Section \ref{subsubsec:diagnostics_C3}. As in Section \ref{section:model_development}, model fitting is carried out using the \verb|evgam| software package \citep{Youngman2022}. For the first stage, $v_{\tau}(\boldsymbol{\omega},\boldsymbol{x}_t)$ is estimated by exploiting a link between the loss function typically used for quantile regression and the asymmetric Laplace distribution \citep{Yu2001}. The spline coefficients associated with $\psi_{\sigma}$ and $\psi_{\xi}$ are estimated subsequently using the obtained threshold exceedances. 

% allowing us to estimate the coefficients associated with $\psi_v$ in a relatively fast and objective manner \CB{Have kept this comment as I think it's important to justify our quantile regression procedure}
% For the first stage, $v_{\tau}(\boldsymbol{\omega},\boldsymbol{x}_t)$ is estimated using a likelihood-based inference procedure which exploits a link between the titled loss function, typically used in quantile regression, and the asymmetric Laplace distribution \citep{Yu2001}, allowing one to estimate the coefficients associated with $\psi_v$ in a relatively fast and objective manner. Estimation of $\psi_v$ is achieved via restricted maximum likelihood estimation (REML), with subsequent estimation of $\psi_{\sigma}$ and $\psi_{\xi}$ is also carried out in a similar manner. In general, REML schemes avoid over-fitting through penalisation of the likelihood function, as noted in Section \ref{section:model_development} \CB{Not removing this stuff yet, we just need to make sure it's covered when GAMs are first introduced}. Furthermore, formulation via likelihood functions avoids the use of MCMC, which can be computationally expensive; see \citet{Wood2017} and \citet{Youngman2019} for further details. 

\subsubsection{Selection of GAM formulations and diagnostics} \label{subsubsec:diagnostics_C3}

Prior to estimation of the threshold and parameter functions, we specify a quantile level $\tau$ and formulations for each of the GAMs. To begin, we fix $\tau=0.9$ and consider a  variety of formulations for each $\psi_v, \psi_{\sigma}$ and $\psi_{\xi}$. By comparing metrics for model selection, namely AIC, BIC and CRPS, we found the following formulations to be sufficient
% \begin{align*}
%     \psi_v(\boldsymbol{z}_t) &= \beta_u + s_v(a_t) + \beta_s \mathbbm{1}(s_t = 2), \\
%     \psi_{\sigma}(\boldsymbol{z}_t) &= \beta_{\sigma} + s_{\sigma}(a_t),\\
%     \psi_{\xi}(\boldsymbol{z}_t) &= \beta_{\xi},
% \end{align*}
\begin{myequation}\label{eqn:C3_gam_formulations}
    \psi_v(\boldsymbol{x}_t) = \beta_u + s_v(a_t) + \beta_s \mathbbm{1}(s_t = 2), \quad  \psi_{\sigma}(\boldsymbol{x}_t) = \beta_{\sigma} + s_{\sigma}(a_t)\quad \text{and}\quad \psi_{\xi}(\boldsymbol{x}_t) = \beta_{\xi},
\end{myequation}
for parts 1 and 2, where $\beta_{u}, \beta_{\sigma}, \beta_{\xi} \in \RR$ denote constant intercept terms, $\mathbbm{1}$ denotes the indicator function with corresponding coefficient $\beta_s \in \RR$, and $s_{u},s_{\sigma}$ denote cubic regression splines of dimension $B = 10$. The shape parameter is set to constant for the reasons outlined in Section \ref{subsubsec:model_sel}. Cubic basis functions are used for $\psi_v$ and $\psi_{\sigma}$ since they have several desirable properties, including continuity and smoothness \citep{Wood2017}. Setting $B = 10$ appears more than sufficient to capture the trends relating to the atmosphere variable. Alternative formulations were tested for both parts, but this made little difference to the resulting model fits. 
% Moreover, the fact the same model selection appeared suitable for both parts of C3 provides evidence of robustness for these GAM formulations. 

% cubic basis functions are selected for $\psi_v$ and $\psi_{\sigma}$ since they have several desirable properties, such as optimality in various respects, continuity and smoothness \citep{Wood2017}. Setting $B = 10$ appears more than sufficient to capture trends over the atmosphere variable. Alternative smooth splines and basis dimensions were tested for both parts 1 and 2, but this made little difference to the resulting model fits. Moreover, the fact the same model selection appeared suitable for both parts of C3 provides evidence of robustness for these GAM formulations. 

% As noted in \citet{Wood2017}, the most important choice here is the basis dimension; this corresponds with the flexibility of the resulting formulation. In practice, it is better to select a higher dimension than one would expect to be necessary, since the REML scheme will adjust estimates of coefficients to avoid over-fitting. Furthermore, as long as the basis dimension is sufficiently high, the locations of knots has little impact on the resulting model fits. \CB{Merge with Conor's} This is commonplace in many non-stationary modelling procedures \citep[e.g.][]{Eastoe2009,Murphy-Barltrop2022}, since the shape parameter is notoriously difficult to estimate \CB{Could we move this/cut this, since it should be already explained in C1/2}.with knots placed at equally spaced quantiles of the atmosphere variable.

We remark that the seasonal covariate is only present with the formulation for $\psi_v$. Once accounted for in the non-stationary threshold, the seasonal covariate appeared to have little influence on the fitted GPD parameters. More complex GAM formulations were tested involving interaction terms between the seasonal and atmospheric covariates, which showed little to no improvement in model fits. Thus, we prefer the simpler formulations on the basis of parsimony. 

With GAM formulations selected, we now consider the quantile level $\tau \in (0,1)$. To assess sensitivity in our formulation, we set $\Tau:= \{0.8,0.81,\hdots,0.99\}$ and fit the GAMs outlined in equation \eqref{eqn:C3_gam_formulations} for each $\tau \in \Tau$. Letting $\delta_{\boldsymbol{\omega},t}$ and $\mathcal{T}_{\tau}:=\{ t \in \{1,\hdots,n\} \mid \delta_{\boldsymbol{\omega},t} > v_{\tau}(\boldsymbol{\omega},\boldsymbol{x}_t)\}$ denote the min-projection observations and indices of threshold-exceeding observations, respectively, we expect the set $\mathcal{E}:=\{ -\log\left\{ 1 - F_{GPD}( \delta_{\boldsymbol{\omega},t} - v_{\tau}(\boldsymbol{\omega},\boldsymbol{x}_t)) \mid \sigma\left(\boldsymbol{\omega} \mid \boldsymbol{x}_t\right),\xi\left(\boldsymbol{\omega} \mid \boldsymbol{x}_t\right) \right\} \mid t \in \mathcal{T}_{\tau}\}$ to follow a standard exponential distribution.

\begin{figure}[h!]
    \centering
    \includegraphics[width=.8\textwidth]{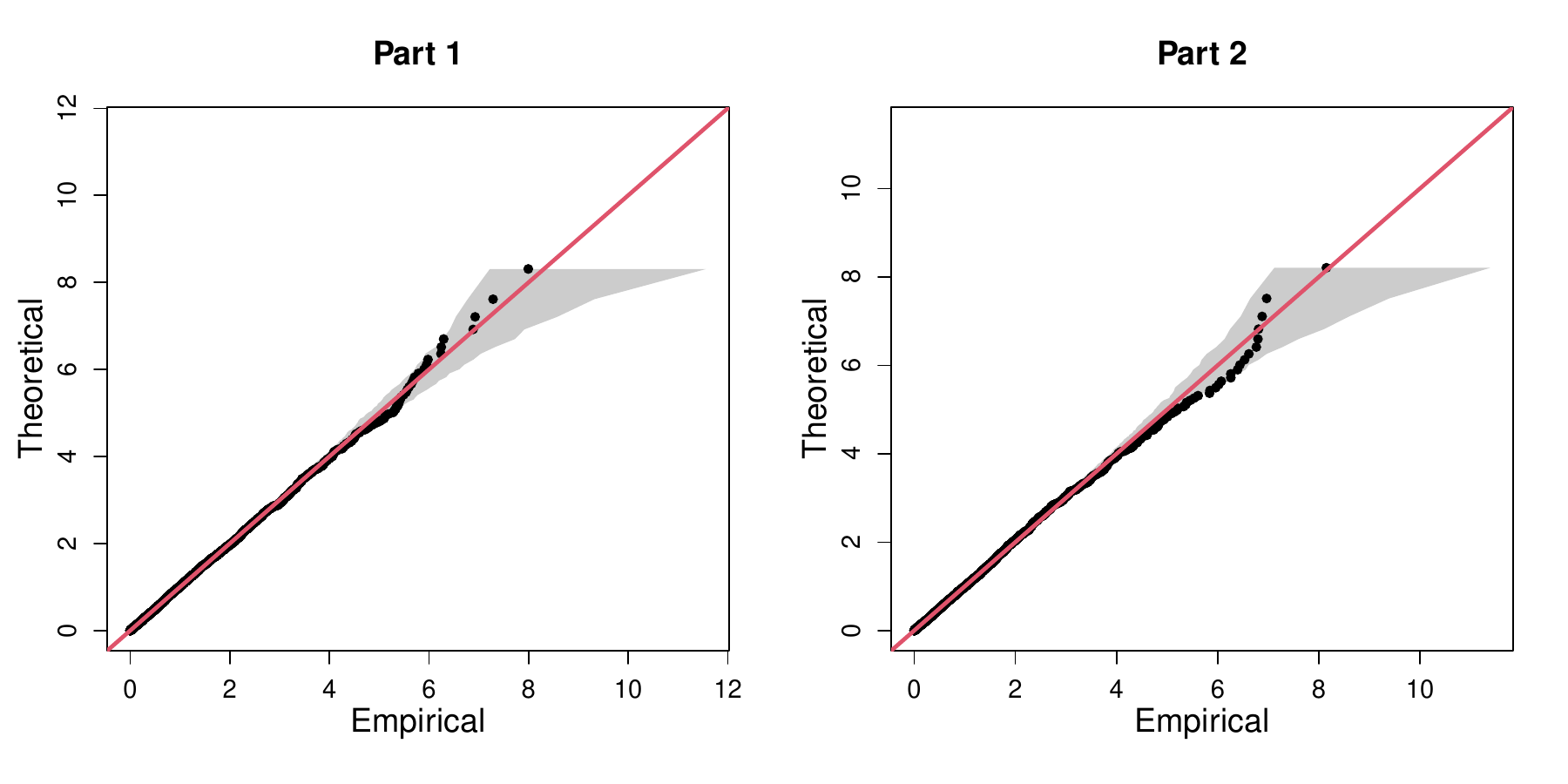}
    \caption{Final QQ plots for parts 1 (left) and 2 (right) of C3, with the $y=x$ line given in red. In both cases, the grey regions represent the 95\% bootstrapped tolerance bounds.}
    \label{fig:c3_evgam_qqplots}
\end{figure}

With all exceedances transformed to a unified scale, we compare the empirical and model exponential quantiles using QQ plots, through which we assess the relative performance of each $\tau \in \Tau$. We then selected $\tau$ values for which the empirical and theoretical quantiles appeared most similar in magnitude. From this analysis, we set $\tau = 0.83$ and $\tau = 0.85$ for parts 1 and 2, respectively. The corresponding QQ plots are given in Figure \ref{fig:c3_evgam_qqplots}, where we observe reasonable agreement between the empirical and theoretical quantiles. However, whilst these values appeared optimal within $\Tau$, we stress that adequate model fits were also obtained for other quantile levels, suggesting our modelling procedure is not particularly sensitive to the exact choice of quantile. Furthermore, we also tested a range of quantile levels below the $0.8$-level, but were unable to improve the quality of model fits. 

Plots illustrating the estimated GPD scale parameter functions are given in the Supplementary Material, with the resulting dependence trends in agreement with the observed trends from Section \ref{subsec:edac3}. We also remark that the estimated GPD shape parameters obtained for parts 1 and 2 were $0.042\, (0.01,0.075)$ and $0.094\,(0.059,0.128)$, respectively, where the brackets denote $95\%$ confidence intervals obtained using posterior sampling \citep{Wood2017}. These estimates, which indicate slightly heavy-tailed behaviour within the min-projection variable, provide insight into why the original exponential modelling framework is not appropriate for C3.  

% To better illustrate the estimated trend in dependence, Figure \ref{fig:c3_sigma_estimates} illustrates the estimated scale functions, $\sigma\left(\boldsymbol{\omega} \mid \boldsymbol{x}_t\right)$, over atmosphere for parts 1 and 2. Under the assumption of asymptotic normality in the spline coefficients, 95\% confidence intervals are obtained via posterior sampling; see \citet{Wood2017} for more details. We observe that $\sigma$ tends to increase and decrease over atmosphere for parts 1 and 2, respectively, although the trend is less pronounced for the latter. Under our modelling framework, we note that higher values of $\sigma$ are associated with less positive extremal dependence in the direction $\boldsymbol{\omega}$ of interest; to see this, observe that the survivor function of the GPD with fixed $\xi$ is negatively associated with $\sigma$. Considering the trend in $\sigma\left(\boldsymbol{\omega} \mid \boldsymbol{x}_t\right)$, our results indicate a decrease in dependence in the regions where all variables are extreme. We also remark that the estimated GPD shape parameters obtained for parts 1 and 2 were $0.042\, (0.01,0.075)$ and $0.094\,(0.059,0.128)$, respectively, where uncertainty intervals were again obtained using posterior sampling. These estimates, which indicate slight heavy-tailed behaviour within the min-projection variable, providing further insight into why the original exponential modelling framework, described in equation \eqref{eqn:ns_wt_model}, was is not appropriate for C3.  

Overall, these results suggest different extremal dependence trends exist for the two simplex points $\bm \omega^{(1)}$ and $\bm \omega^{(2)}$, illustrating the importance of the flexibility in our model. These findings are also in agreement with empirical trends observed in Section \ref{subsec:edac3}, suggesting our modelling framework is successfully capturing the underlying extremal dependence structures.

% ultimately aiming to find the 
% Assuming the modelling assumption is satisfied at a given quantile level, we would expect these quantiles to be in good agreement. Consequently, we used quantile-quantile (QQ) plots as a means to compare different quantile levels. Uncertainty in these plots was characterised non-parametrically via bootstrapping. In summary, this gave us a visual means to compare between different quantile levels. Examples of the resulting QQ plots are given in the supplementary material. 

% To select the quantile level $\tau$, we considered values in the set $\{0.8,0.81,\hdots,0.99\}$. For each quantile level, we obtained estimates of the threshold and parameter functions, as specified above. We considered the quality of each model fit using the following procedure. First,

% \LA{Is this the one that will go on the Results \& Discussion section?} \CB{Would probably opt to have it as it's own section - need to talk about how we define diagnostics, and how we test assumptions. Results can just be for the answers we obtain.}

\subsection{Results} \label{subsec:resultsc3}

%\LA{I think this section needs to be structured differently, I just don't know why. Haven't changed anything. Should the plots we got with the confidence bands go here or in the SM?}

%\CB{Suggested: 1) detail how we obtain probability estimates from the non-stationary estimates for part 1. 2) Note that the procedure is analogous for part 2. Give probability estimates for both in one paragraph. }

Given estimates of threshold and parameter functions, probability estimates can be obtained via Monte Carlo techniques. Taking $p_1$, for instance, we have 
% \begin{myalign*}
%     p_1 &=& \Pr(Z_1>\Tilde{y},Z_2>\Tilde{y},Z_3>\Tilde{y}) = \Pr(\min\left(Z_1/w^{(1)}_1, Z_2/w^{(1)}_2, Z_3/w^{(1)}_3 \right) > r^{(1)}) \\
%     &=& \displaystyle{\int_{\boldsymbol{X}_t} \Pr(T_{\boldsymbol{\omega}^{(1)},\;t} > r^{(1)} \mid \boldsymbol{X}_t = \boldsymbol{x}_t) f_{\boldsymbol{X}_t}(\boldsymbol{x}_t) \mathrm{d}\boldsymbol{x}_t} \\
%     &=& \displaystyle{(1-\tau)\int_{\boldsymbol{X}_t} \Pr(T_{\boldsymbol{\omega}^{(1)},\;t} > r^{(1)} \mid T_{\boldsymbol{\omega}^{(1)},\;t} > v_{\tau}(\boldsymbol{\omega}^{(1)},\boldsymbol{x}_t),\boldsymbol{X}_t = \boldsymbol{x}_t)  f_{\boldsymbol{X}_t}(\boldsymbol{x}_t) \mathrm{d}\boldsymbol{x}_t} \\
%     &\approx& \displaystyle{(1-\tau) \frac{1}{n} \sum_{t=1}^n  \left( 1 + \frac{\xi\left(\boldsymbol{\omega}^{(1)} \mid \boldsymbol{x}_t\right) (r^{(1)} - v_{\tau}(\boldsymbol{\omega}^{(1)},\boldsymbol{x}_t))}{\sigma\left(\boldsymbol{\omega}^{(1)} \mid \boldsymbol{x}_t\right)} \right)^{-1/\xi\left(\boldsymbol{\omega}^{(1)} \mid \boldsymbol{x}_t\right)}},
% \end{myalign*}
\begin{myequation*}
    \begin{array}{rcl}
        p_1 &=& \Pr(Z_1>\Tilde{y},Z_2>\Tilde{y},Z_3>\Tilde{y}) \\
        &=& \Pr\left(\min\left(Z_1/w^{(1)}_1, Z_2/w^{(1)}_2, Z_3/w^{(1)}_3 \right) > r^{(1)}\right) \\
        &=& \displaystyle{\int_{\boldsymbol{X}_t} \Pr(T_{\boldsymbol{\omega}^{(1)},\;t} > r^{(1)} \mid \boldsymbol{X}_t = \boldsymbol{x}_t) f_{\boldsymbol{X}_t}(\boldsymbol{x}_t) \mathrm{d}\boldsymbol{x}_t} \\
        &=& \displaystyle{(1-\tau)\int_{\boldsymbol{X}_t} \Pr(T_{\boldsymbol{\omega}^{(1)},\;t} > r^{(1)} \mid T_{\boldsymbol{\omega}^{(1)},\;t} > v_{\tau}(\boldsymbol{\omega}^{(1)},\boldsymbol{x}_t),\boldsymbol{X}_t = \boldsymbol{x}_t)  f_{\boldsymbol{X}_t}(\boldsymbol{x}_t) \mathrm{d}\boldsymbol{x}_t} \\
        &\approx& \displaystyle{\frac{1-\tau}{n} \sum_{t=1}^n  \left( 1 + \frac{\xi(\boldsymbol{\omega}^{(1)} \mid \boldsymbol{x}_t) \left(r^{(1)} - v_{\tau}(\boldsymbol{\omega}^{(1)},\boldsymbol{x}_t)\right)}{\sigma\left(\boldsymbol{\omega}^{(1)} \mid \boldsymbol{x}_t\right)} \right)^{-1/\xi\left(\boldsymbol{\omega}^{(1)} \mid \boldsymbol{x}_t\right)}},
    \end{array}
\end{myequation*}
assuming $\{ \boldsymbol{x}_t: t \in \{1,\hdots,n\}\}$ is a representative sample from $\boldsymbol{X}_t$. The procedure for $p_2$ is analogous. We note that this estimation procedure is only valid when $r^{(1)} > v_{\tau}(\boldsymbol{\omega}^{(1)},\boldsymbol{x}_t)$, or $r^{(2)} > v_{\tau}(\boldsymbol{\omega}^{(2)},\boldsymbol{x}_t)$, for all $t$: however, for each $\tau \in \Tau$, this inequality is always satisfied, owing to the very extreme nature of the probabilities in question. Through this approximation, we obtain $\hat p_1=1.480449\times 10^{-5}$ and $\hat p_2=2.460666\times 10^{-5}.$

%\CB{Should we mentioned these were close to the true values?} \LA{from the challenge results?}\CB{Yes - though on reflection, we can just summarise in the discussion}

%\CB{This section is finished now - check you're happy with it and everything makes sense}. 

\section{Challenge C4} \label{sec:c4}

\subsection{Exploratory data analysis} \label{subsec:edac4}

Challenge C4 entails estimating survival probabilities across $50$ locations on the island of Utopula. As stated in \citet{Rohrbeck2023}, the Utopula island is split in two administrative areas, for which the respective regional governments 1 and 2 have collected data concerning the variables $Y_{i,t}$, $i \in I = \{1,\ldots,50\},\,  t \in \{1,\ldots,10,000$\}. Index $i$ denotes the $i^{\text{th}}$ location, with locations $i \in \{1, \ldots,25\}$ and $i \in \{26, \ldots,50\}$ belonging to the administrative areas of governments $1$ and $2$, respectively. Index $t$ denotes the time point in days; however, since $Y_{i,t}$ are IID for all $i$, we drop the subscript $t$ for the remainder of this section.

\begin{figure}[t]
      \centering
        \includegraphics[width=0.5\textwidth,keepaspectratio]{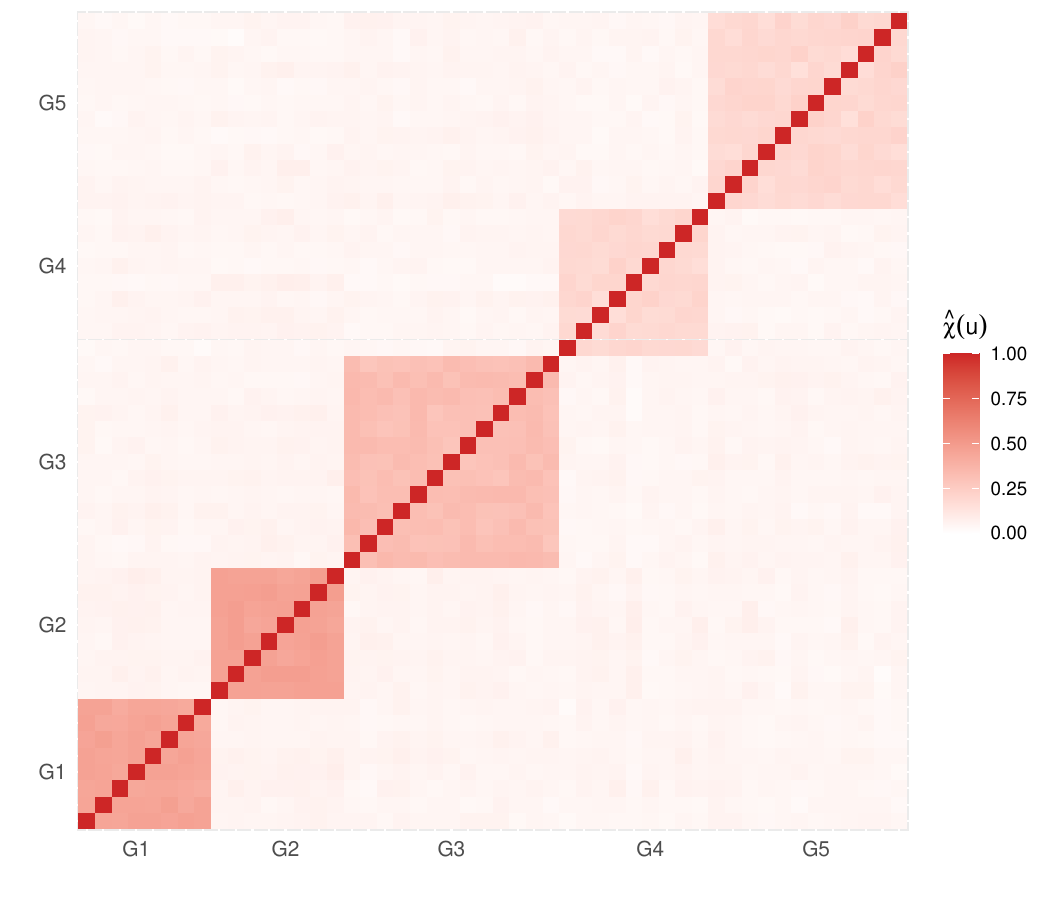}
      \caption{Heat map of estimated empirical pairwise $\chi(u)$ extremal dependence coefficients with $u=0.95$.}
      \label{fig:chi heat}
\end{figure}

Since many multivariate extreme value models are only applicable in low-to-moderate dimensions, we consider dimension reduction based on an exploration of the extremal dependence structure of the data. In particular, we analyse pairwise estimates of the extremal dependence coefficient $\chi(u)$, introduced in equation \eqref{eq:chi}, for all possible pairwise combinations of sites; the resulting estimates, using $u=0.95$, are presented in the heat map of Figure \ref{fig:chi heat}. Identification of any dependence clusters is achieved through visual investigation, which seems appropriate for this data. We note, however, that should visual considerations not suffice, alternative more sophisticated clustering methods are available and can be applied; see for example \citet{Bernard2013}. 

Figure \ref{fig:chi heat} suggests the existence of $5$ distinct subgroups where all variables within each subgroup have similar extremal dependence characteristics, while variables in different subgroups appear to be approximately independent of each other in the extremes. It is worth mentioning that the same clusters are identified when we analyse pairwise estimates of the extremal dependence coefficient $\eta(u)$; the resulting estimates can be found in the Supplementary Material. Moreover, examining the magnitudes of $\chi(\cdot)$ and $\eta(\cdot)$ estimates, it does not appear reasonable to assume asymptotic dependence between variables in the same group. We therefore consider models that can be applied to data structures that do not take their extreme values simultaneously. The indices of the five aforementioned subgroups are $G_1=\{4, 14, 19, 28, 30, 38, 43, 44\}$, $G_2=\{3, 10, 15, 18, 22, 29, 45, 47\}$, \linebreak $G_3=\{8, 21, 25, 26, 32, 33, 34, 40, 41, 42, 48, 49, 50\}$, $G_4=\{1, 2, 5, 7, 9, 17, 20, 31, 46\}$ and $G_5=\{6, 11, 12, 13, 16, 23, 24, 27, 35, 36, 37, 39\}$.
Groups $G_1$ and $G_2$ include the most strongly dependent variables (shown by the darkest color blocks in Figure \ref{fig:chi heat}), followed by group $G_3$, while groups $G_4$ and $G_5$ contain the most weakly dependent variables. We henceforth assume independence between these groups of variables, i.e., $\Pr((Y_i)_{i \in G_k} \in A_k,(Y_i)_{i \in G_{k'}} \in A_{k'}) = \Pr((Y_i)_{i \in G_k} \in A_k)\Pr((Y_i)_{i \in G_{k'}} \in A_{k'})$, $A_k\subset \RR^{\vert G_k \vert},A_{k'}\subset \RR^{\vert G_{k'} \vert}$, for any $k \neq k' \in \{1,\hdots,5\}$. 

Challenge C4 requires us to estimate the probabilities $p_1=\text{Pr}\left(Y_{i}>s_i;\,i\in I\right)$ and $p_2=\text{Pr}(Y_{i}>s_1;\, i  \in I)$, where $s_i := \mathbbm{1}(i \in \{1,2,\hdots,25\})s_1 + \mathbbm{1}(i \in \{26,27,\hdots,50\})s_2$ and $s_1$ ($s_2$) denotes the marginal level exceeded once every year (month) on average. Under the assumption of independence between groups, the challenge can be broken down to $5$ lower-dimensional challenges involving the estimation of joint tail probabilities for each $G_k,$ $k \in \{1,\ldots,5\}.$ These can then be multiplied together to obtain the required overall probabilities due to (assumed) between-group independence; specifically, we have $p_1= \prod_{k=1}^5\text{Pr}\left(Y_{i} >s_i ;\ i\in G_k\right)$ 
%$p_1=\text{Pr}\left(Y_{i_j}>s_j,\,i_j=1,\ldots,25,\, j=1,2 \right)=\prod_{k=1}^5\text{Pr}\left(Y_{i_j} >s_j ;\ Y_{i_j,t}\in G_k\right)$ 
%$p_1=\text{Pr}\left(Y_{i_j,t}>s_j,\,i_j=1,\ldots,25,\, j=1,2 \right)=\prod_{k=1}^5\text{Pr}\left(Y_{i_j,t} >s_j ;\ Y_{i_j,t}\in G_k\right)$
and 
%$p_2=\text{Pr}\left(Y_{i_j,t}>s_1,\,i_j=1,\ldots,25,\, j=1,2\right)=\prod_{k=1}^5\text{Pr}\left(Y_{i_j,t} >s_1 ;\ Y_{i_j,t}\in G_k\right)$
%$p_2=\text{Pr}\left(Y_{i_j}>s_1,\,i_j=1,\ldots,25,\, j=1,2\right)=\prod_{k=1}^5\text{Pr}\left(Y_{i_j} >s_1 ;\ Y_{i_j}\in G_k\right)$
$p_2=\prod_{k=1}^5\text{Pr}\left(Y_{i} >s_1 ;\ i\in G_k\right)$.% We now consider estimation of within group probabilities. %\CB{$s_j$ needs defining - would recommend doing this in terms of indices, e.g., $s_j := \mathbbm{1}(j \in \{1_1,2_2,\hdots,25_1\})s_1 + \mathbbm{1}(j \in \{1_2,2_2,\hdots,25_2\})s_2$. }
%\LK{Specifically, we have \allowbreak $p_1=\text{Pr}\left(Y_{i_j}>s_j,\,i_j=1,\ldots,25,\, j=1,2 \right)=\prod_{k=1}^5\text{Pr}\left(Y_{i_j} >s_j ;\ i_j\in G_k\right)$ 
%$p_1=\text{Pr}\left(Y_{i_j,t}>s_j,\,i_j=1,\ldots,25,\, j=1,2 \right)=\prod_{k=1}^5\text{Pr}\left(Y_{i_j,t} >s_j ;\ Y_{i_j,t}\in G_k\right)$
%and \allowbreak 
%$p_2=\text{Pr}\left(Y_{i_j,t}>s_1,\,i_j=1,\ldots,25,\, j=1,2\right)=\prod_{k=1}^5\text{Pr}\left(Y_{i_j,t} >s_1 ;\ Y_{i_j,t}\in G_k\right)$
%$p_2=\text{Pr}\left(Y_{i_j}>s_1,\,i_j=1,\ldots,25,\, j=1,2\right)=\prod_{k=1}^5\text{Pr}\left(Y_{i_j} >s_1 ;\ i_j\in G_k\right)$, where $s_j := \mathbbm{1}(j \in \{1_1,2_1,\hdots,25_1\})s_1 + \mathbbm{1}(j \in \{1_2,2_2,\hdots,25_2\})s_2$ and $s_1$ ($s_2$) is the marginal level exceeded once every year (month) on average. We now consider estimation of within group probabilities.}
%\LA{I think I'm confused with the notation - discuss in the meeting. Overall prefer the $i_j$ one}

\subsection{Conditional extremes} \label{subsec:methodc4}

The conditional multivariate extreme value model (CMEVM) of \cite{Heffernan2004} provides a flexible multivariate extreme value framework capable of capturing a range of extremal dependence forms without making assumptions about the specific form of joint dependence structure. Consider a $d$-dimensional random variable $\boldsymbol{W}=(W_1, \ldots, W_d)$ on standard Laplace margins. For $i\in\{1,\ldots, d\}$, the CMEVM approach assumes the existence of parameter vectors $\boldsymbol{\alpha}_{-|i} \in [-1, 1]^{d-1}$ and $\boldsymbol{\beta}_{-|i} \in (-\infty, 1]^{d-1}$ such that 
\begin{myequation}\label{eq:cond_assumption} 
    \lim_{u_i \rightarrow \infty}\operatorname{Pr}\left\{\boldsymbol{W}_{-i} \leq\boldsymbol{\alpha}_{-|i} W_i+W_i^{\boldsymbol{\beta}_{-|i}} \boldsymbol{z}_{\mid i}, W_i - u_i > w \mid W_i>u_i\right\}=e^{-w}H_{\mid i}\left(\boldsymbol{z}_{\mid i}\right), \quad w >0, 
\end{myequation}
with non-degenerate distribution function $H_{\mid i},$ vector operations are applied componentwise, and conditional threshold $u_i$. The vector $\boldsymbol{W}_{-i}$ denotes $\boldsymbol{W}$ excluding its $i^{\text{th}}$ component and $\boldsymbol{z}_{\mid i}$ is within the support of the residual random vector $\boldsymbol{Z}_{\mid i}=(\boldsymbol{W}_{-i}-\boldsymbol{\alpha}_{-|i}w_i)/{w_i^{\boldsymbol{\beta}_{-|i}}}\sim H_{\mid i}$. We apply this model to data where $W_i>u_i$, for some finite conditioning threshold $u_i$, to estimate the probabilities $p_1$ and $p_2$ defined in Section~\ref{subsec:edac4}, using the inference procedure of \cite{Keef2013a}. %\AF{For estimation of the probabilities should we mention the resampling approach in the original paper?} %\CB{Note that I found a few typos in this section - all have been corrected now, and I changed the ordering slightly to help it flow better:)}

\label{subsec:condextremes}

\subsection{Results} \label{subsec:resultsc4}

Let $\boldsymbol{W}:=(W_1, \ldots, W_{50})$ denote the random vector after transformation to standard Laplace margins. This vector is divided into the five subgroups identified in Section~\ref{subsec:edac4}, and the subgroup probabilities are estimated using extreme predictions obtained from the sampling method of \cite{Heffernan2004}. We condition on the first variable of each subgroup being extreme, and simulate $10^8$ predictions from each of the resulting fitted conditional extremes models. To account for uncertainty in the estimates, we perform a parametric bootstrapping procedure with $100$ samples.

Sensitivity analyses of the estimated probabilities to the choice of conditioning variable suggest no significant effect. Furthermore, we consider a range of conditioning thresholds; the corresponding estimates of subgroup probabilities defined in Section~\ref{subsec:edac4} appear relatively stable with respect to the conditioning threshold quantile. We ultimately select 0.85-quantiles for the conditioning thresholds of our final probability estimates. These are given by $\hat{p}_1 = 1.093634\times 10^{-26}$ $(2.149591\times 10^{-36}, 1.359469\times 10^{-24})$ and $\hat{p}_2 = 1.075787\times 10^{-31}$ $(1.596381\times 10^{-46}, \linebreak1.850425\times10^{-29})$, with 95\% confidence intervals obtained from parametric bootstrapping given in parentheses.

\section{Discussion} \label{sec:discussion}
In this paper, we have proposed a range of statistical methods for estimating extreme quantities for challenges C1-C4. 
% \AF{I have edited the below - Ryan feel free to make updates} 
For the univariate challenge C1, we estimated the 0.9999-quantile, and the associated 50\% confidence intervals, of $Y\mid\boldsymbol{X} = \boldsymbol{x_{i}}$, $i \in \{1,\hdots,n\}$. For challenge C2, we estimated a quantile, corresponding to a once in 200 year level, of the marginal distribution $Y$ whilst incorporating the loss function in equation \eqref{eq:loss_fn}. Overall we ranked $6^{\text{th}}$ and and $4^{\text{th}}$ for challenges C1 and C2, respectively.

For challenge C1, our final model (model 7 in Table \ref{tab:C1_Models}) was chosen to minimise the model selection criteria; however, QQ plots showed over-estimation of the most extreme values of the response (see Figure \ref{fig:C1_Results}). As a result, the conditional quantiles calculated for C1 are generally over-estimated when compared with the true quantiles. If we ignore the model selection criteria and chose the model based on a visual assessment of QQ plots, we would have chosen model 5 in Table \ref{tab:C1_Models} and this would have covered the true quantile on fewer occasions than our chosen model. Therefore, the main issue with our results concerns the width of the confidence intervals.

Narrow confidence intervals are an indication of over-fitting and this could have arisen in several places. For instance, \citet{Rohrbeck2023} suggested all the seasonality is captured in the threshold, while our model includes a seasonal threshold and a covariate for seasonality in the scale parameter of the GPD model. As well as over-fitting, the model may not have been flexible enough; this could be, in part, due to our model missing covariates. For instance, the true model contained $V_{2}$ as a covariate \citep{Rohrbeck2023} whilst our model did not. In addition, the basis dimensions for our splines are low. In practice, a higher dimension than we would expect should be considered and, although we chose the dimension using a model-based approach, it may have resulted in the splines not being flexible enough to capture all of the trends in the data. 

Narrow confidence intervals may have also resulted from the choice of uncertainty quantification procedure. Changing the block length $l$ in our stationary bootstrap procedure would alter the confidence interval widths, although this was carefully chosen to reflect the temporal dependence in the data. Alternative methods, such as the standard bootstrap procedure or the delta method, could be implemented to investigate how this affects the confidence interval widths. We expect that such confidence intervals will be wider than those presented here since the dependence in the data is not accounted for, but assuming temporal independence would be inaccurate. Therefore, whilst adopting an alternative procedure may widen confidence intervals, thus improving our performance, such intervals may not be well calibrated for this data set.

As the same model for C1 and C2 was used, the over-fitting and over-estimation problems in the model for the first challenge are carried through to the second challenge. The adjustment for the loss function only had a slight effect on parameter estimates and subsequent quantile estimation as in equation \eqref{eq:penalised_likelihood} since $S(\theta)$ is dominated by the log-likelihood. However, as the loss function is biased towards over-estimation, this small adjustment resulted in further over-estimation of the true quantile in C2.

For the first multivariate challenge C3, we employed an extension of the method proposed by \cite{Wadsworth2013} to estimate probabilities of three variables lying in extremal sets. Our extension accounts for non-stationarity in the extremal dependence structure, with GAMs used to represent covariate relationships. The QQ plots for the resulting model suggested reasonable fits. For this challenge, we ranked 5\textsuperscript{th} and our estimates are on the same order of magnitude as the truth~\citep{Rohrbeck2023}.

% For the first multivariate challenge C3, we employed an extension of the method proposed by \cite{Wadsworth2013} to estimate the probability of three variables lying in an extremal set. Our extension to this framework accounts for non-stationarity in the data when estimating the associated coefficient which relies on GAMs to obtain the parameters. The QQ plots for the resulting models suggested reasonable fits. For this challenge, we ranked 5\textsuperscript{th} and our estimates are on the same order of magnitude as the truth~\citep{Rohrbeck2023}.

% \ED{Can we of wider applications for some of our methods?} \CB{Have written an example below - but I'm in two minds over whether it is worth including, given our position overall. Could cut to save space, and focus more on limitations/discrepencies}

% While the techniques proposed in this paper were developed with the Coputopia dataset in mind, there is scope to use such techniques more generally within the modelling of environmental extremes. \CB{Too restrictive?} For example, the model proposed in equation \eqref{eqn:wt_model_gpd} has not been previously considered within the literature, and could allow for more accurate estimation of joint tail probabilities for data structures exhibiting non-stationary dependence. Applications and development of the methods proposed in this paper represent open areas for future research. 

We note similarities in the methodologies presented for the challenges C1, C2, and C3. Specifically, each of the proposed methods used the EVGAM framework for capturing non-stationary tail behaviour via a generalised Pareto distribution. We acknowledge that the model selection tool proposed for C1 and C2 could also be applied for C3. However, we opted not to use this tool for several reasons. Firstly, unlike the univariate setting, there is no guarantee of convergence to a GPD in the limit, and the GPD tail assumption thereby needs to be tested. Moreover, in exploratory analysis, we tested the model selection tool for C3 but found the selected models and quantiles to not be satisfactory, particularly in the upper tail of the min-projection variable. We therefore selected a model manually, using QQ plots to evaluate performance. Exploring threshold and model selection techniques for multivariate extremes represents an important area of research. 
% \AF{CB we don't discuss the different methods for the threshold here or in C1/C2 discussion - do we need to?}  

In the final multivariate challenge C4, we estimated very high-dimensional joint survival probabilities. To do so, we split the probability into 5 lower-dimensional asymptotically independent components, then estimated each using the CMEVM of \cite{Heffernan2004}. In the final rankings of \cite{Rohrbeck2023}, we ranked $3^{\text{rd}}$ for this challenge. A more prudent method could have been implemented, as groups of variables were never truly independent. Alternatively, our use of the conditional model could be improved by estimating individual group probabilities across varying thresholds and then taking an average value as our final result. Despite the relative stability of probability estimates with respect to threshold discussed in Section~\ref{subsec:methodc4}, this may have improved our overall ranking. We also do not report the effect of choice of conditioning variable on our estimates. Preliminary analysis suggested this to be negligible. However, conditioning on each site in a given subgroup and then taking a weighted sum of the resulting probabilities \citep[e.g.,][]{Keef2013} may have resulted in more robust estimates.

\section*{Declarations}

\subsection*{Ethical Approval and Consent to Participate}
Not Applicable

\subsection*{Consent for Publication}
Not Applicable

\subsection*{Materials Availability}
Materials that support the findings of this study are available from the corresponding author upon reasonable request.

\subsection*{Code Availability}
Code supporting the findings of this study is available from the corresponding author upon reasonable request.

\subsection*{Data Availability}
The data that support the findings of this study are available from the corresponding author upon reasonable request.

\subsection*{Competing interests}
The authors have no relevant financial or non-financial interests to disclose.

\subsection*{Funding}
This work was supported by EPSRC grant numbers EP/L015692/1, EP/S022252/1, EP/W523811/1 and EP/W524438/1, and SFI grant number 18/CRT/6049.

\subsection*{Authors' contributions}
All authors contributed equally to this work. 

\subsection*{Acknowledgments}
This paper is based on work completed while L\'idia Andr\'e, Eleanor D'Arcy, Conor Murphy, Callum Murphy-Barltrop and Matthew Speers were part of the EPSRC funded STOR-i centre for doctoral training (EP/L015692/1, EP/S022252/1), Ryan Campbell, Aiden Farrell and Lydia Kakampakou were part of EPSRC funded projects (EP/W523811/1, EP/W524438/1), and Dáire Healy was part of the Science Foundation Ireland funded project (18/CRT/6049). We are grateful to the two referees and editors for constructive comments and suggestions that have improved this article. We would also like to thank Ben Youngman for his assistance with the \verb|evgam| package in the \verb|R| computing language, as well as Christian Rohrbeck, Emma Simpson and Jonathan Tawn for their hard work in organising the data challenge.

\subsection*{Supplementary Information}
\textbf{Supplementary Material for ``Extreme value methods for estimating rare events in Utopia"}: File containing additional figures. (supplementary.pdf)

% \newpage

\bibliography{library}

\pagebreak

\begin{center}
\textbf{\large Supplementary Material to ``Extreme value methods for estimating rare events in Utopia"}
\end{center}
%%%%%%%%%% Merge with supplemental materials %%%%%%%%%%
%%%%%%%%%% Prefix a "S" to all equations, figures, tables and reset the counter %%%%%%%%%%
\setcounter{equation}{0}
\setcounter{figure}{0}
\setcounter{table}{0}
\setcounter{page}{1}
\setcounter{section}{0}
\makeatletter
\renewcommand{\theequation}{S\arabic{equation}}
\renewcommand{\thefigure}{S\arabic{figure}}
\renewcommand{\thesection}{S\arabic{section}}
\renewcommand{\bibnumfmt}[1]{[S#1]}
\renewcommand{\citenumfont}[1]{S#1}

\section{Additional figures for Section 3}\label{appendix_S3}
\begin{figure}[h]
    \centering
    \includegraphics[width=0.8\textwidth]{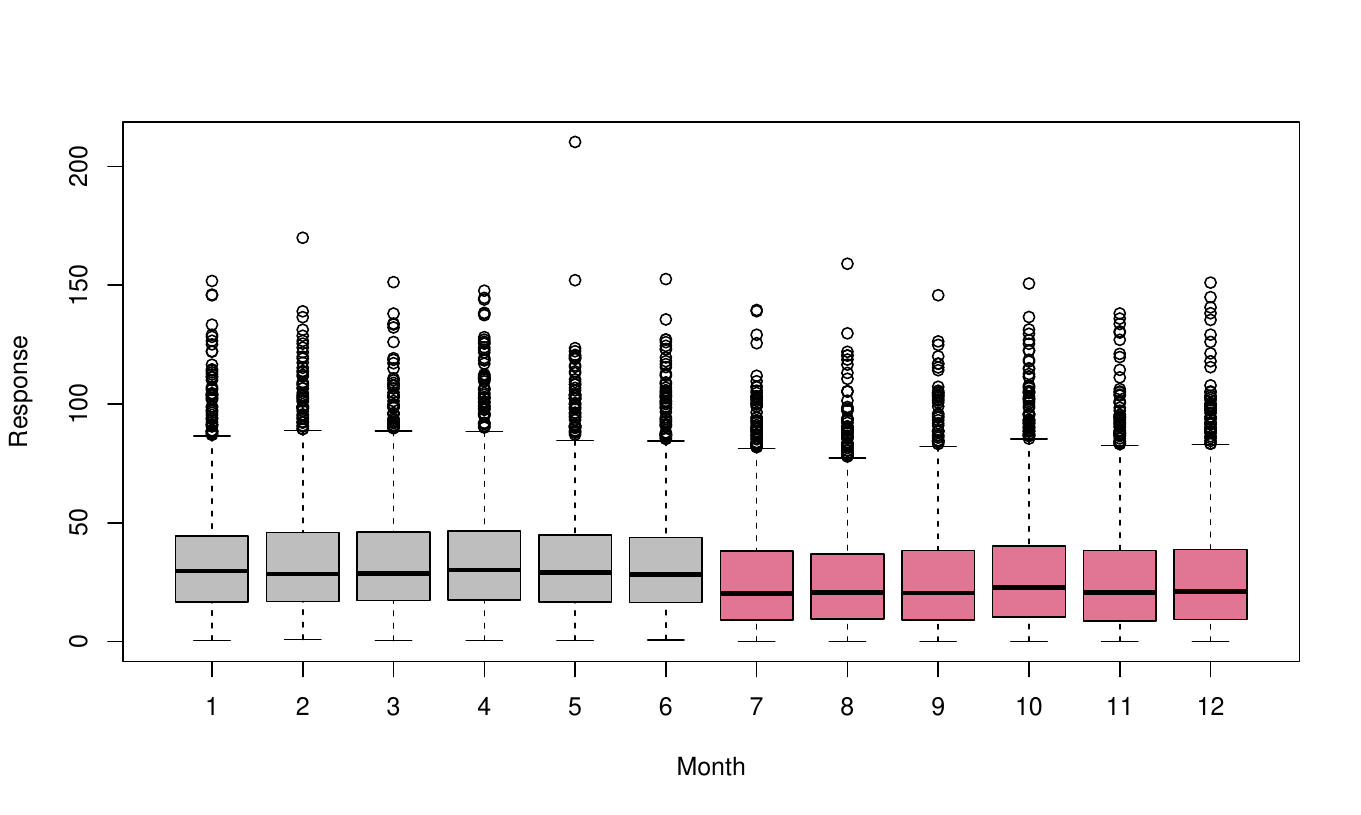}
    \caption{Box plot of the response variable $Y$ with each month and season (season 1 in grey and season 2 in red).}
    \label{fig:EA_mnth}
\end{figure}

\begin{figure}[h]
    \centering
    \includegraphics[width=0.245\textwidth]{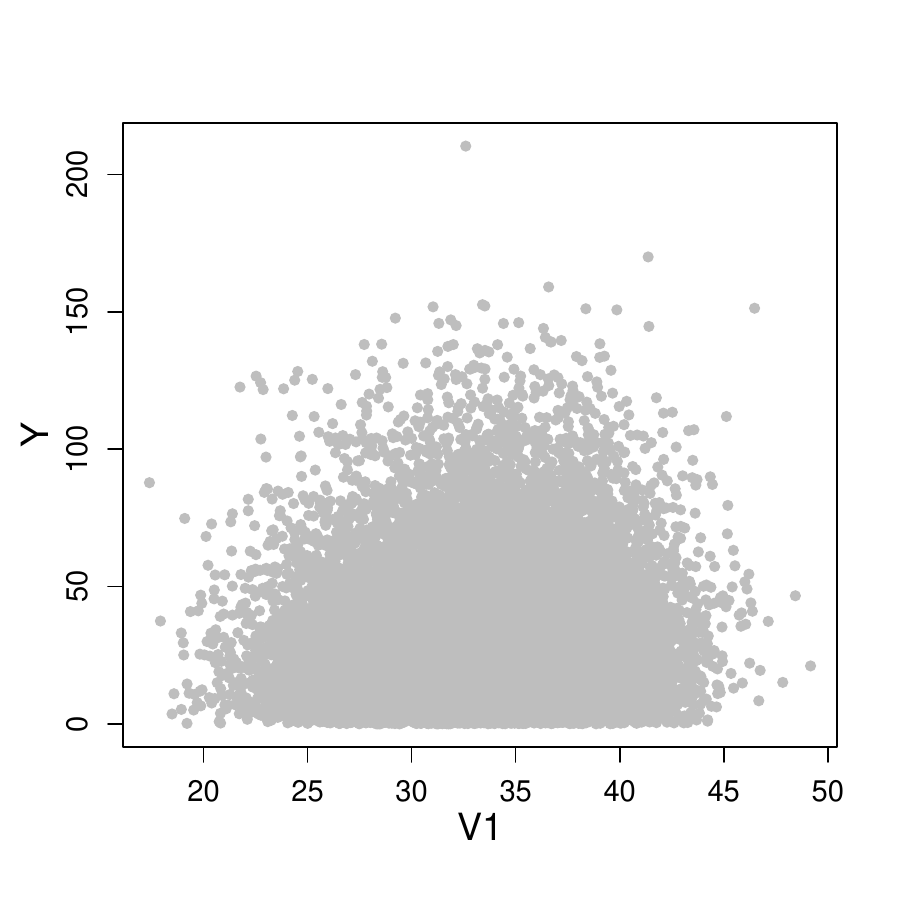}    \includegraphics[width=0.245\textwidth]{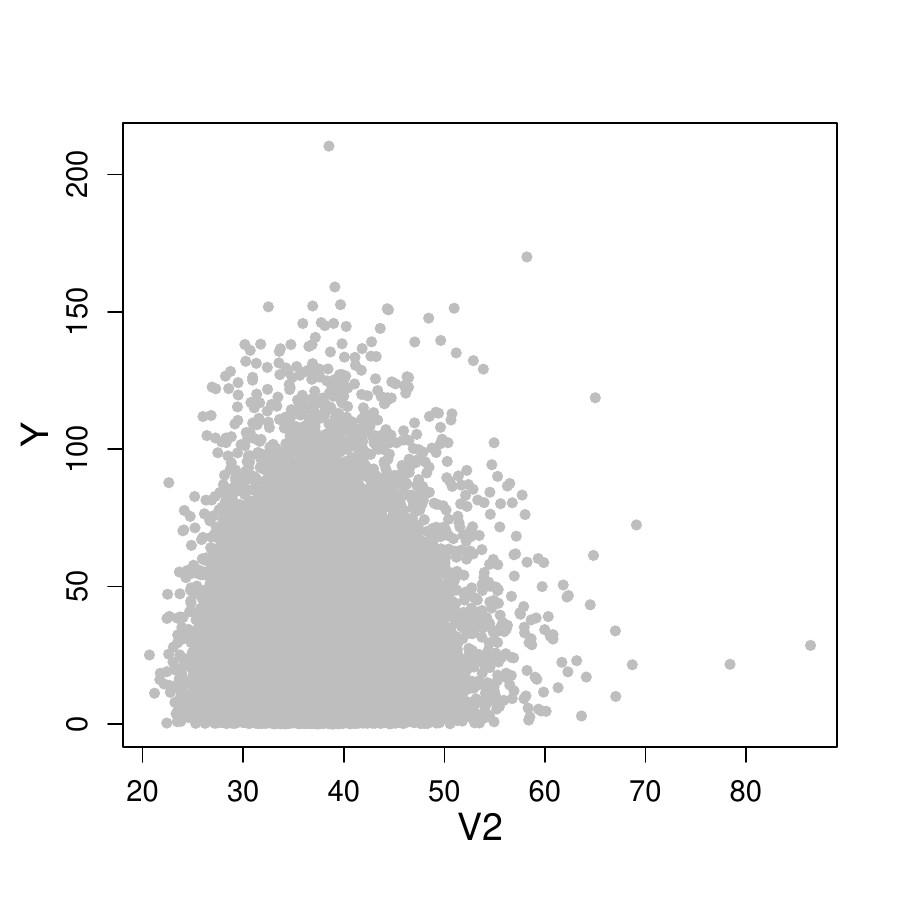}
    \includegraphics[width=0.245\textwidth]{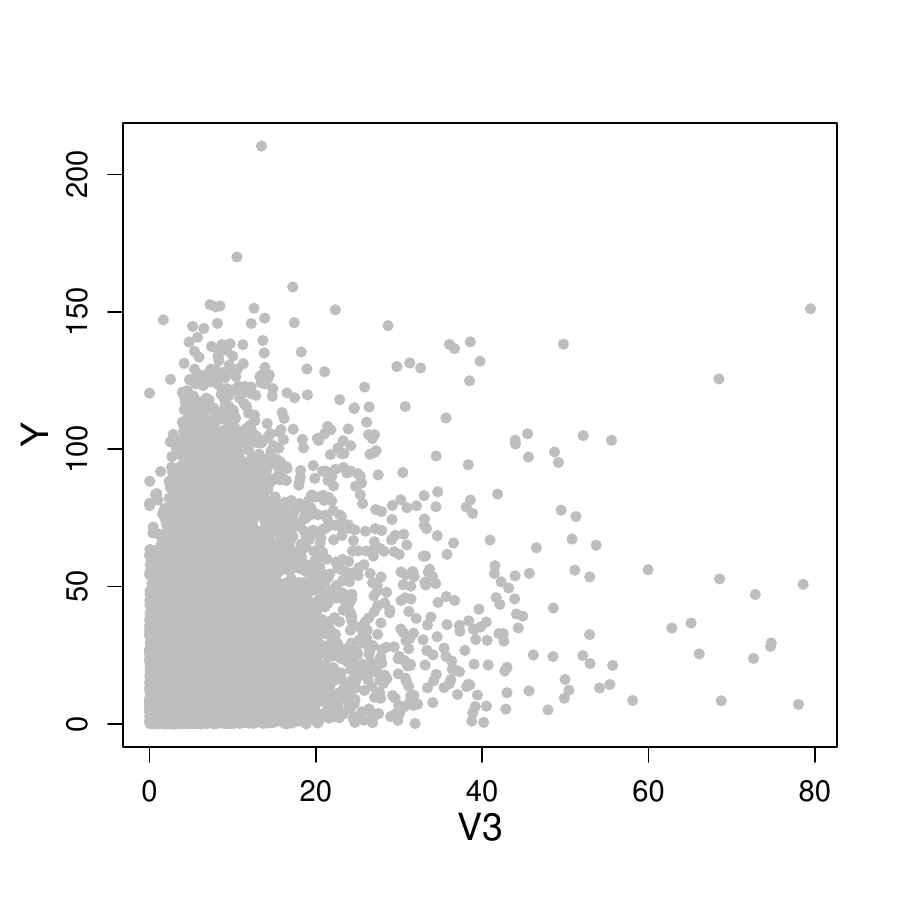}
    \includegraphics[width=0.245\textwidth]{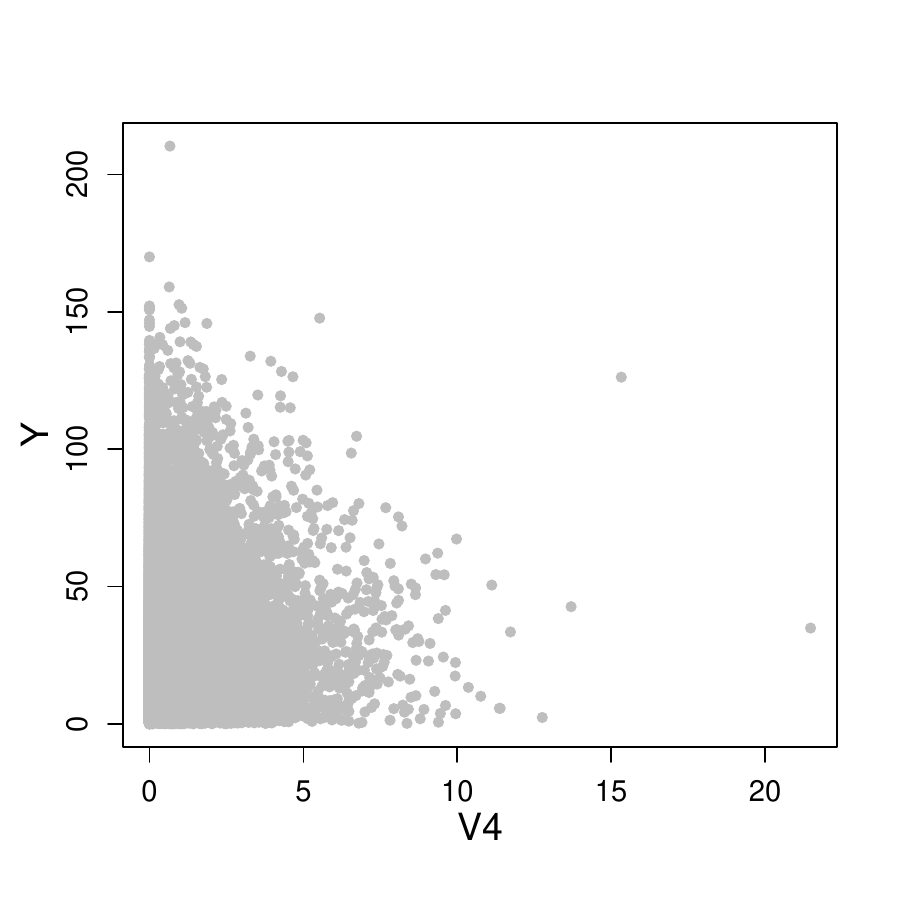}\\
    \includegraphics[width=0.245\textwidth]{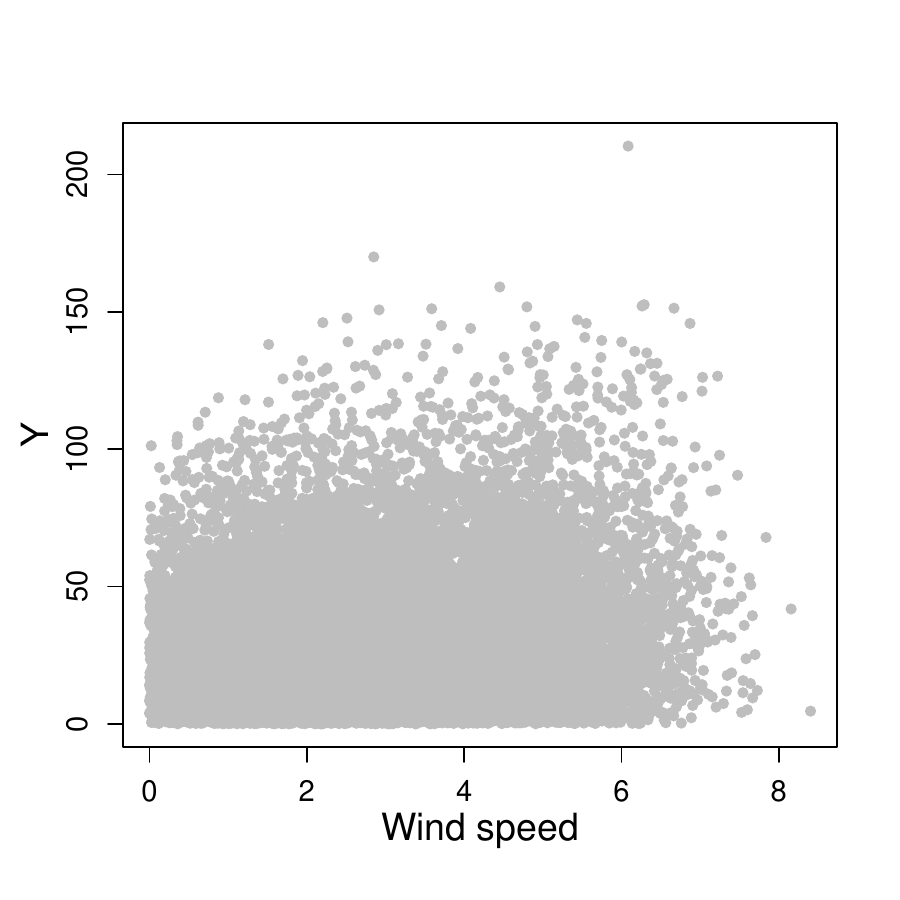}
    \includegraphics[width=0.245\textwidth]{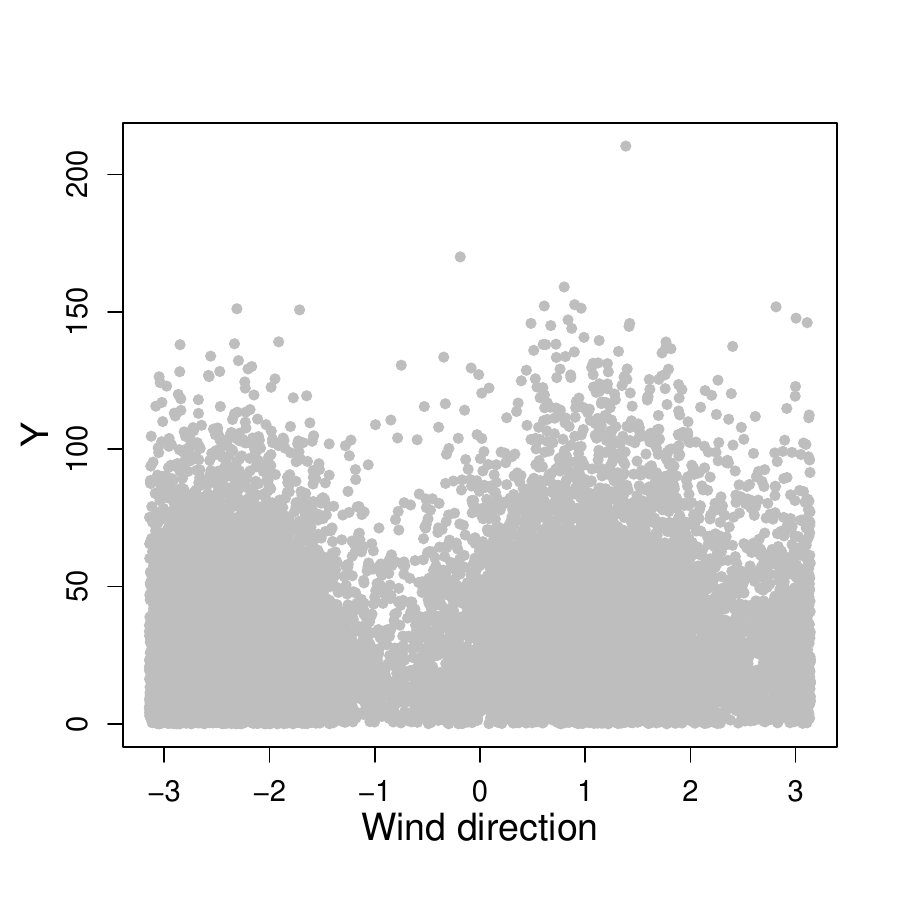}
    \includegraphics[width=0.245\textwidth]{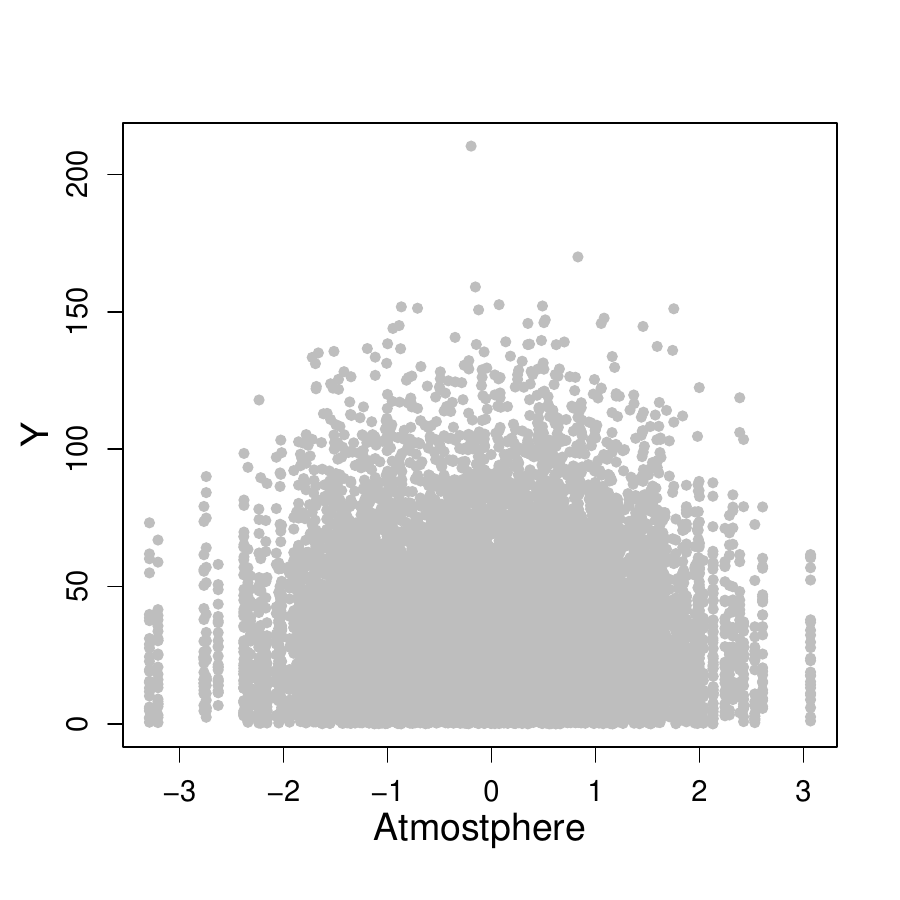}

    \caption{Scatter plots of explanatory variables $V_1,\ldots,V_4$,  wind speed ($V_6$), wind direction ($V_7$) and atmosphere ($V_8$), from top-left to bottom-right (by row), against the response variable $Y$.}
    \label{fig:scatter_allcov}
\end{figure}

\begin{figure}[h!]
    \centering
    \includegraphics[width=0.245\textwidth]{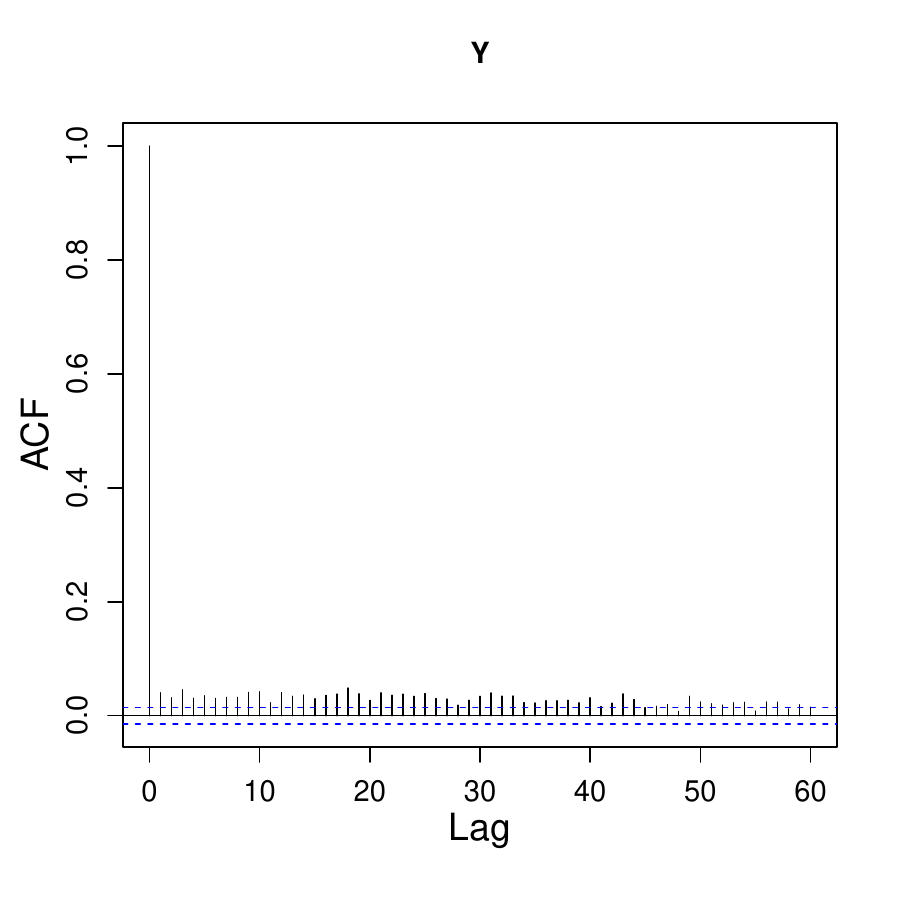}    
    \includegraphics[width=0.245\textwidth]{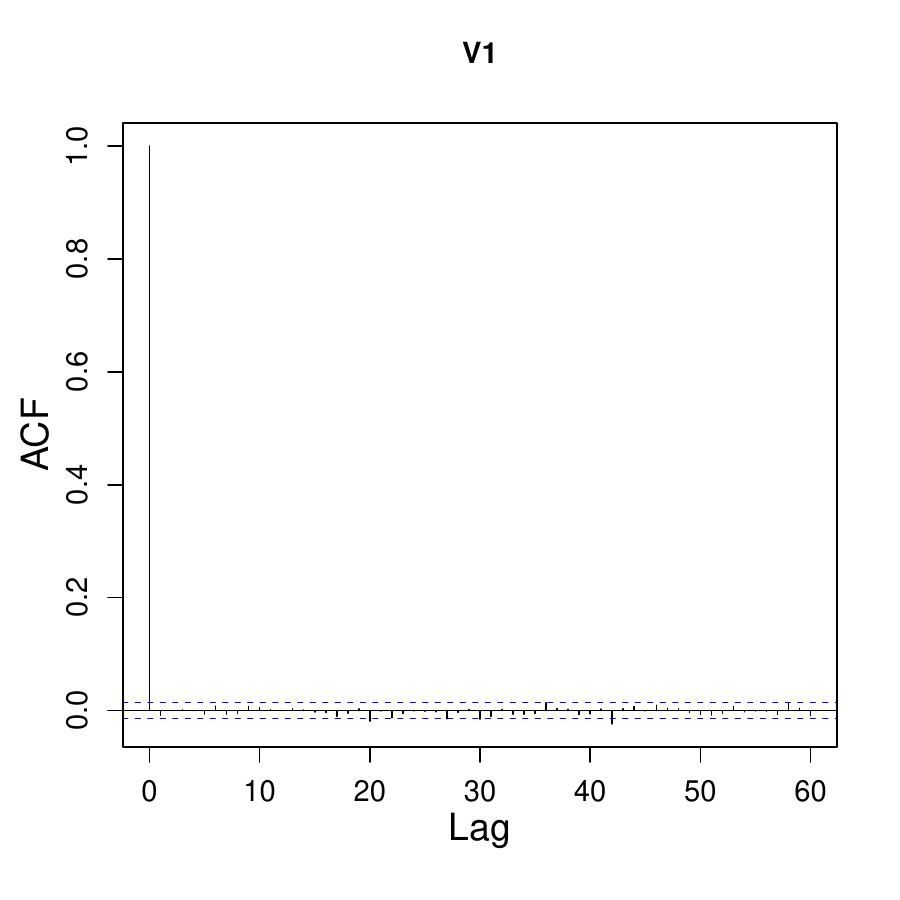}    \includegraphics[width=0.245\textwidth]{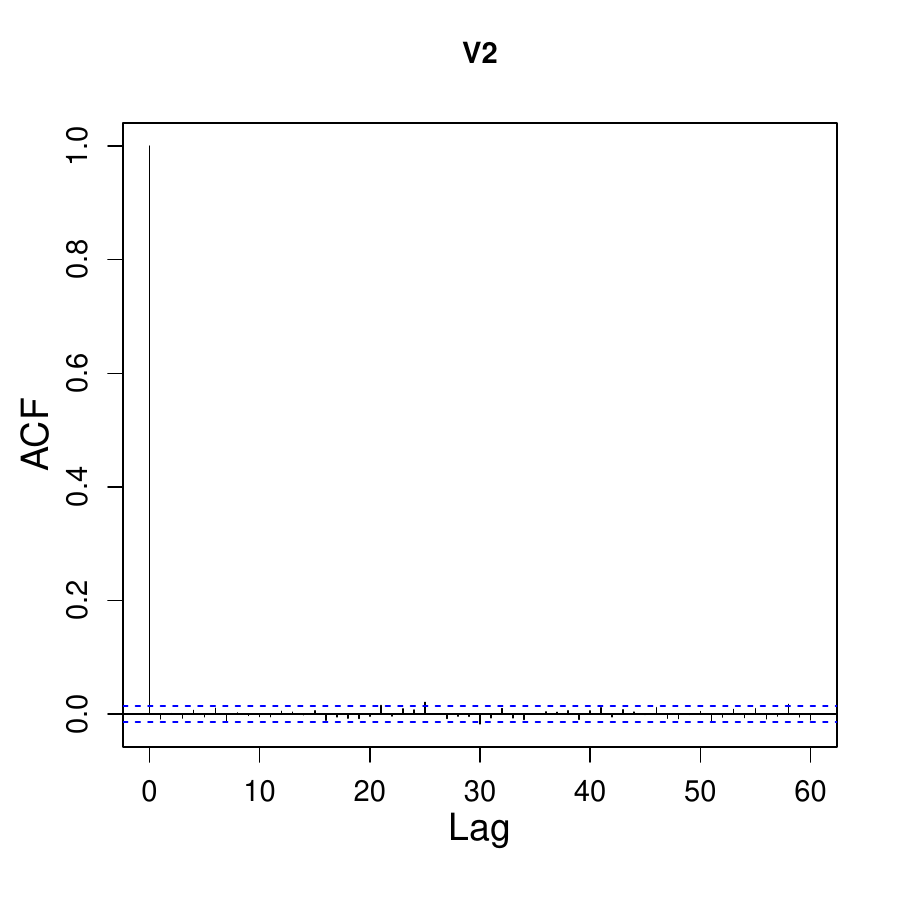}
    \includegraphics[width=0.245\textwidth]{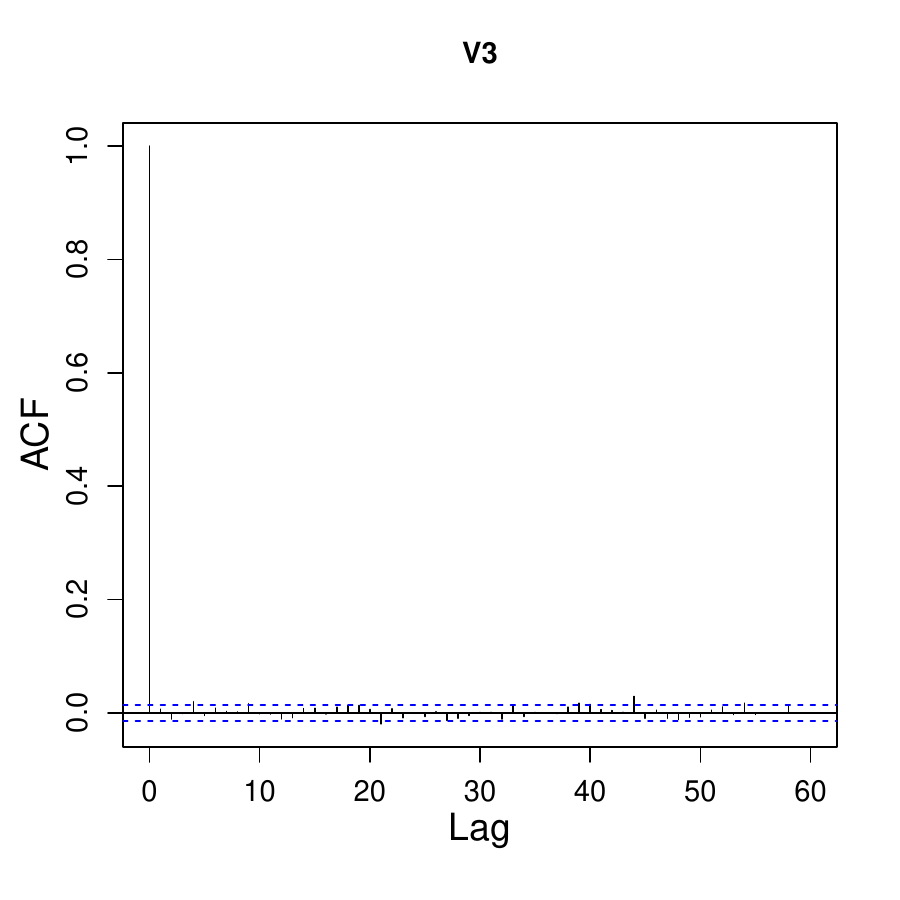}\\
    \includegraphics[width=0.245\textwidth]{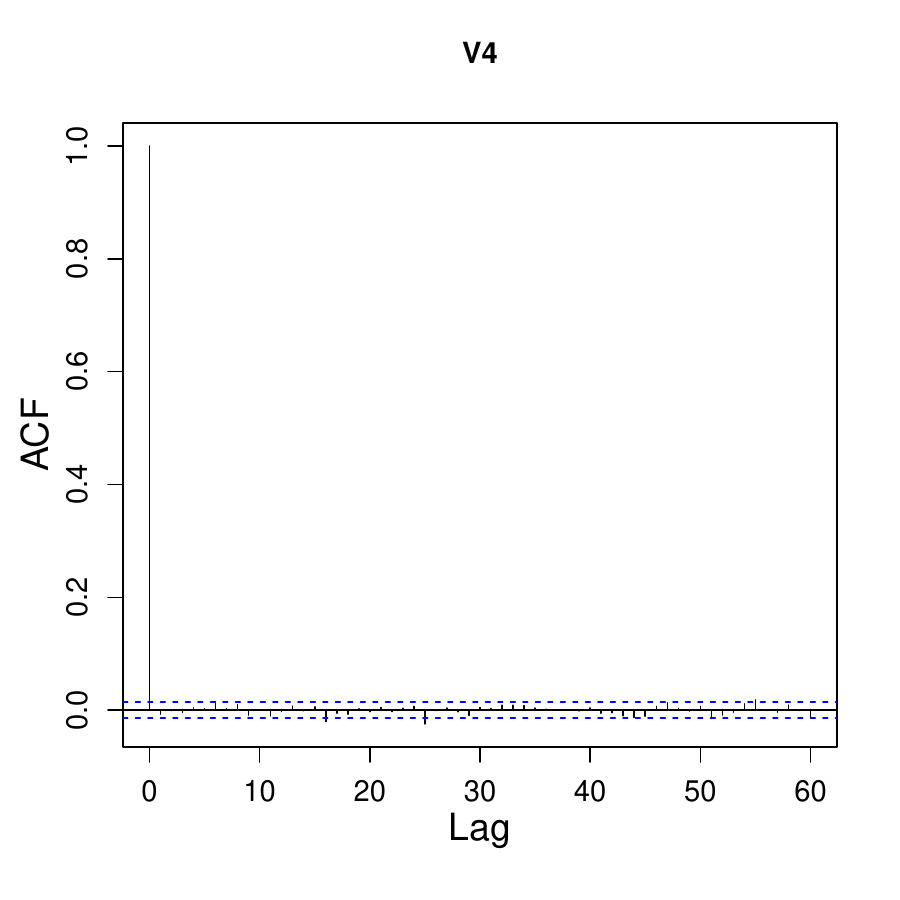}
    \includegraphics[width=0.245\textwidth]{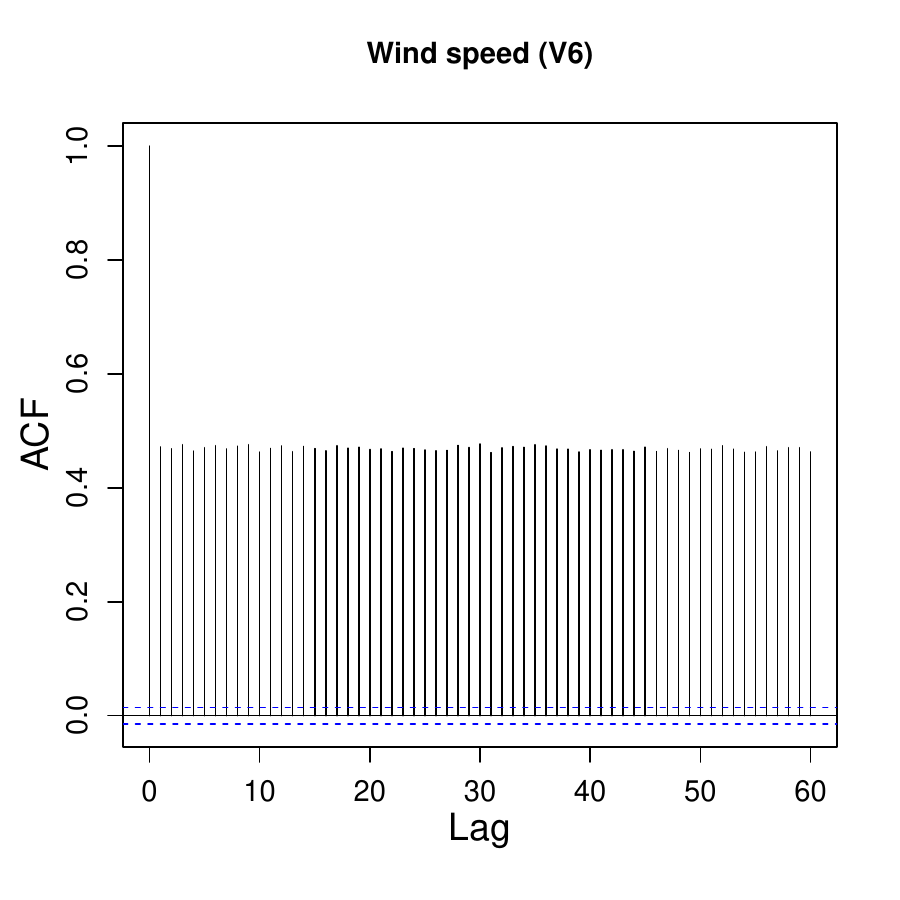}
    \includegraphics[width=0.245\textwidth]{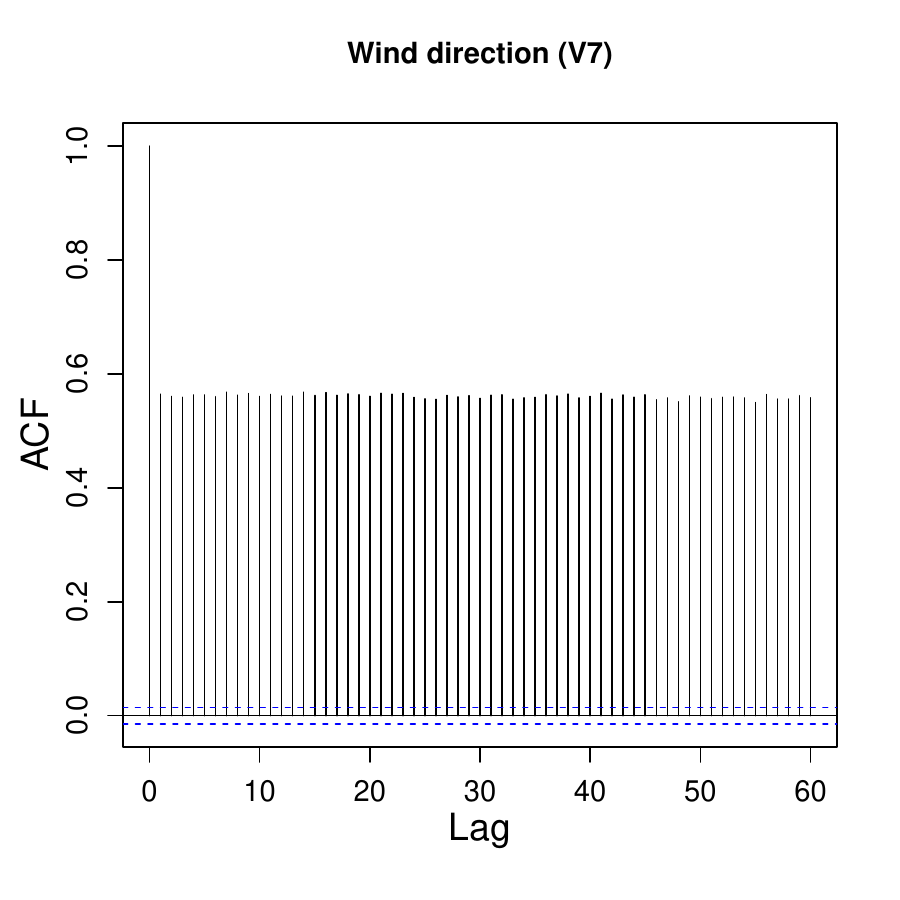}
    \includegraphics[width=0.245\textwidth]{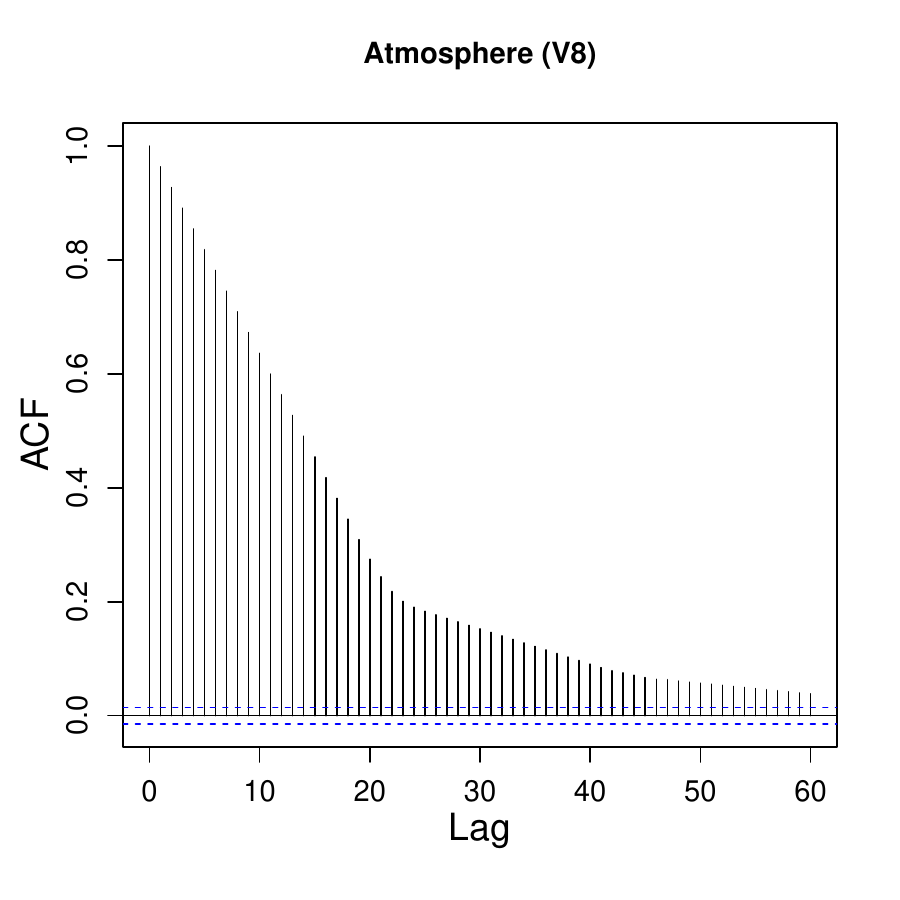}

    \caption{Autocorrelation function plots for the response variable $Y$ and explanatory variables $V1,\ldots,V4$, wind speed ($V6$), wind direction ($V7$) and atmosphere ($V8$), from top-left to bottom-right (by row).}
    \label{fig:acf_allcov}
\end{figure}

In this section, we present additional figures for Section 3 of the main paper, concerned with challenges C1 and C2. Figures~\ref{fig:EA_mnth}-\ref{fig:acf_allcov} support the exploratory analysis for challenges C1 and C2. We explore the within-year seasonality of the response variable $Y$ in Figure~\ref{fig:EA_mnth}, looking at the distribution of $Y$ per month and across the two seasons. This shows that there is a significant difference in the distribution of $Y$ between seasons 1 and 2, but within each season there is little difference across months. 

Figure~\ref{fig:scatter_allcov} shows a scatter plot of $Y$ against each covariate $V_1,\ldots,V_8$, excluding $V_6$ which corresponds to season. Covariates $V_1,V_2$ and $V_8$ do not seem to have a relationship with $Y$, whilst there seems to be dependence for the remaining covariates. These observed relationships appear complex and non-linear. 

\FloatBarrier

We also explore temporal dependence in Figure~\ref{fig:acf_allcov} that details the auto-correlation function (acf) values for the response $Y$ and explanatory variables $V_{1},\ldots,V_4,V_6,\ldots,V_8$, up to a lag of 60. All variables have negligible acf values beyond lag 0, except $V_6$ (wind speed), $V_7$ (wind direction) and $V_8$ (atmosphere).

\FloatBarrier

\begin{figure}[H]
    \centering
    \includegraphics[width=0.8\textwidth]{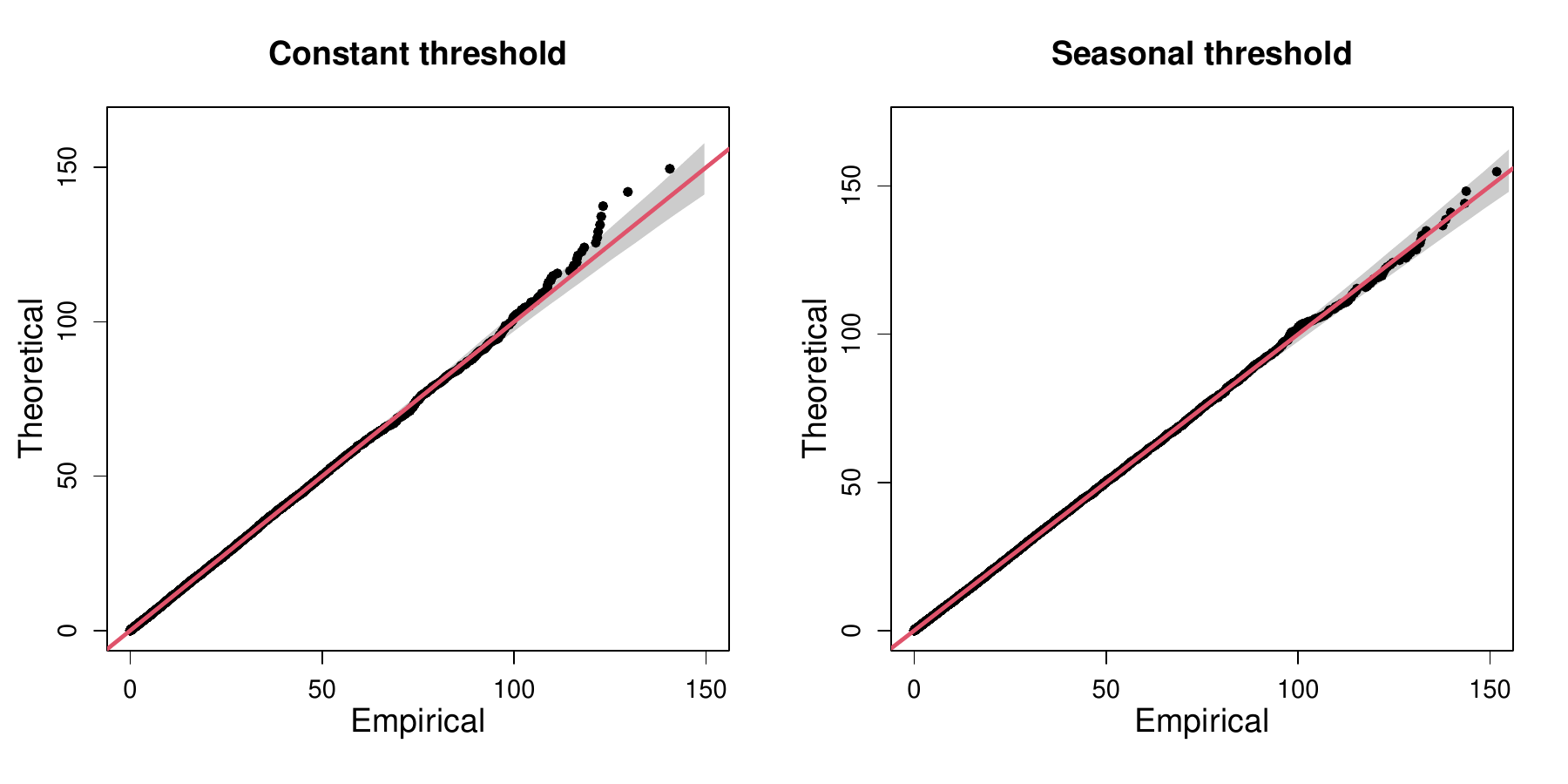}
    \caption{QQ-plots showing standard GPD model fits with 95\% tolerance bounds (grey) above a constant (left) and stepped-seasonal (right) threshold.}
    \label{fig: qqplot_const_vs_seasonal}
\end{figure}

\FloatBarrier
\newpage
Figure~\ref{fig: qqplot_const_vs_seasonal} shows the QQ-plots corresponding to a standard GPD model fitted to the excesses of $Y$ above a constant (left) and seasonally-varying threshold (right). 95\% tolerance bounds (grey) show a lack of agreement between observations and the standard GPD model above a constant threshold. The second plot demonstrates a significant improvement in model fit. 

\begin{figure}[h]
    \centering
    \includegraphics[width=\textwidth]{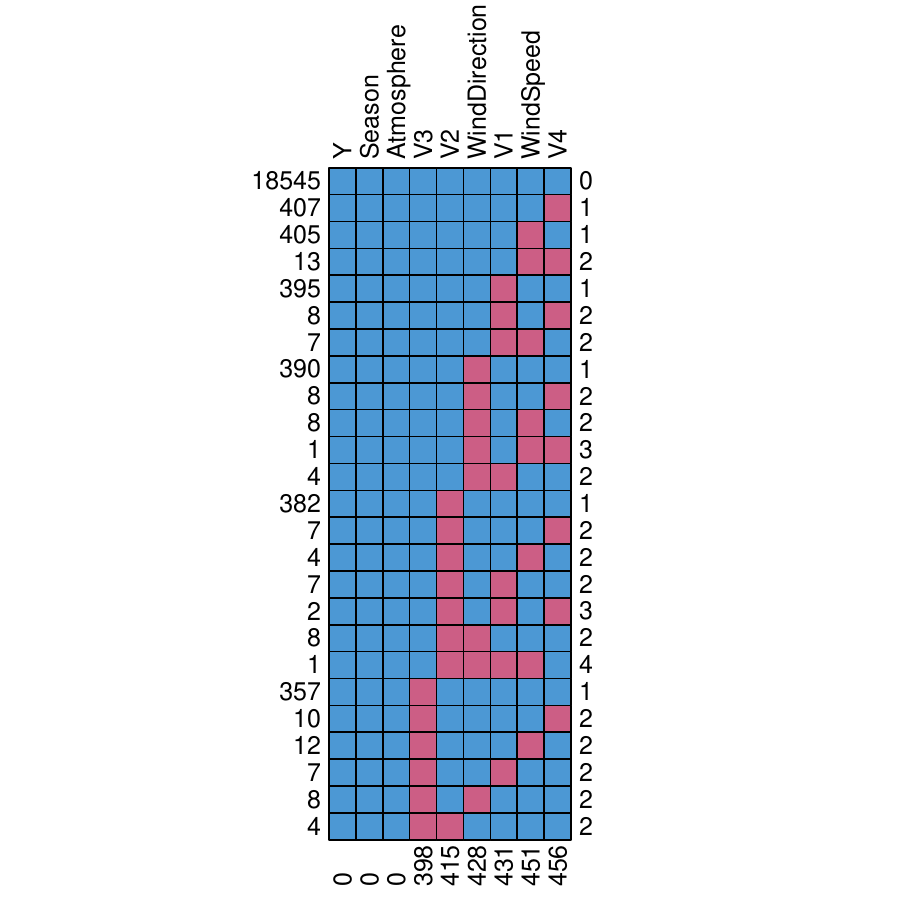}
    \caption{Detailed pattern of missing predictor variables in the Amaurot data set.}
    \label{fig:231106_Missing_Data_Summary}
\end{figure}

Figure \ref{fig:231106_Missing_Data_Summary} shows a detailed summary of the pattern of missing data in the data and can be produced using the \verb|missing_pattern| function in the \verb|finalfit| package in \verb|R| \citep{Finalfit2023}. To interpret the figure note that blue and red squares represent observed and missing variables, respectively. The number on the right indicates the number of missing predictor variables (i.e., the number of red squares in the row), while the number on the left is the number of observations that fall into the row category. On the bottom, we have the number of observations that fall into the column category. For example, 18,545 observations are fully observed (denoted by the first row); there are 407 observations where only $V4$ is missing (denoted by the second row), 13 observations where both $V4$ and $V6$ are missing (denoted by the fourth row), 456 observations where $V4$ and at least one other predictor is missing (denoted by the last column in the table), etc. It can be seen that there are very few observations where more than one predictor is missing.

\FloatBarrier

\section{Additional figures for Section 4}\label{appendix_S4}
\begin{figure}[h!]
    \centering
    \includegraphics[width=\textwidth]{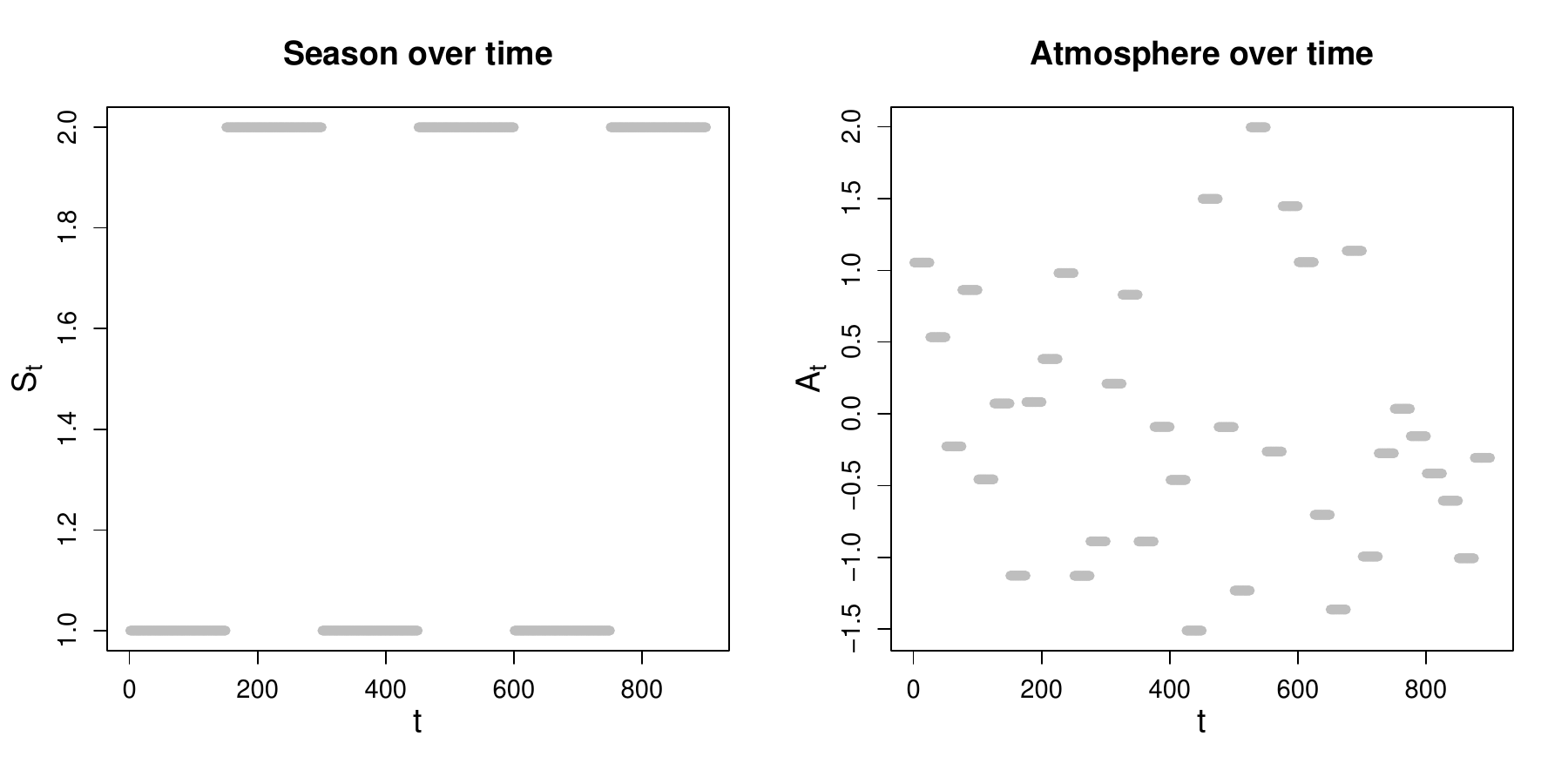}
    \caption{Plots of $S_t$ (left) and $A_t$ (right) against $t$ for the first $3$ years of the observation period.} %\CB{Move to supplementary? I feel like I've described these enough already}
    \label{fig:c3_data_illustration}
\end{figure}

In this section, we present additional plots related to Section 4 of the main article. Figure~\ref{fig:c3_data_illustration} illustrates the time series of both covariates for the first 3 years of the observation period. It can be seen how the seasons vary periodically over each year, as well as the discrete nature of the atmospheric covariate.

\begin{figure}[h!]
    \centering
    \includegraphics[width=\textwidth]{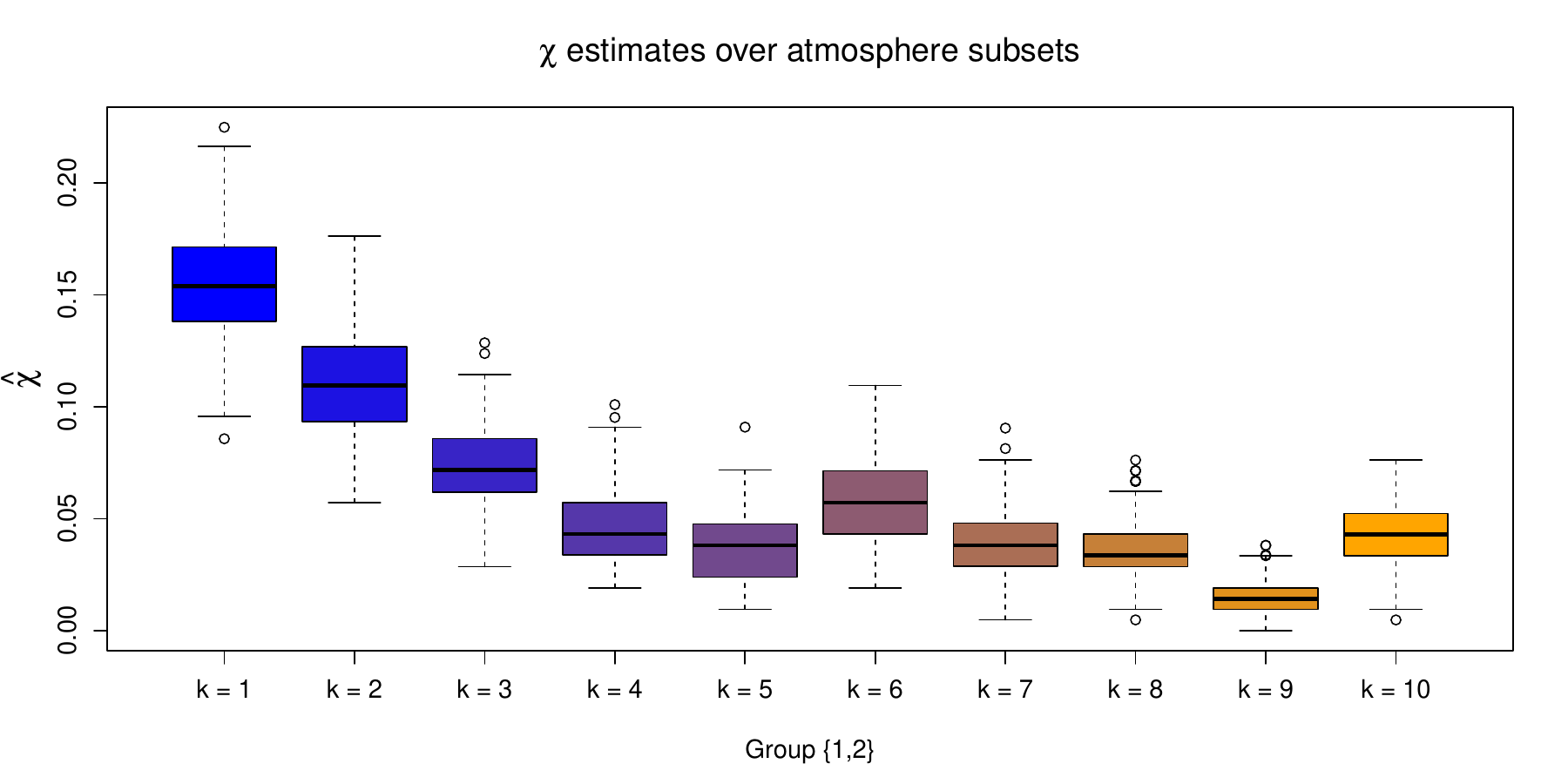}
    \caption{Boxplots of empirical $\chi$ estimates obtained for the subsets $G^A_{I,k}$, with $k = 1, \hdots, 10$ and $I=\{1,2\}$. The colour transition (from blue to orange) over $k$ illustrates the trend in $\chi$ estimates as the atmospheric values are increased.}  
    \label{fig:c3_chi_estimates_group1}
\end{figure}

\begin{figure}[h!]
    \centering
    \includegraphics[width=\textwidth]{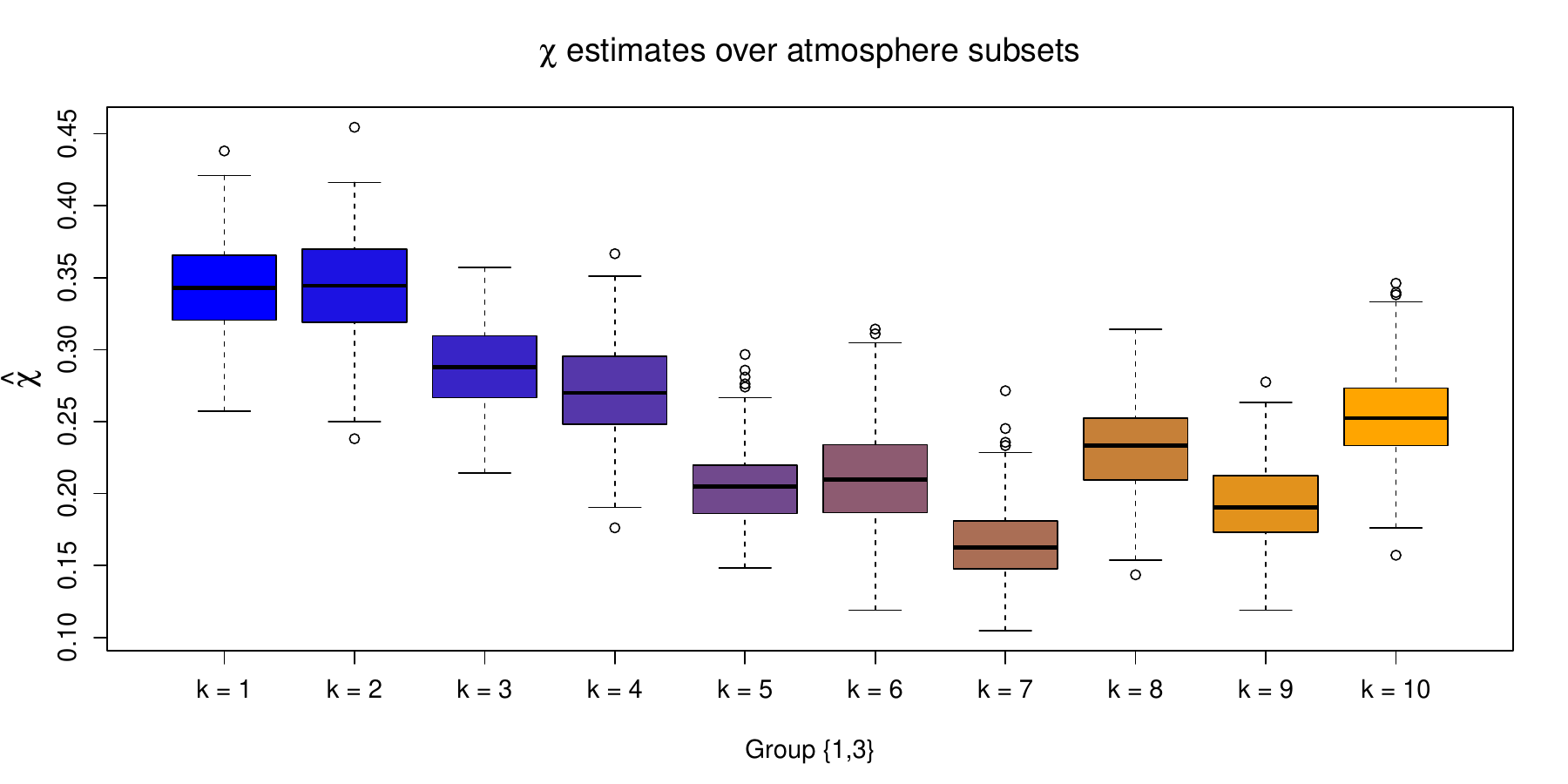}
    \caption{Boxplots of empirical $\chi$ estimates obtained for the subsets $G^A_{I,k}$, with $k = 1, \hdots, 10$ and $I=\{1,3\}$. The colour transition (from blue to orange) over $k$ illustrates the trend in $\chi$ estimates as the atmospheric values are increased.}  
    \label{fig:c3_chi_estimates_group2}
\end{figure}

\begin{figure}[h!]
    \centering
    \includegraphics[width=\textwidth]{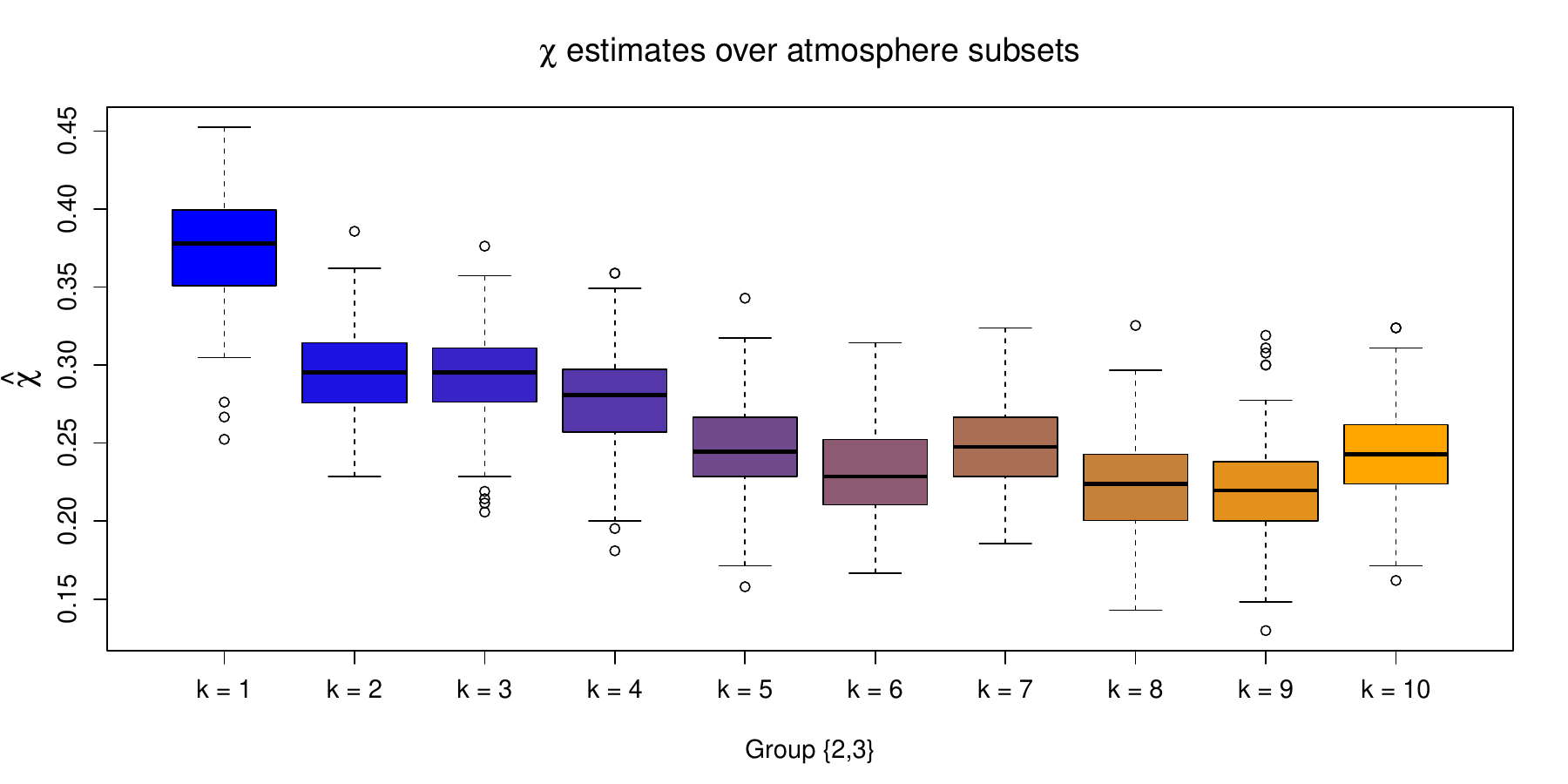}
    \caption{Boxplots of empirical $\chi$ estimates obtained for the subsets $G^A_{I,k}$, with $k = 1, \hdots, 10$ and $I=\{2,3\}$. The colour transition (from blue to orange) over $k$ illustrates the trend in $\chi$ estimates as the atmospheric values are increased.}  
    \label{fig:c3_chi_estimates_group3}
\end{figure}

\begin{figure}[h!]
    \centering
    \includegraphics[width=\textwidth]{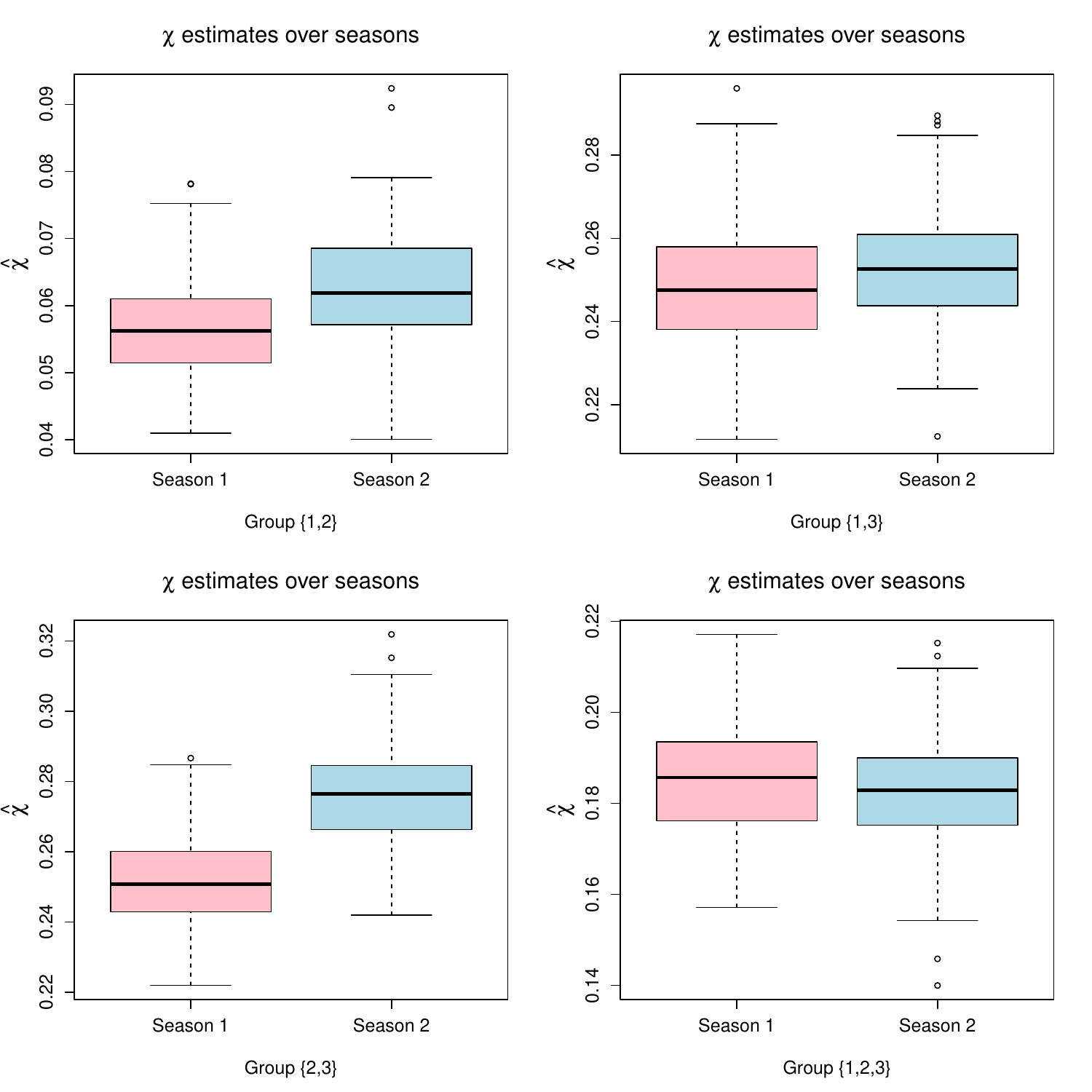}
    \caption{Boxplots of empirical $\chi$ estimates obtained for the subsets $G^S_{I,k}$, with $k = 1, 2$. In each case, pink and blue colours illustrate estimates for seasons 1 and 2, respectively. From top left to bottom right: $I=\{1,2,3\}$, $I=\{1,2\}$, $I=\{1,3\}$, $I=\{2,3\}$.}  
    \label{fig:chi_seasons}
\end{figure}

\newpage
Bootstrapped $\chi$ estimates for the groups $G^A_{I,k}, k\in \{1,\hdots,10\}$, $I \in \mathcal{I}\setminus \{1,2,3\}$ and $G^S_{I,k}, k\in \{1,2\}$, $I \in \mathcal{I}$ are given in Figures \ref{fig:c3_chi_estimates_group1} - \ref{fig:chi_seasons}. These estimates illustrate the impact of atmosphere on the dependence structure.

\FloatBarrier

For a $3$-dimensional random vector, the angular dependence function, denoted $\lambda$, is defined on the unit-simplex $\boldsymbol{S}^2$ and describes extremal dependence along different rays $\boldsymbol{\omega} \in \boldsymbol{S}^2$. As noted in Section 4.2 of the main manuscript, we can associate each of the probabilities from C3, $p_1$ and $p_2$, with points on $\boldsymbol{S}^2$, denoted $\boldsymbol{\omega}^1$ and $\boldsymbol{\omega}^2$ respectively. With $I=\{1,2,3\}$, we consider $\lambda(\boldsymbol{\omega}^1)$ and $\lambda(\boldsymbol{\omega}^2)$ over the subsets $G^S_{I,k}, \; k\in\{1,2\}$ and $G^A_{I,k}, \; k\in\{1,\hdots,10\}$. We note that $\lambda(\boldsymbol{\omega}^1)$ is analogous with the coefficient of tail dependence $\eta \in (0,1]$ \citep{Ledford1996}, with $\eta = 1/3\lambda(\boldsymbol{\omega}^1)$; this corresponds with the region where all variables are simultaneously extreme. Furthermore, $\lambda(\boldsymbol{\omega}^2)$, which corresponds to a region where only two variables are extreme, is only evaluated after an additional marginal transformation of the third Coputopia time series; see Section 4.2 of the main manuscript. 

\begin{figure}[h!]
    \centering
    \includegraphics[width=\textwidth]{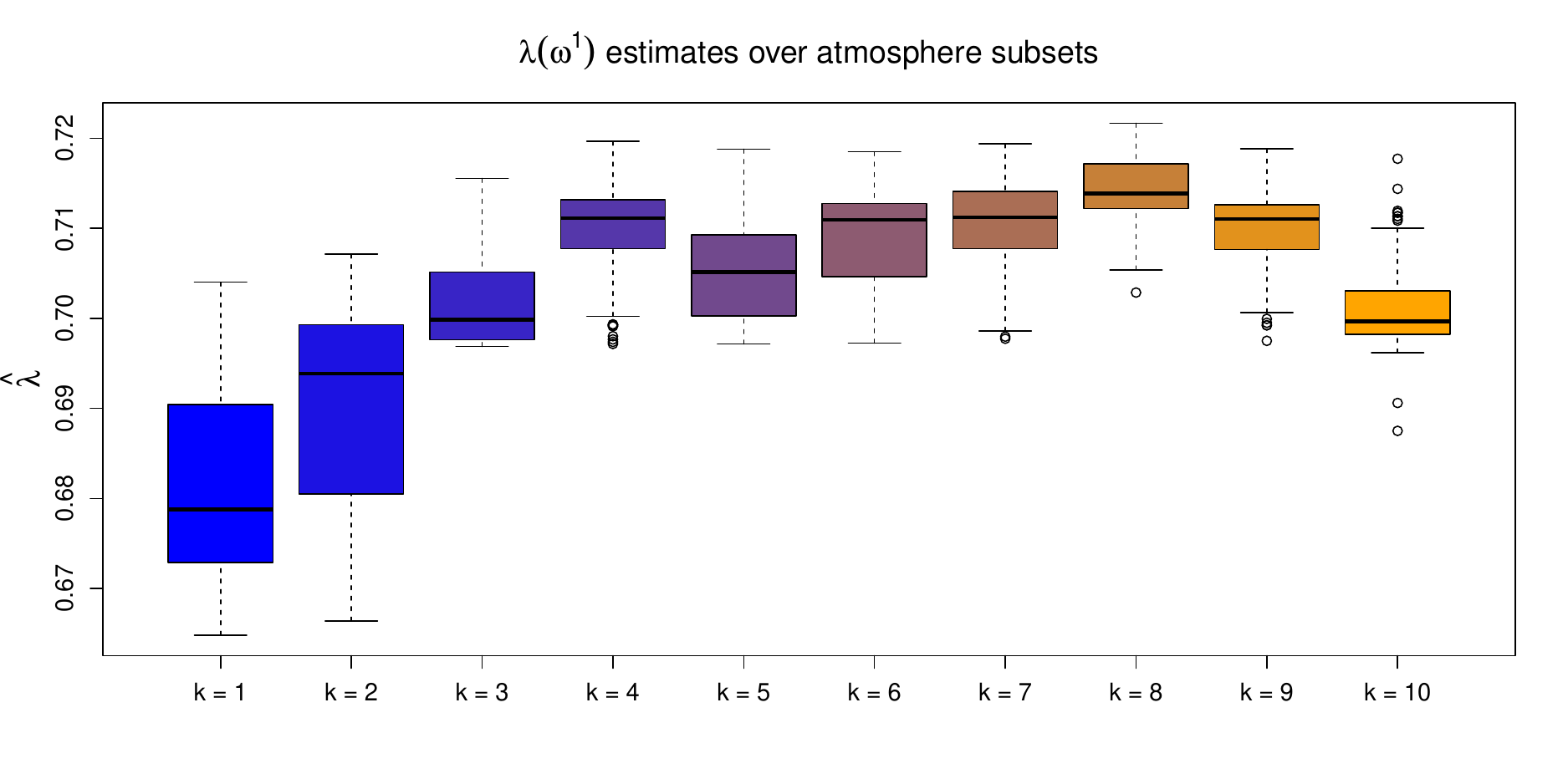}
    \caption{Boxplots of empirical $\lambda(\boldsymbol{\omega}^1)$ estimates obtained for the subsets $G^A_{I,k}$, with $k = 1, \hdots, 10$ and $I=\{1,2,3\}$. The colour transition (from blue to orange) over $k$ illustrates the trend in $\lambda$ estimates as the atmospheric values are increased.}  
    \label{fig:c3_atmos_lam_i}
\end{figure}

\begin{figure}[h!]
    \centering
    \includegraphics[width=.6\textwidth]{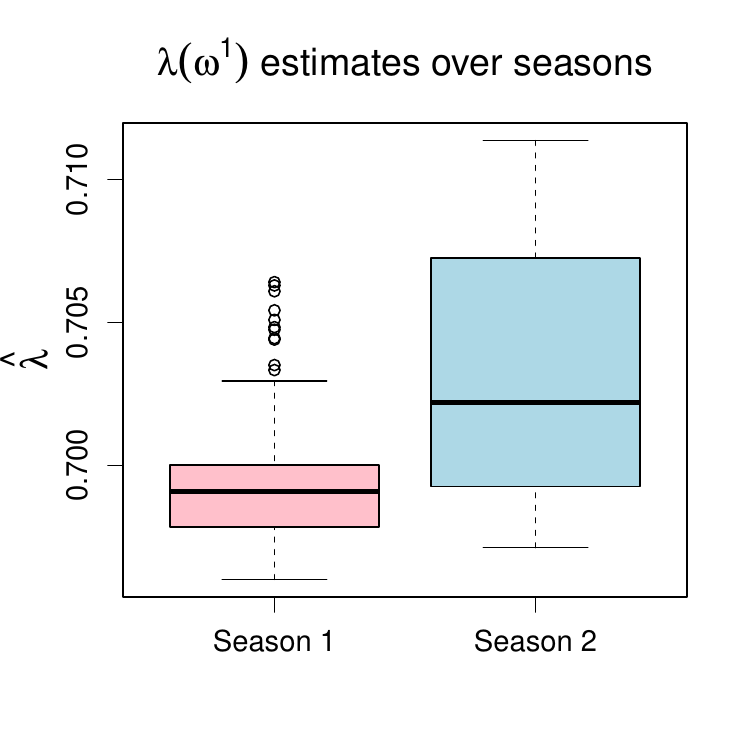}
    \caption{Boxplots of empirical $\lambda(\boldsymbol{\omega}^1)$ estimates obtained for the subsets $G^S_{I,k}$, with $k = 1, 2$ and $I=\{1,2,3\}$. In each case, pink and blue colours illustrate estimates for seasons 1 and 2, respectively.}  
    \label{fig:c3_season_lam_i}
\end{figure}

\begin{figure}[h!]
    \centering
    \includegraphics[width=\textwidth]{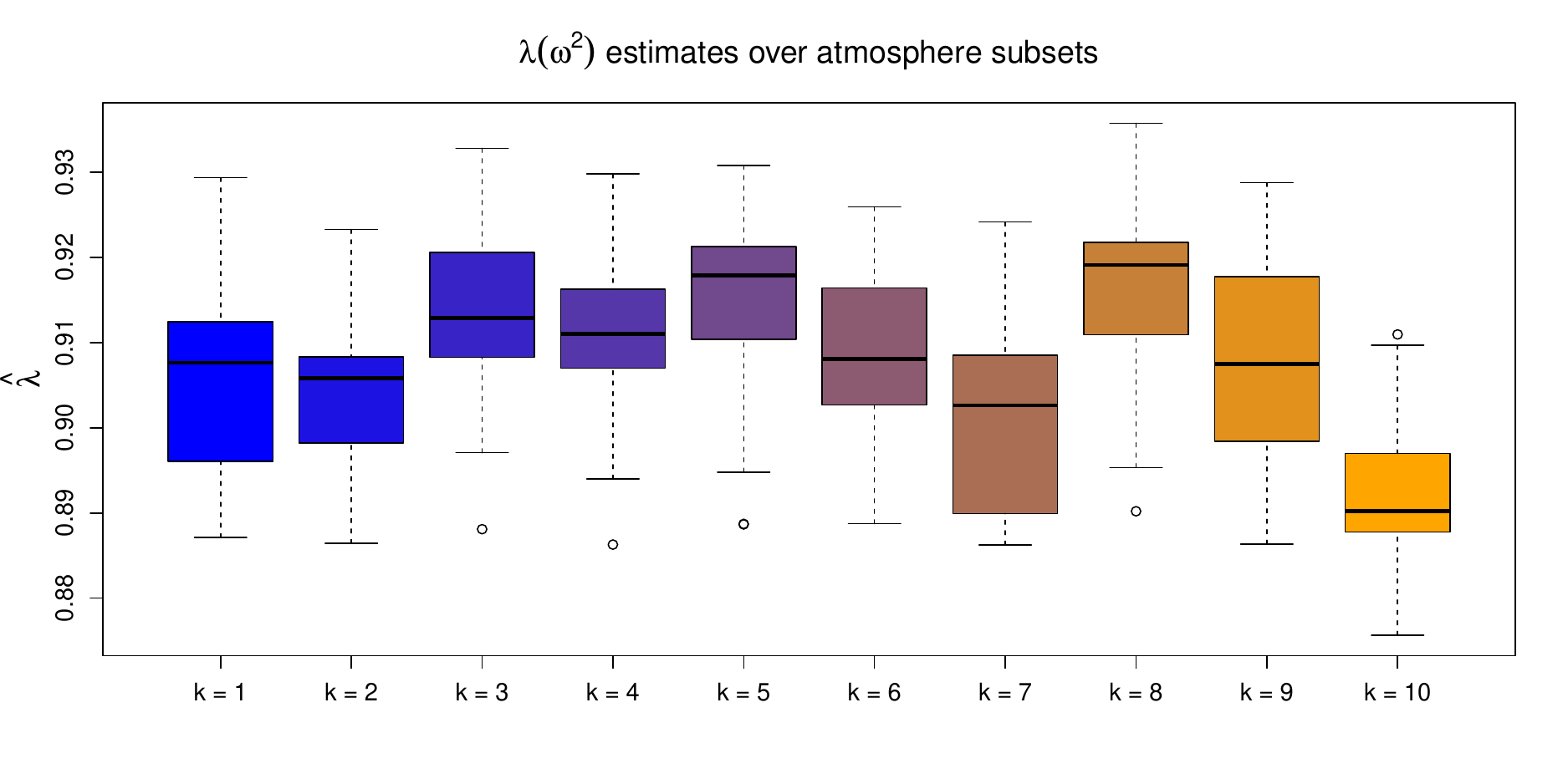}
    \caption{Boxplots of empirical $\lambda(\boldsymbol{\omega}^2)$ estimates obtained for the subsets $G^A_{I,k}$, with $k = 1, \hdots, 10$ and $I=\{1,2,3\}$. The colour transition (from blue to orange) over $k$ illustrates the trend in $\lambda$ estimates as the atmospheric values are increased.}  
    \label{fig:c3_atmos_lam_ii}
\end{figure}

\begin{figure}[h!]
    \centering
    \includegraphics[width=.6\textwidth]{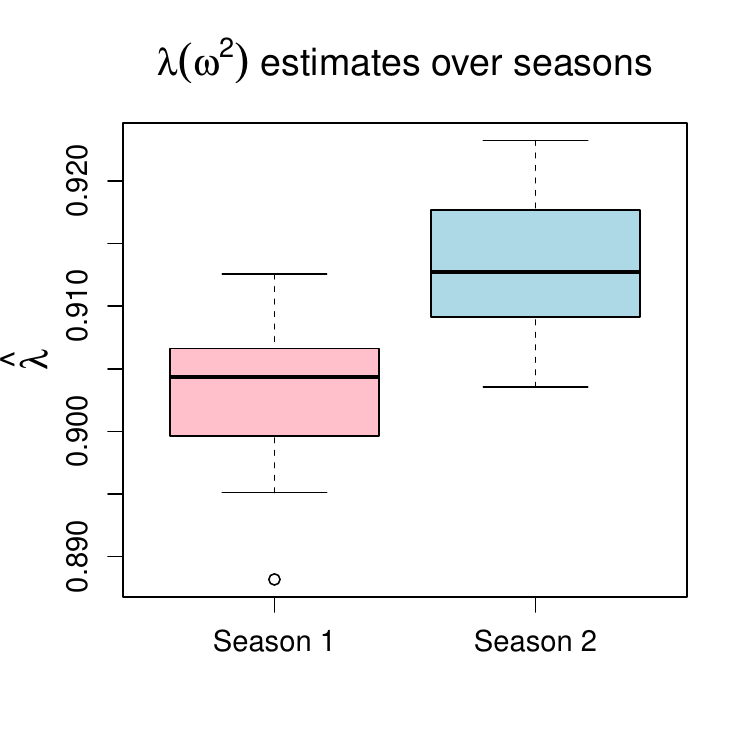}
    \caption{Boxplots of empirical $\lambda(\boldsymbol{\omega}^2)$ estimates obtained for the subsets $G^S_{I,k}$, with $k = 1, 2$ and $I=\{1,2,3\}$. In each case, pink and blue colours illustrate estimates for seasons 1 and 2, respectively.}  
    \label{fig:c3_season_lam_ii}
\end{figure}

Estimation of $\lambda$ for each simplex point and subset was achieved using the Hill estimator \citep{Hill1975} at the $90\%$ level, with uncertainty subsequently quantified via bootstrapping. These results are given in Figures \ref{fig:c3_atmos_lam_i} - \ref{fig:c3_season_lam_ii}. These plots provide further evidence of a relationship between the extremal dependence structure and the covariates. 

% From these plots, one can observe clear trends in $\lambda$ over both covariates. Note that different trends were observed over atmosphere for $\lambda(\boldsymbol{\omega}^1)$ and $\lambda(\boldsymbol{\omega}^2)$, with the former trend appearing more pronounced. Furthermore, the observations for $\boldsymbol{\omega}^1$ were in close agreement with what was observed for $\chi$. 

% \textcolor{red}{AF[is this meant to be $\lambda(\boldsymbol{\omega}_i)$? Should it be $\lambda(\boldsymbol{\omega}^{1})$]}

% \textcolor{red}{See above comment}

\begin{figure}[h!]
    \centering
    \includegraphics[width=.8\textwidth]{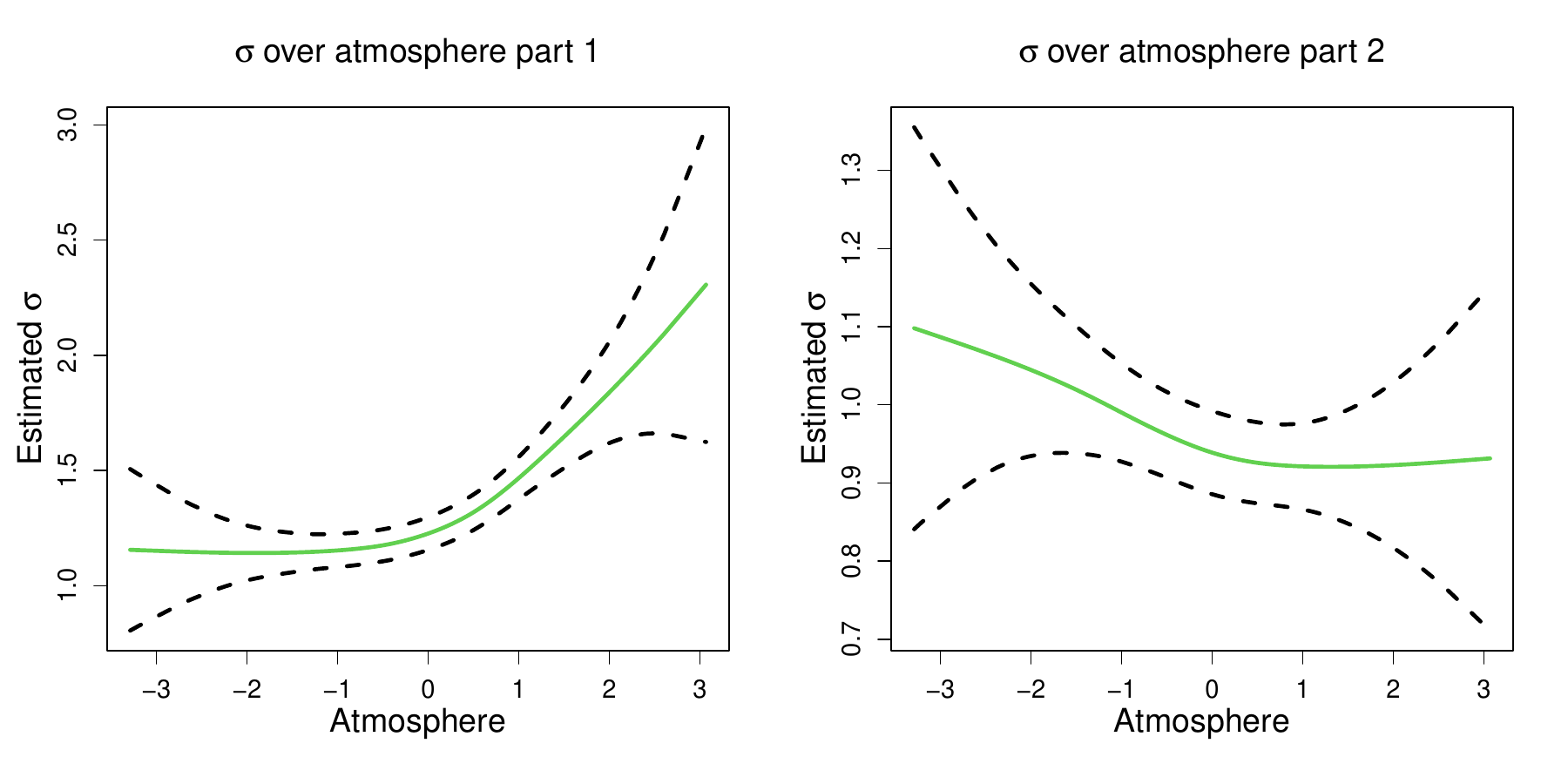}
    \caption{Estimated $\sigma$ functions (green) over atmosphere for part 1 (left) and 2 (right). In both cases, the regions defined by the black dotted lines represent 95\% confidence intervals obtained using posterior sampling.} %\CB{Maybe also move this to supplementary to save space? Or combine with previous figure?}\LA{I think having read the paper until now, this should go to the SM}
    \label{fig:c3_sigma_estimates}
\end{figure}

To illustrate the estimated trend in dependence, Figure \ref{fig:c3_sigma_estimates} shows the estimated scale functions, $\sigma\left(\boldsymbol{\omega} \mid \boldsymbol{x}_t\right)$, over atmosphere for parts 1 and 2. Under the assumption of asymptotic normality in the spline coefficients, 95\% confidence intervals are obtained via posterior sampling; see \citet{Wood2017} for more details. We observe that $\sigma$ tends to increase and decrease over atmosphere for parts 1 and 2, respectively, although the trend is less pronounced for the latter. Under our modelling framework, we note that higher values of $\sigma$ are associated with less positive extremal dependence in the direction $\boldsymbol{\omega}$ of interest; to see this, observe that the survivor function of the GPD with fixed $\xi$ is negatively associated with $\sigma$. Considering the trend in $\sigma\left(\boldsymbol{\omega} \mid \boldsymbol{x}_t\right)$, our results indicate a decrease in dependence in the region where all variables are extreme. 

% We also remark that the estimated GPD shape parameters obtained for parts 1 and 2 were $0.042\, (0.01,0.075)$ and $0.094\,(0.059,0.128)$, respectively, where uncertainty intervals were again obtained using posterior sampling. These estimates, which indicate slight heavy-tailed behaviour within the min-projection variable, providing further insight into why the original exponential modelling framework, described in equation \eqref{eqn:ns_wt_model}, was is not appropriate for C3.  

\clearpage

\section{Additional figures for Section 5}\label{appendix_S5}

\begin{figure}[h]
      \centering
      \includegraphics[width=0.6\textwidth,keepaspectratio]{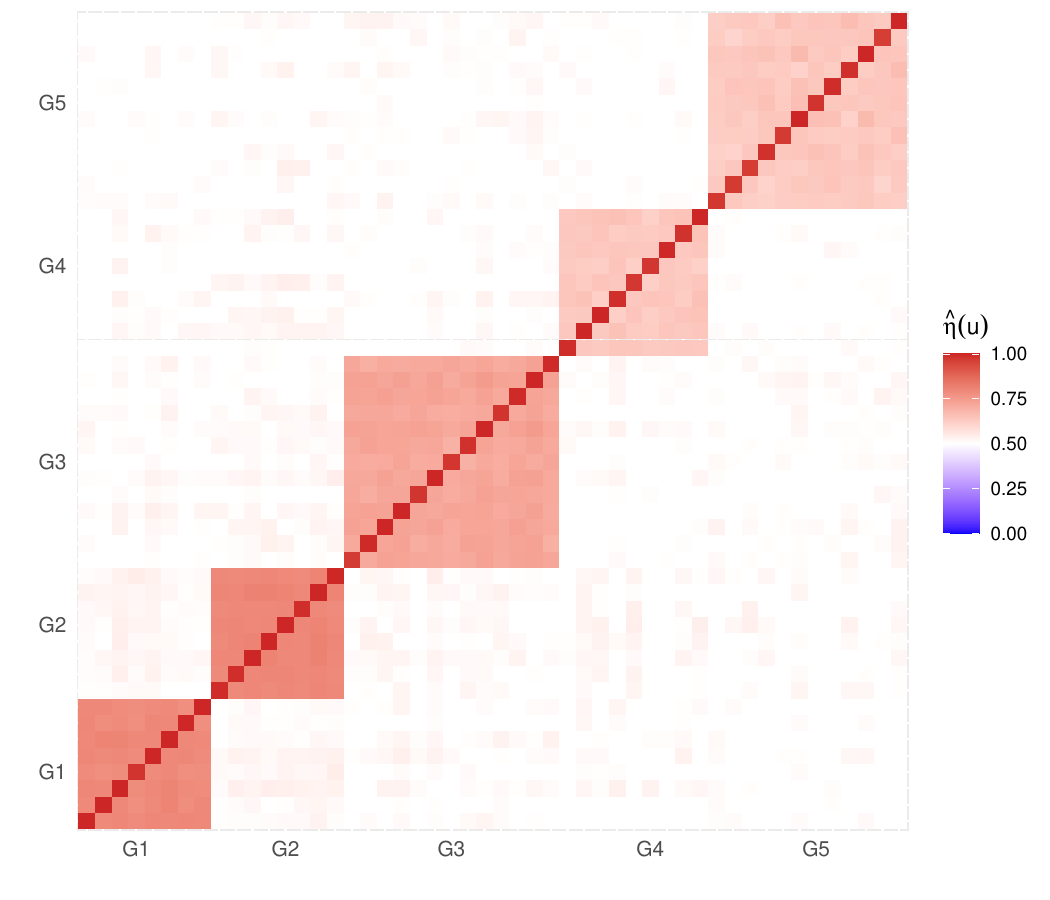}
      \caption{Heat map of estimated empirical pairwise $\eta(u)$ extremal dependence coefficients with $u=0.95.$}
      \label{fig:eta_heat.pdf}
\end{figure}

In this section, we present additional plots related to Section 5 of the main article an we refer to $p_1$ and $p_2$ as parts 1 and 2 of C4, respectively. Figure \ref{fig:eta_heat.pdf} shows a heat map of empirically estimated $\eta(\cdot)$ dependence coefficients and provides further evidence of the existence of the $5$ dependence subgroups identified in our exploratory analysis for challenge C4. It also suggests that our modelling assumptions are reasonable; specifically that  there is in-between group independence, and that the extremes within each group do not occur simultaneously.

\begin{figure}[h]
    \centering
    \includegraphics[width=1\textwidth]{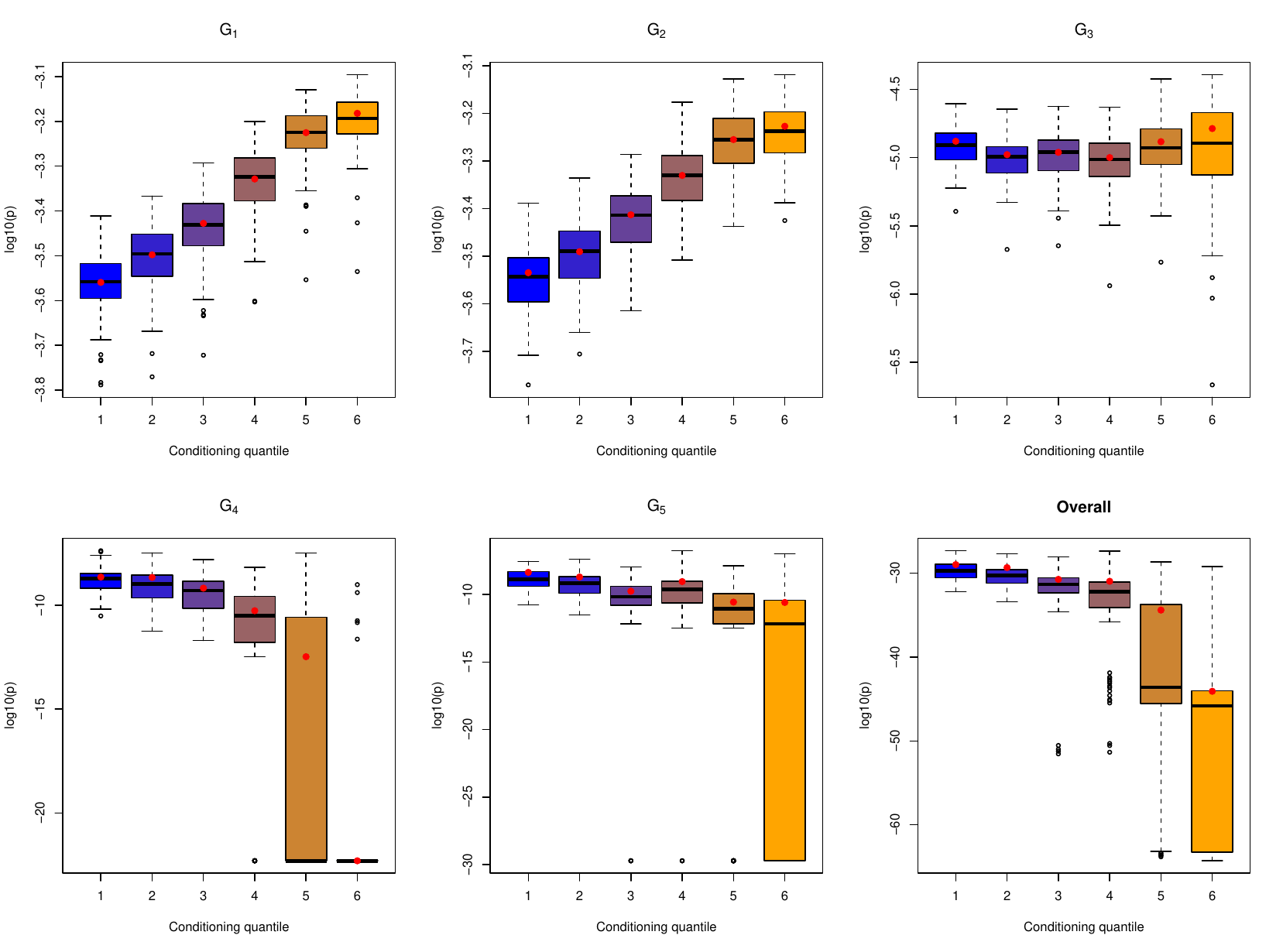} 
    \caption{ Part 1 subgroup and overall bootstrapped probability estimates on the log scale. The red points indicate the original sample estimates and the colouring of the boxplots indicates the choice of conditioning threshold, with the conditioning quantile indices 1-6 referring to the quantile levels $\{0.7,0.75,0.8,0.85,0.9,0.95\}$, respectively.}
    \label{fig:c4partiresults}
\end{figure}

\begin{figure}[h]
    \centering
    \includegraphics[width=1\textwidth]{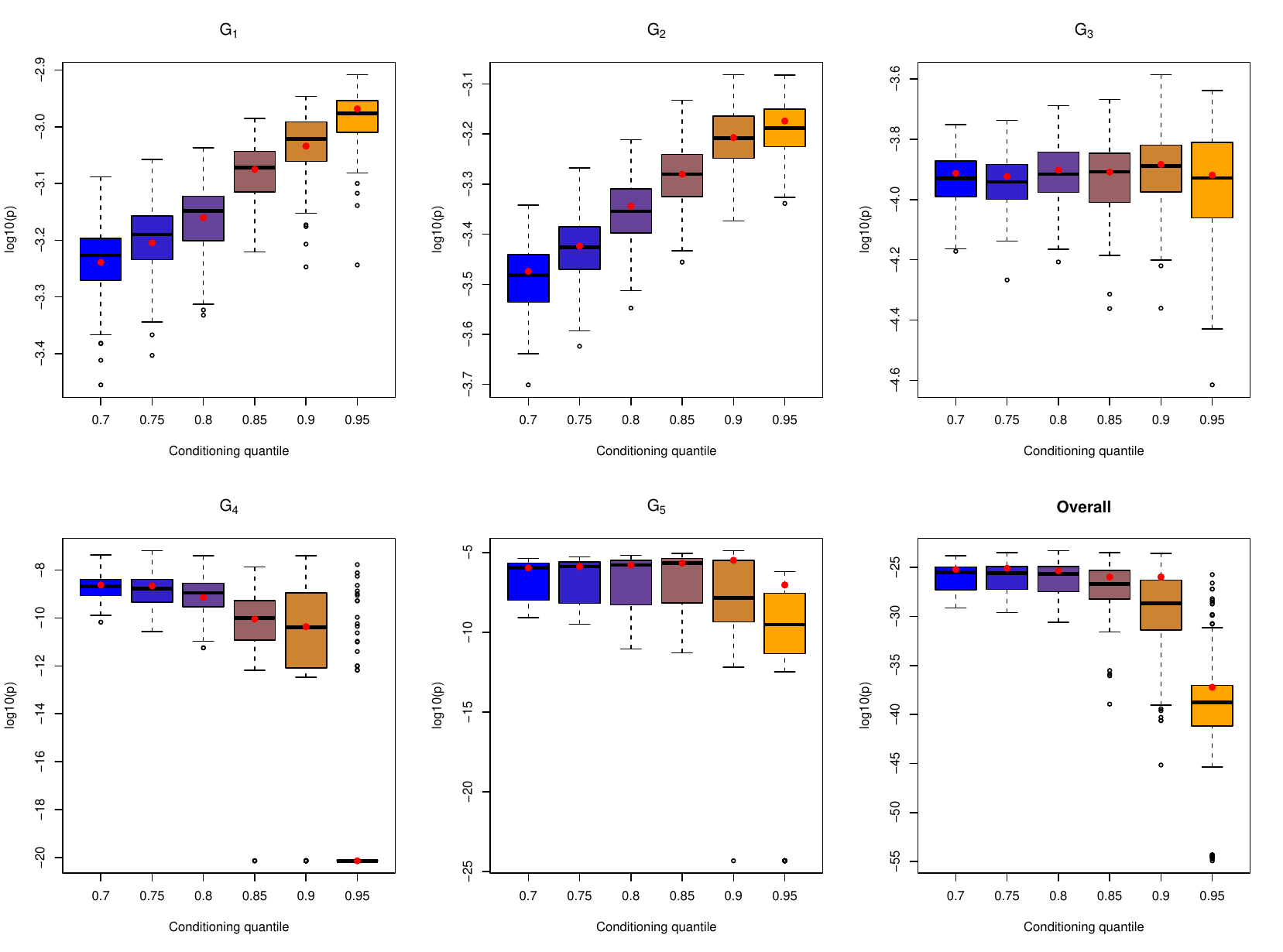}
\caption{Part 2 subgroup and overall bootstrapped probability estimates on the log scale for C4. The red points indicate the original sample estimates and the colouring of the boxplots indicates the choice of conditioning threshold, with the conditioning quantile indices 1-6 referring to the quantile levels $\{0.7,0.75,0.8,0.85,0.9,0.95\}$, respectively.}
    \label{fig:c4partiilogprob}
\end{figure}

\noindent Figure \ref{fig:c4partiresults} shows the bootstrapped estimated individual group and overall probabilities with respect to conditioning threshold quantile for part 1 of challenge C4. Similarly, Figure \ref{fig:c4partiilogprob} shows the bootstrapped estimated individual group and overall probabilities with respect to conditioning threshold quantile for part 2 of challenge C4.

\end{document}